\RequirePackage{fix-cm}

\documentclass[journal]{IEEEtran}

\usepackage{graphicx}
\usepackage{mathptmx}      % use Times fonts if available on your TeX system
\usepackage{amsmath}
\usepackage{url}

\usepackage{lipsum}
\usepackage{color}
\usepackage{graphicx}
\usepackage{ragged2e}
\usepackage{cite}
\usepackage{footnote}
\makesavenoteenv{tabular}
\usepackage[caption=false,font=footnotesize]{subfig}
\usepackage{rotating}
\begin{document}

\title{Compressed-domain visual saliency models: A comparative study}

\author{Sayed Hossein Khatoonabadi  \and \\
        Ivan V. Baji\'{c}	\and \\
				Yufeng Shan
\thanks{This work was supported in part by the Cisco Research Award CG\# 573690 and NSERC Grant RGPIN 327249.}}


\author{Sayed Hossein Khatoonabadi, Ivan V. Baji\'{c}, and Yufeng Shan
\thanks{S. H. Khatoonabadi and I. V. Baji\'{c} are with the School of Engineering Science, Simon Fraser University,
Burnaby, BC, Canada. E-mail: skhatoon@sfu.ca, ibajic@ensc.sfu.ca}
\thanks{Y. Shan is with Cisco Systems, Boxborough, MA, USA. E-mail: yshan@cisco.com}
\thanks{This work was supported by the Cisco Research Award CG\# 573690.}}

\maketitle

\begin{abstract}
Computational modeling of visual saliency has become an important research problem in recent years, with applications in video quality estimation, video compression, object tracking, retargeting, summarization, and so on. While most visual saliency models for dynamic scenes operate on raw video, several models have been developed for use with compressed-domain information such as motion vectors and transform coefficients. This paper presents a comparative study of eleven such models as well as two high-performing pixel-domain saliency models on two eye-tracking datasets using several comparison metrics. The results indicate that highly accurate saliency estimation is possible based only on a partially decoded video bitstream.
The strategies that have shown success in compressed-domain saliency modeling are highlighted, and certain challenges are identified as potential avenues for further improvement. 
\keywords{Visual saliency, fixation points, compressed-domain processing, motion vectors}
\end{abstract}

%%%%%%%%%%%%%%%%%%%%%%
% Section I
%%%%%%%%%%%%%%%%%%%%%%
\section{Introduction}

Visual saliency estimation is a process of finding certain parts in an image (or video) that are likely to draw attention compared to their spatial (and temporal) surroundings. The Human Visual System (HVS) is able to automatically shift the focus of attention to salient regions in the pre-attentive, early vision phase. This ability allows the brain to restrict high-level processing of a scene to a relatively small part at any given time. Many models have been introduced based on physiological and psychophysical findings to imitate the HVS in order to predict human visual attention~\cite{itti98model}. Visual saliency models find a large number of applications in image processing and computer vision, such as quality assessment~\cite{wang11information,culibrk11salient,feng11saliency,liu11visual,larson08can,moorthy09visual}, compression~\cite{itti04foveation,wang2005foveated,guo10novel,li11vagba,hadizadeh11saliency,hadizadeh14saliency}, retargeting~\cite{fang12saliency,lu10video}, segmentation~\cite{mishra12active,fukuchi09saliency}, object recognition~\cite{han10biologically}, object tracking~\cite{mahadevan13biologically}, abstraction~\cite{ji12video}, guiding visual attention~\cite{hagiwara11guiding,mateescu13guiding}, and so on. 

Many computational models have been introduced during the past 25 years to estimate visual saliency. Despite the existence of numerous models, their high computational complexity is a serious drawback when it comes to practical applications that need to run in real time, or for devices with restricted complexity and memory requirements, such as mobile devices. One way to reduce the computational cost of saliency estimation is to use compressed-domain features, such as motion vectors (MVs), motion-compensated prediction residuals or their transform coefficients, and so on. This way, part of decoding can be avoided, a smaller amount of data needs to be processed compared to pixel-domain methods, and some of the information produced during encoding (e.g., MVs and transform coefficients) can be reused~\cite{khatoonabadi13video}. Compressed-domain algorithms for visual saliency estimation have been developed for various applications such as image retargeting~\cite{fang12saliency}, video transcoding~\cite{xie08region,liu09motion,sinha04region}, quality estimation~\cite{lin12no}, video retrieval~\cite{ma01new}, video skimming~\cite{ma02model}, salient motion detection~\cite{muthuswamy13salient}, and so on. Although there are relatively few compressed-domain saliency models compared to their pixel-domain counterparts, their potential for practical deployment makes them an important research topic. 

The purpose of this paper is to provide a comprehensive comparison among compressed-domain visual saliency models for video, similar to what has been done for pixel-domain models in~\cite{borji13quantitative}. The present paper is an extension of our preliminary study in~\cite{khatoonabadi14comparison} and takes into consideration two well-known pixel-domain algorithms as benchmarks for comparison, as well as two more recent compressed-domain algorithms that have appeared since~\cite{khatoonabadi14comparison}. It also provides a more extensive comparison involving a larger number of videos from two ground truth data sets, as well as a number of different accuracy metrics. In the literature, existing models have been developed for different applications and their evaluation was based on different datasets and quantitative criteria. Furthermore, models are often tailored to a particular video coding standard, and the encoding parameter settings used in the evaluation are often not reported. All of this makes a fair and comprehensive comparison more challenging. To enable meaningful comparison, in this work we reimplemented all compared methods on the same platform, and evaluated them under the same encoding conditions on two popular eye-tracking datasets. A number of different metrics has been employed in the comparison in order to illuminate various aspects of the models' performance. The results of the comparison indicate which strategies seem promising in the context of compressed-domain saliency estimation for video, and point the way towards improving existing models and developing new ones. 
Last but not least, this study has been performed in a reproducible research manner~\cite{vandewalle09reproducible}. The MATLAB code and data used in this study are available online~\cite{CSEsupplementary}.

The paper is organized as follows. Section~\ref{sec:models_data} reviews the visual saliency models used in the study and the two ground truth gaze point datasets. Section~\ref{sec:evaluation} describes the evaluation framework, including the accuracy metrics employed in the evaluation and the procedures used to correct for center bias and border effects. Section~\ref{sec:results} presents the results of the evaluation, while Sections~\ref{sec:discussion} and~\ref{sec:conclusions} provide discussion and conclusions, respectively.

%%%%%%%%%%%%%%%%%%%
% Section II
%%%%%%%%%%%%%%%%%%%
\section{Models and Data}
\label{sec:models_data}

Our study includes eleven compressed-domain saliency models. Their performance is compared amongst themselves, and also against two high-performing pixel-domain models in order to gain insight into the relationship between the accuracy of the current state of the art in pixel-domain and compressed-domain saliency estimation for video. Among the pixel-domain models, we chose AWS (Adaptive Whitening Saliency)~\cite{AWS}, which takes only spatial information into account, and GBVS (Graph-Based Visual Saliency)~\cite{harel07graph} with DIOFM channels (DKL-color, Intensity, Orientation, Flicker, and Motion), which takes both spatial and temporal information into account for estimating the saliency. AWS is frequently reported as one of the top performing models on still natural images~\cite{borji13quantitative,kim13visual}. GBVS is another well-known model, often used as a benchmark for comparison. Since MATLAB implementations of both these models are available, it makes the computational comparison with MATLAB implementations of compressed-domain models meaningful.
This section briefly describes the eleven compressed-domain visual saliency models included in the study and the datasets used to evaluate them. 

%%%%%%%%%%%%%%%%%%%%%%%%%%%%%%%%%%%%%%%%%%%%%%%%%%%%%
\subsection{Compressed-Domain Visual Saliency Models}
\label{sec:models}
%%%%%%%%%%%%%%%%%%%%%%%%%%%%%%%%%%%%%%%%%%%%%%%%%%%%%
In this study, our goal is to evaluate visual saliency models for video that have been designed explicitly for, or have the potential to work in, the compressed domain. This means that they should operate with the kind of information found in a compressed video bitstream, such as block-based Motion Vector Field (MVF), prediction residuals or their transforms, block coding modes, etc. We surveyed the literature on the topic and found eleven prominent models listed in Table~\ref{tab:Models}, sorted according to the publication year. Different models assume different coding standards, for example MPEG-1, MPEG-2, MPEG-4 SP (Simple Profile), MPEG-4 ASP (Advanced Simple Profile), and MPEG-4 part 10, better known as H.264/AVC (Advanced Video Coding). For each model, the data used from the compressed bitstream, their intended application, as well as data and evaluation method, if any, are also included in the table. As seen in the table, only a few of the most recent models have been evaluated using gaze data from eye-tracking experiments, which is thought to be the ultimate test for a visual saliency model. This fact makes the present study all the more relevant. 
Interested readers are referred to the supplementary material~\cite{CSEsupplementary} for a brief description of various models used in the study.

In addition to the visual saliency models described above, two benchmark models were used in the evaluation: IO and GAUSS. These are derived from the ground truth data itself and will be described in Section~\ref{sec:benchmark_models}, after the two eye-tracking datasets employed in the study are introduced. 

\begin{table*}
%\begin{sidewaystable}

%{\vskip 11cm}

\renewcommand{\arraystretch}{2}
	\centering
	\footnotesize \addtolength{\tabcolsep}{-3pt}
	
	\caption{\small Compressed-domain visual saliency models included in the study \newline 
	(MVF: Motion Vector Field;
	DCT-R: Discrete Cosine Transformation of residual blocks; DCT-P: Discrete Cosine Transformation of pixel blocks; OBDL: Operational Block Description Length; KLD: Kullback-Leibler Divergence; AUC: Area Under Curve; ROC: Receiver Operating Characteristic; NSS: Normalized Scanpath Saliency; JSD: Jensen-Shannon Divergence)}
	%\begin{sideways}
	%\begin{rotate}{-90}
	
	\begin{tabular}{cccccccccc}
	\hline \hline
	\textbf{\#} & \textbf{Model} & \textbf{First Author} & \textbf{Year} & \textbf{Codec} & \textbf{Data} & \textbf{Application} & \textbf{Sequences} & \textbf{Gaze data} & \textbf{Metric(s)} \\ 
	\hline \hline
	1 & \textbf{PMES}    & Ma~\cite{ma01new}  & 2001 & MPEG-1/2 & MVF & Video Retrieval & MPEG-7~\cite{Mpeg7output} & -  & - \\ 
	2 & \textbf{MAM}     & Ma~\cite{ma02model} & 2002 & MPEG-1/2 & MVF & Video Skimming & Specific~\cite{ma02model}  & - & Human Score \\
	3 & \textbf{PIM-ZEN} & Agarwal~\cite{agarwal03fast}   & 2003 & MPEG-1/2 & MVF+DCT-R & ROI Detection & QCIF Standard & - & - \\
	4 & \textbf{PIM-MCS} & Sinha~\cite{sinha04region} & 2004 & MPEG-4 SP  & MVF+DCT-R & Video Transcoding  & QCIF Various &  - & - \\
	5 & \textbf{MCSDM}   & Liu~\cite{liu09motion} & 2009 & H.264/AVC  & MVF & Rate Control & QCIF Standard & - & - \\
	6 & \textbf{GAUS-CS} & Fang~\cite{fang12video} & 2012 & MPEG-4 ASP & MVF+DCT-P & Saliency Detection & CRCNS~\cite{itti06bayesian},\cite{itti09eye} &  Yes & KLD \\
	7 & \textbf{MSM-SM}  & Muthuswamy~\cite{muthuswamy13salient} & 2013 & MPEG-2 & MVF+DCT-P/R & Saliency Detection & Mahadevan~\cite{mahadevan10spatiotemporal} & - & ROC \\
	8 & \textbf{APPROX}  & Hadizadeh~\cite{hadizadeh13visual} & 2013 & - & MVF+DCT-P & Video Compression  & SFU~\cite{hadizadeh12eye} & Yes & KLD+AUC \\
	9 & \textbf{PNSP-CS} & Fang~\cite{fang14video} & 2014 & MPEG-4 ASP & MVF+DCT-P & Saliency Detection & CRCNS~\cite{itti06bayesian},\cite{itti09eye} & Yes  & KLD+AUC \\		
	10 & \textbf{OBDL-MRF} & Khatoonabadi~\cite{khatoonabadi15how} & 2015 & H.264/AVC & OBDL & Saliency Detection & SFU~\cite{hadizadeh12eye}+DIEM~\cite{DIEM} & Yes & AUC+NSS \\
	11 & \textbf{MVE+SRN} & Khatoonabadi~\cite{khatoonabadi15correlates}\cite{khatoonabadi14compressed} & 2015 & H.264/AVC & MVF+DCT-R & Saliency Detection & SFU~\cite{hadizadeh12eye}+DIEM~\cite{DIEM} & Yes & AUC+NSS+JSD \\
	
	\hline \hline
	
	\end{tabular}
	%\end{sideways}
	%\end{rotate}
	
	\label{tab:Models}
	
\end{table*}
%\end{sidewaystable}

%%%%%%%%%%%%%%%%%%%%%%%%%%%%%%%%%%%%%%%%
\subsection{Eye-Tracking Video Datasets}
%%%%%%%%%%%%%%%%%%%%%%%%%%%%%%%%%%%%%%%%
Eye-tracking data is the most typical psychophysical ground truth for visual saliency models~\cite{engelke11visual}. To evaluate saliency models, each model's saliency map is compared with recorded gaze locations of the subjects. Two recent publicly available eye-tracking datasets were used in the study. The reader is referred to~\cite{winkler13overview} for an overview of other existing datasets in the field. 

%%%%%%%%%%
\subsubsection{The SFU Dataset} 
\begin{figure}
\centering
\tiny
\captionsetup[subfigure]{style=default,captionskip=1pt}
\begin{tabular}{cccc}
\vspace{-2em}
\subfloat{\includegraphics[height=.66in]{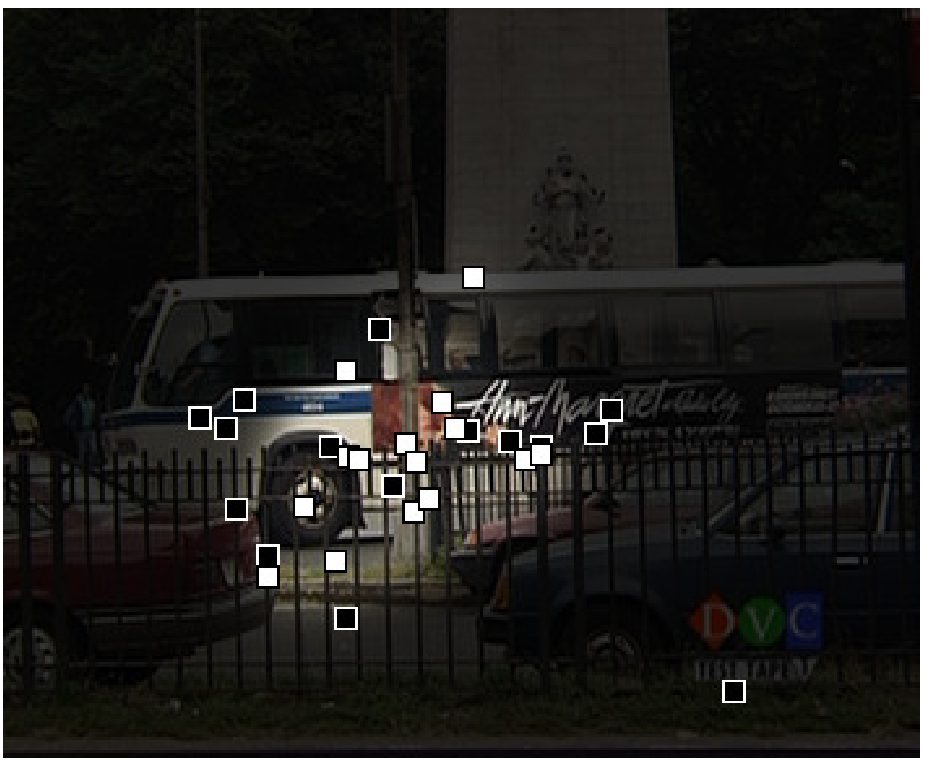}} \hspace*{-2em}& 
\subfloat{\includegraphics[height=.66in]{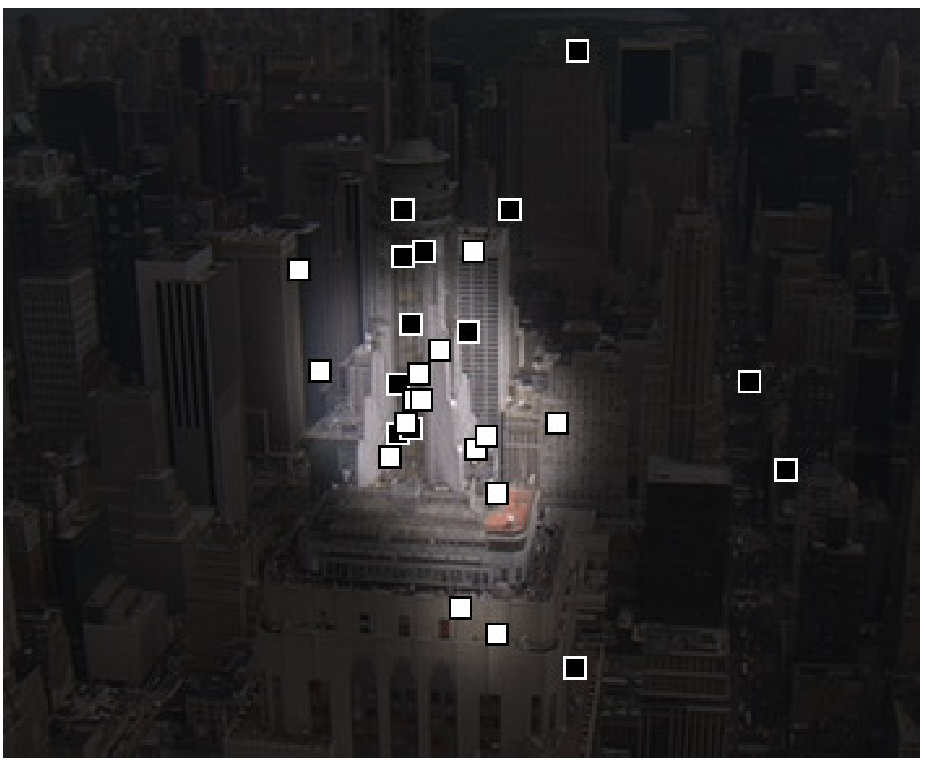}} \hspace*{-2em}& 
\subfloat{\includegraphics[height=.66in]{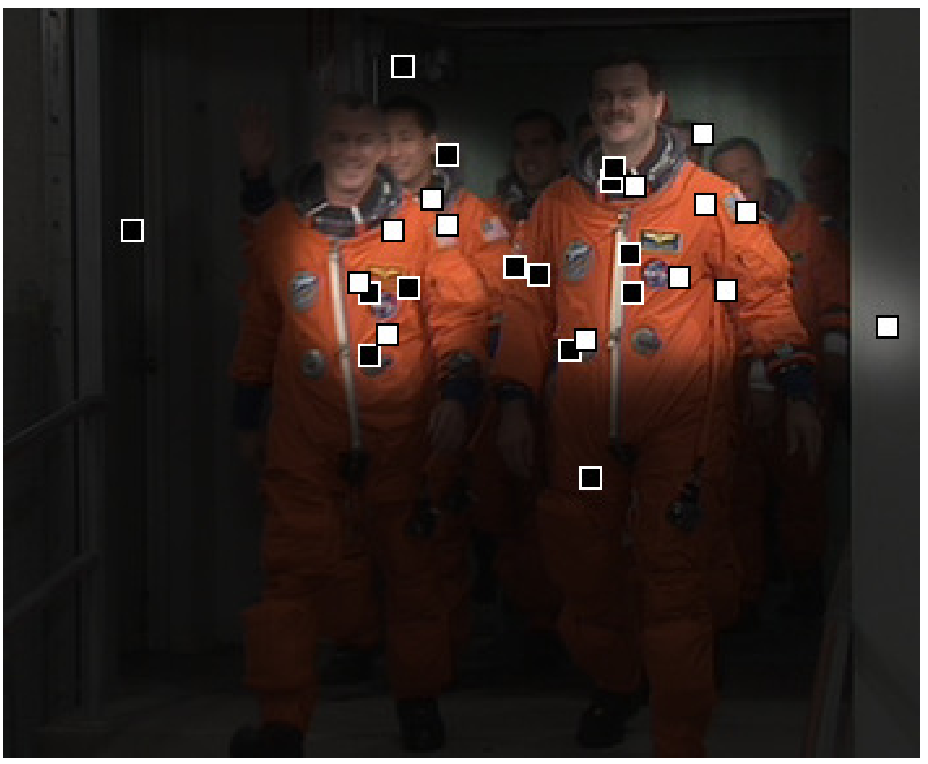}} \hspace*{-2em}& 
\subfloat{\includegraphics[height=.66in]{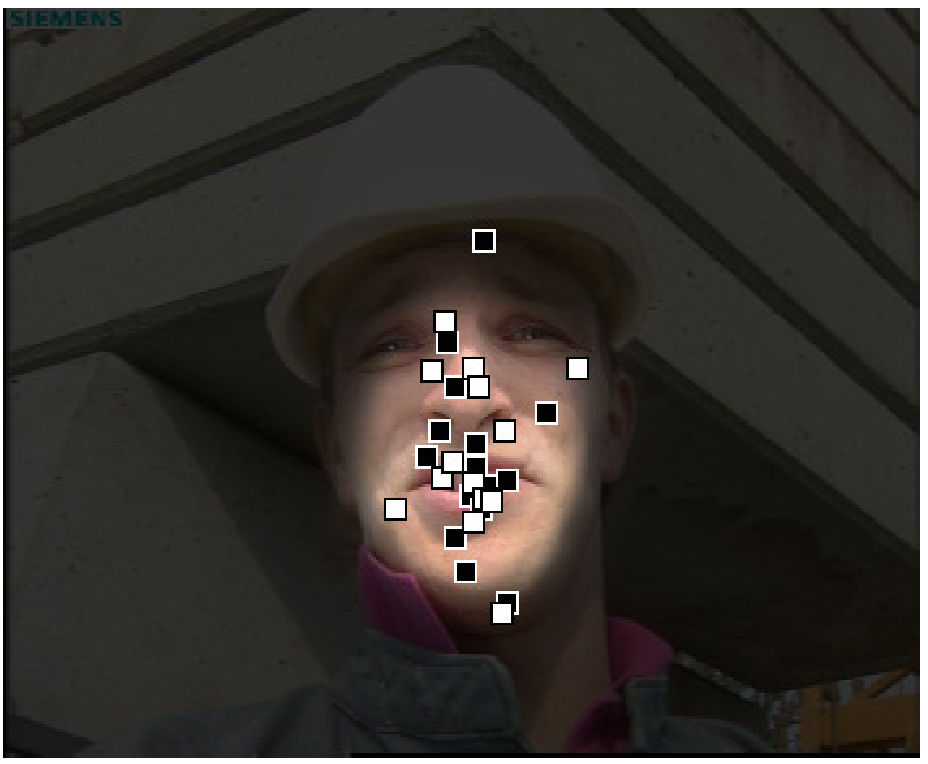}} \\
\vspace{-2em}
\subfloat{\includegraphics[height=.66in]{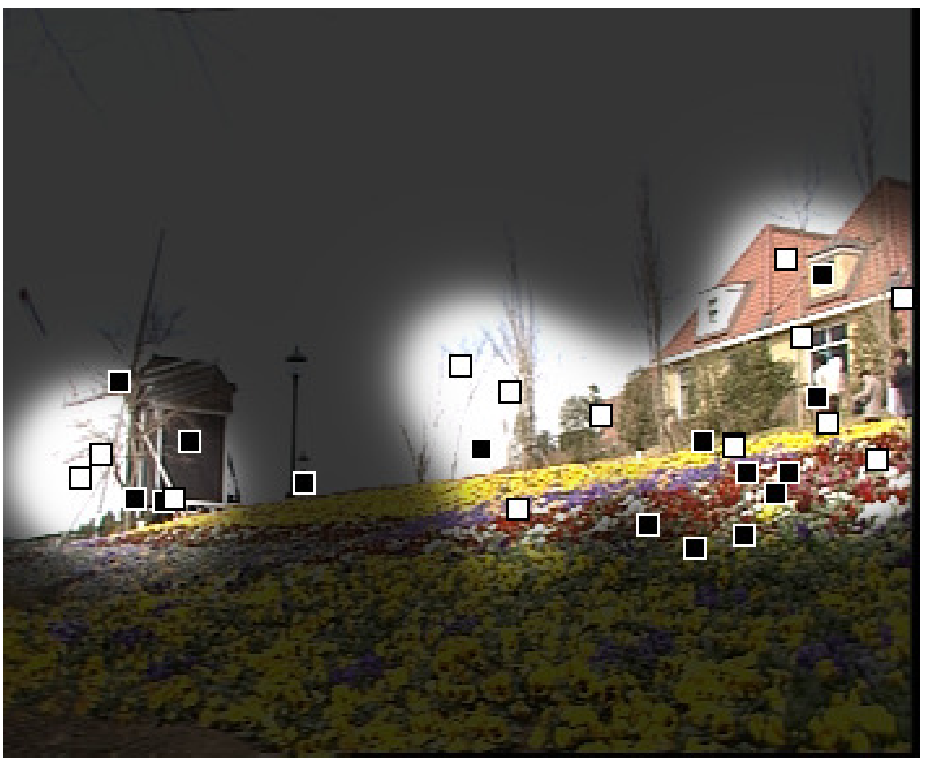}} \hspace*{-2em}&
\subfloat{\includegraphics[height=.66in]{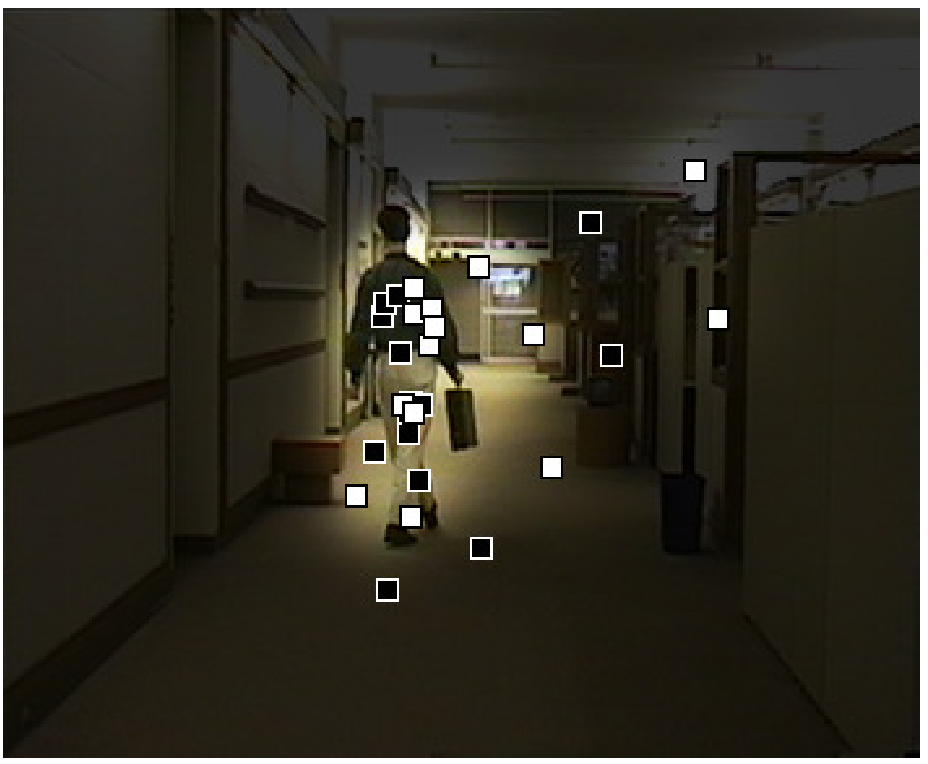}} \hspace*{-2em}&
\subfloat{\includegraphics[height=.66in]{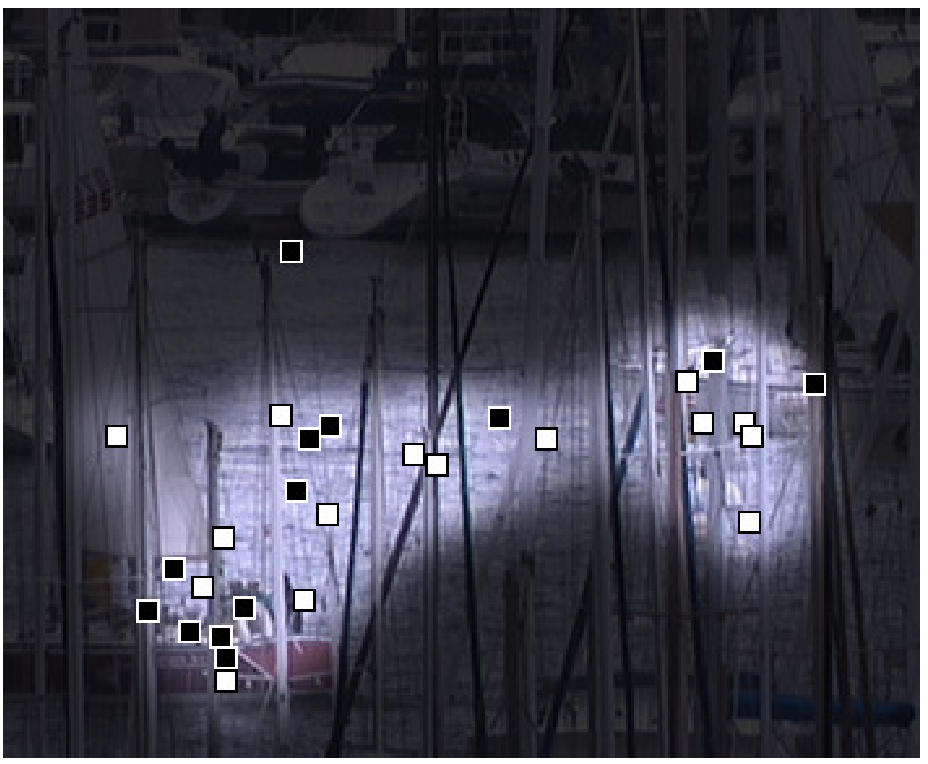}} \hspace*{-2em}&
\subfloat{\includegraphics[height=.66in]{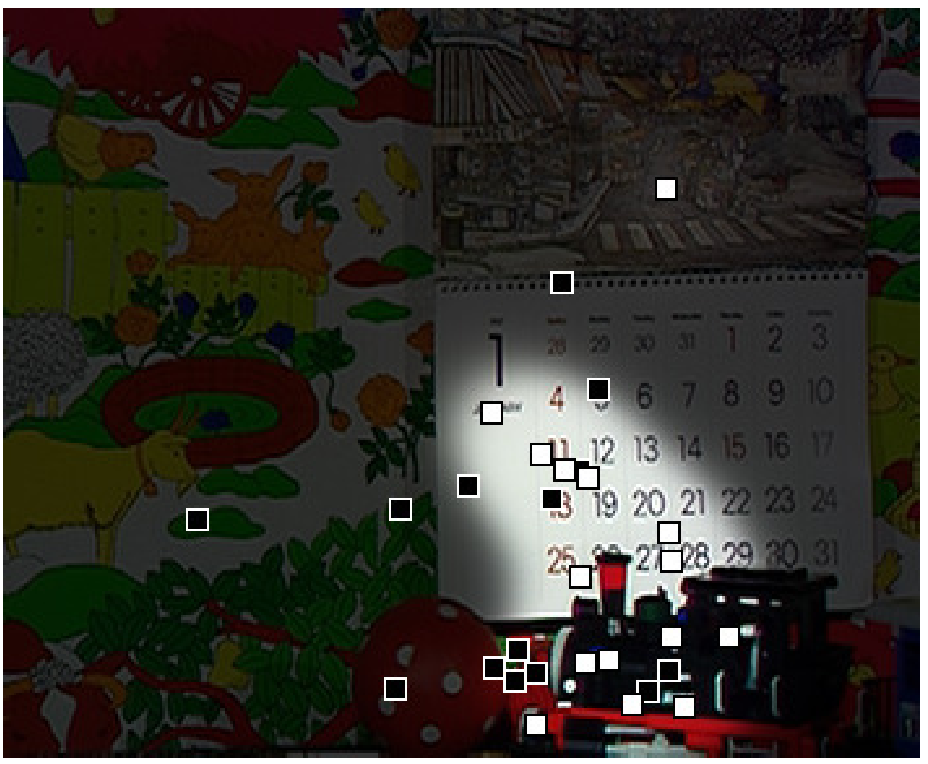}} \\
\vspace{-2em}
\subfloat{\includegraphics[height=.66in]{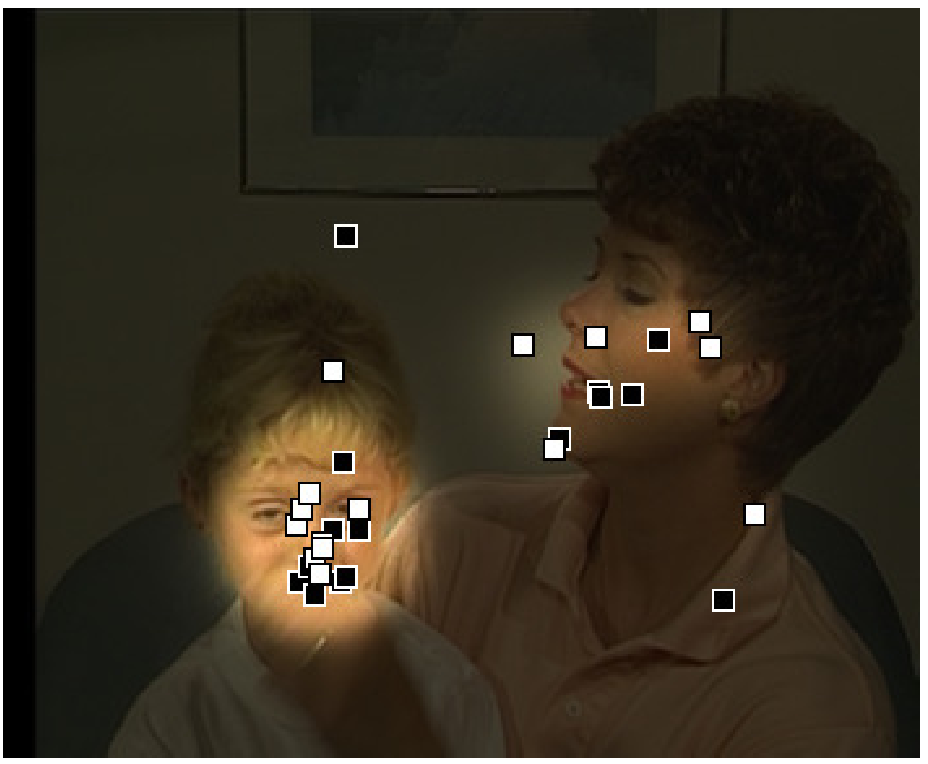}} \hspace*{-2em}&
\subfloat{\includegraphics[height=.66in]{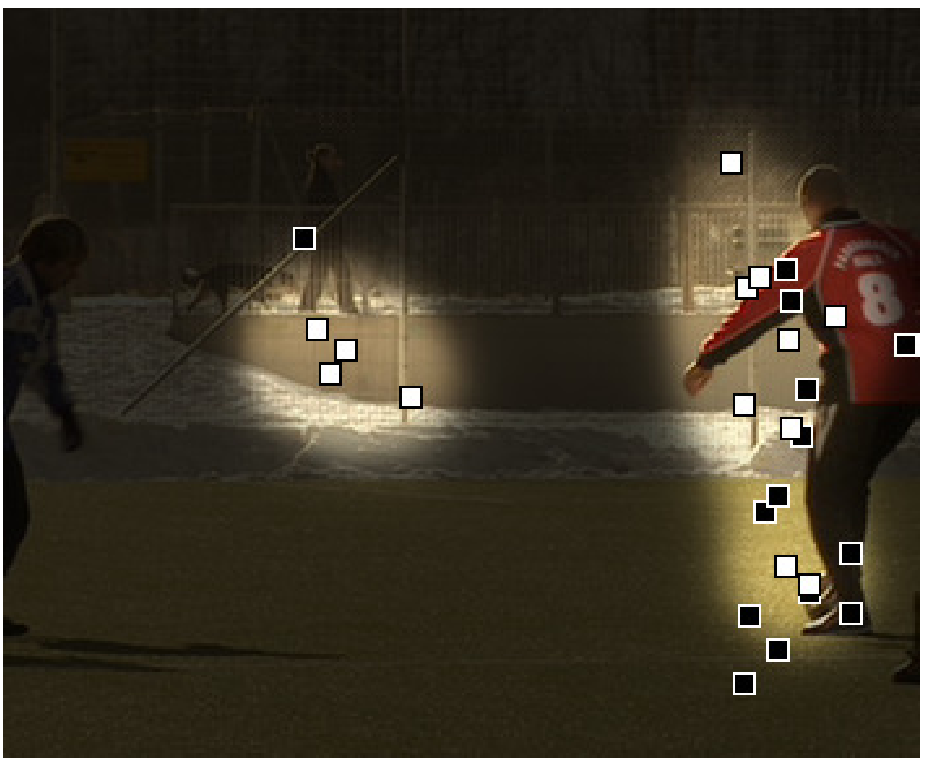}} \hspace*{-2em}&
\subfloat{\includegraphics[height=.66in]{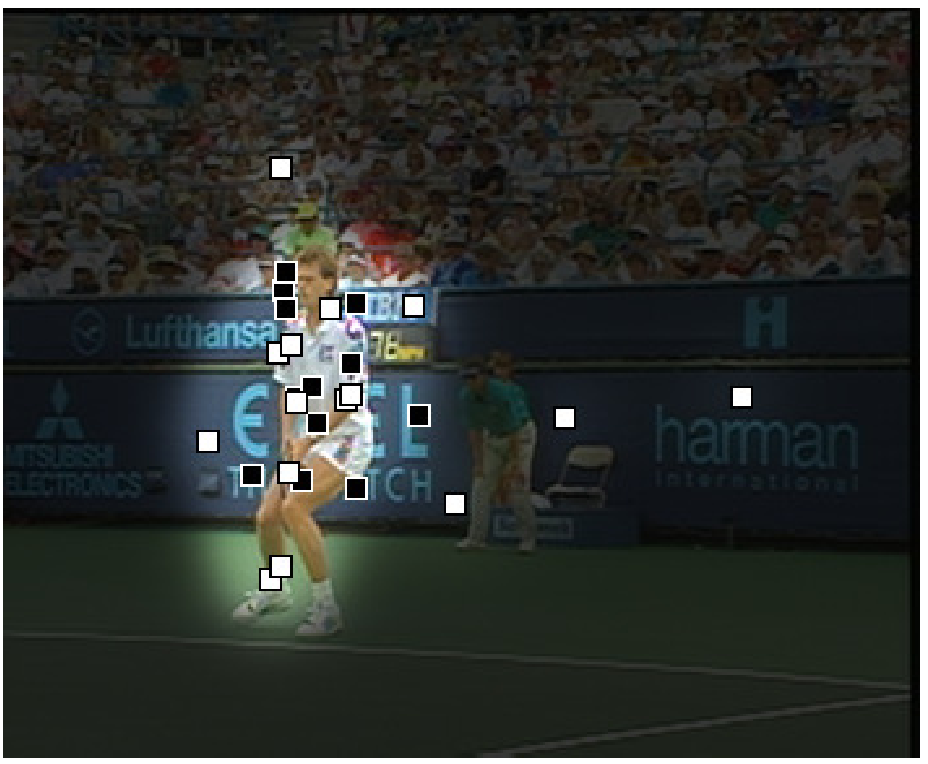}} \hspace*{-2em}&
\subfloat{\includegraphics[height=.66in]{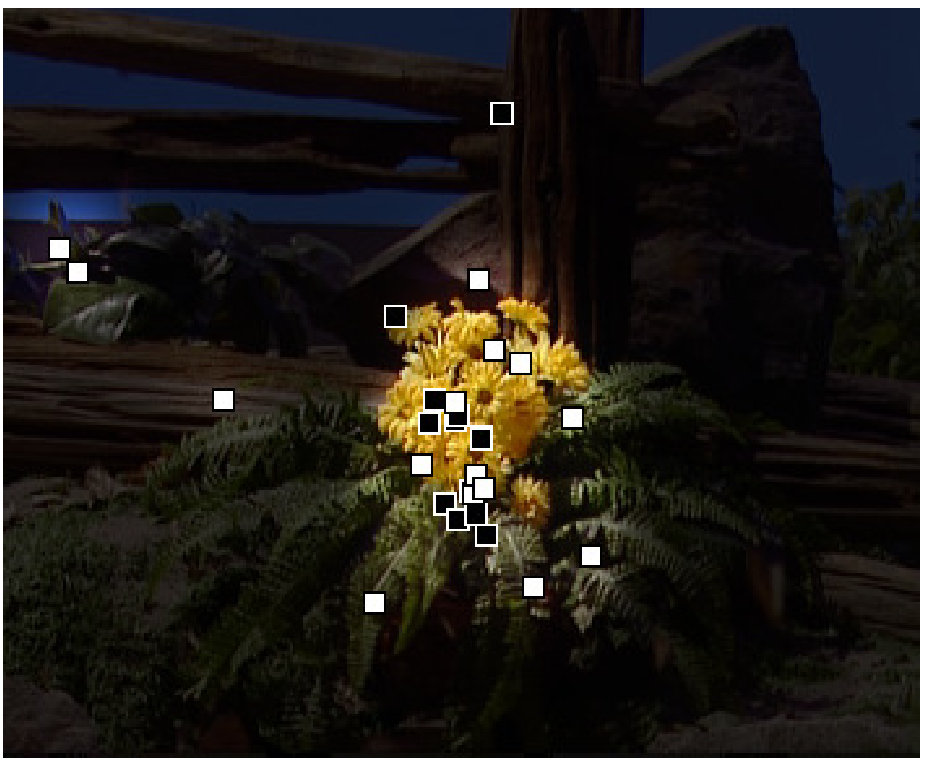}} \\
\end{tabular}
\vspace{2em}
\caption{Sample gaze visualization from the SFU Dataset. The gaze points from the first viewing are indicated as white squares, those from the second viewing as black squares.}
\label{fig:gazevisualSFU}
\end{figure}

The SFU eye-tracking dataset~\cite{hadizadeh12eye} consists of twelve CIF ($352 \times 288$) sequences that have become popular in the video compression and communications community: \textit{Bus}, \textit{City}, \textit{Crew},  \textit{Foreman}, \textit{Flower Garden}, \textit{Hall Monitor}, \textit{Harbour}, \textit{Mobile Calendar}, \textit{Mother and Daughter}, \textit{Soccer}, \textit{Stefan}, and \textit{Tempete}. A total of 15 participants watched all 12 videos while wearing a Locarna Pt-mini head-mounted eye tracker. Each participant took part in the test twice, resulting in two sets of viewings per participant for each video. The first viewing is used as ground truth for evaluating the performance of saliency models, whereas the data from the second viewing is used to construct benchmark models, as described in Section~\ref{sec:benchmark_models}. The results in~\cite{hadizadeh12eye} showed that gaze locations in the first and second viewings can differ notably, however they remain relatively close to each other when there is a single dominant salient region in the scene (for example, the face in the \textit{Foreman} sequence.) As a result, it is reasonable to expect that good saliency models will produce high scores for those frames where the first and second viewing data agree. 
A sample frame from each video has been shown in Fig.~\ref{fig:gazevisualSFU}, overlaid with the  gaze locations from both viewings. The visualization is such that the less-attended regions (according to the first viewing) are indicated by darker colors. Further details about this dataset are shown in Table~\ref{tab:Datasets}.

\begin{table}
	\centering
	%\footnotesize \addtolength{\tabcolsep}{-1pt}
	\caption{\small Datasets used to evaluate compressed-domain visual saliency models.}
	\begin{tabular}{lcc}
	\hline \hline
	\textbf{Dataset} & \textbf{SFU} & \textbf{DIEM} \\
	\textbf{Year} &  2012 & 2011 \\
	\textbf{Sequences} &  12 & 85 \\
	\textbf{Display Resolution} & $704 \times 576^\ast$ & varying \\
	\textbf{Format} & RAW & MPEG-4 \\
	\textbf{FPS} & 30 & 30 \\
	\textbf{Frames} & 90-300   & 888-3401 \\
	\textbf{Participants} & 15  & 35-53$^\S$ \\
	\textbf{Viewings}	& 2$^\dagger$ &  2$^\ddagger$ \\
	\textbf{Screen Resolution} &  $1280 \times 1024$ & $1600 \times 1200$ \\
	\textbf{Screen Diagonal} & ${19\verb+"+}$ & ${21.3\verb+"+}$\\	
	\textbf{Viewing Distance} & 80 cm & 90 cm \\
		\hline \hline
	\end{tabular}
	\newline
	\par
	\raggedright
	$^\ast$The original video resolution ($352 \times 288$) was doubled during the presentation to the participants\newline
	$^\dagger$Each participant watched each sequence twice, after several minutes\newline
	$^\ddagger$Viewings for the left/right eye are available\newline
	$^\S$A total of 250 subjects participated in the study, but not all of them viewed each video; the number of viewers per video was 35-53
	\label{tab:Datasets}
\end{table}

\begin{figure}
\centering
\tiny
\captionsetup[subfigure]{style=default,captionskip=1pt}
\begin{tabular}{cccc}
\vspace{-2em}
\subfloat{\includegraphics[height=.45in]{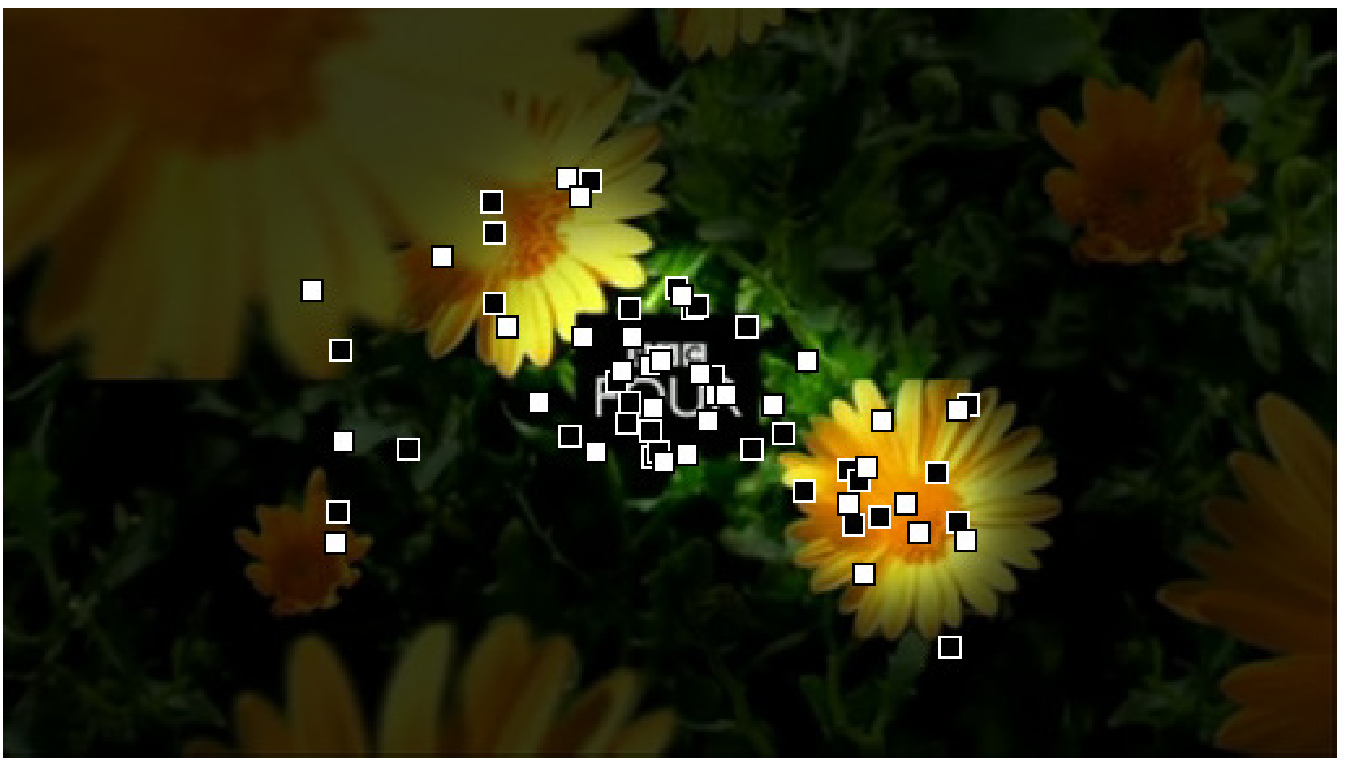}}\hspace{-2em} \hfill &
\subfloat{\includegraphics[height=.45in]{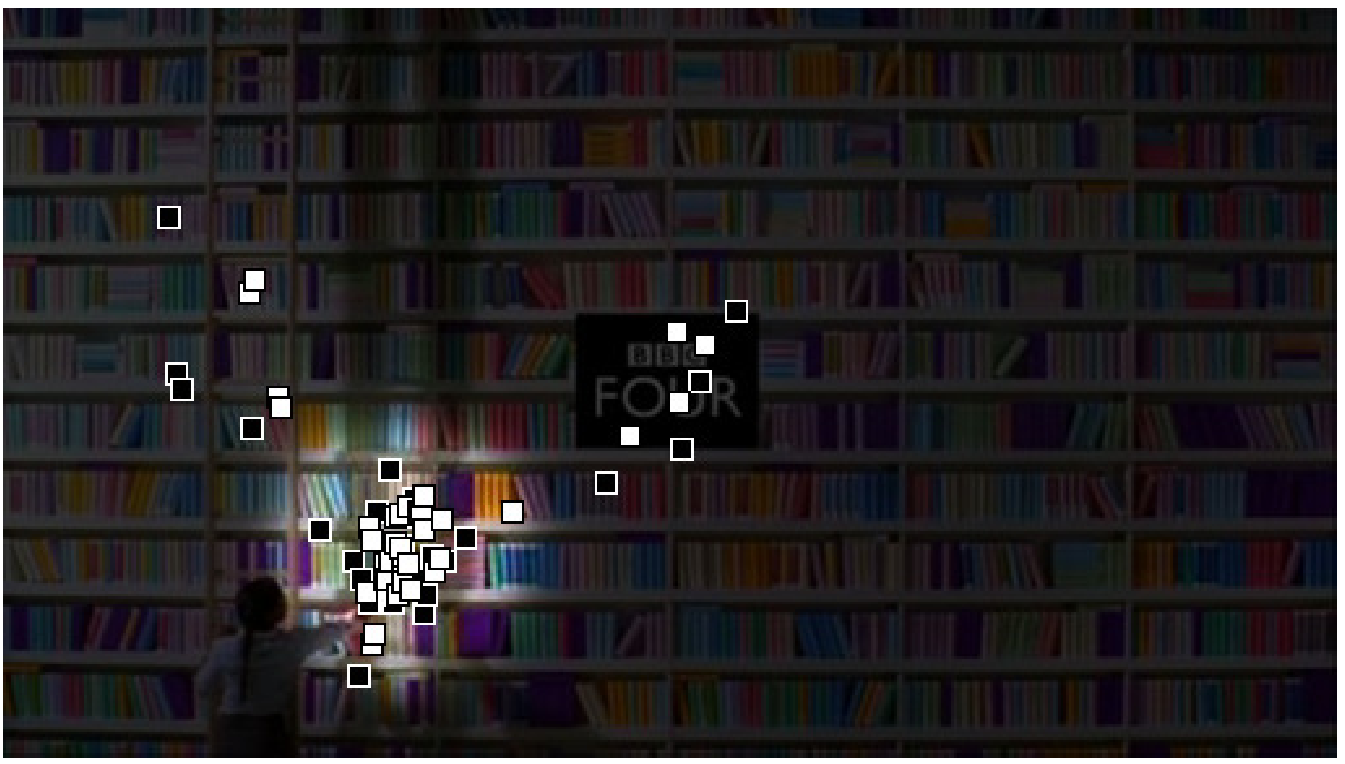}} \hspace{-2em}\hfill &
\subfloat{\includegraphics[height=.45in]{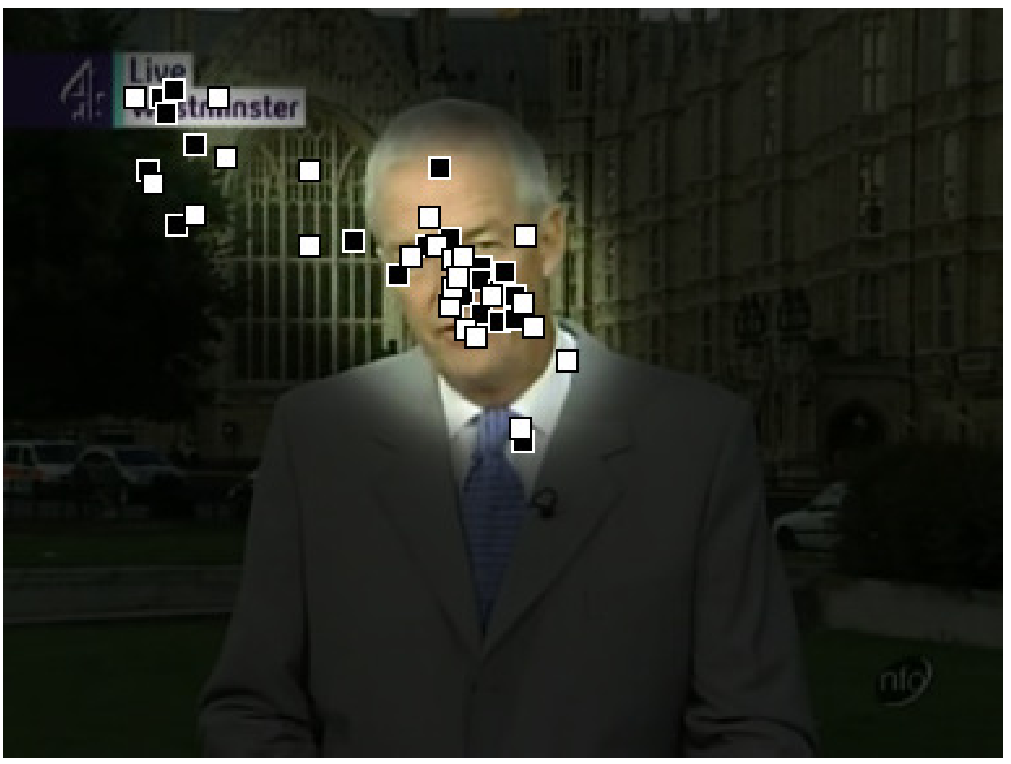}} \hspace{-2em}\hfill &
\subfloat{\includegraphics[height=.45in]{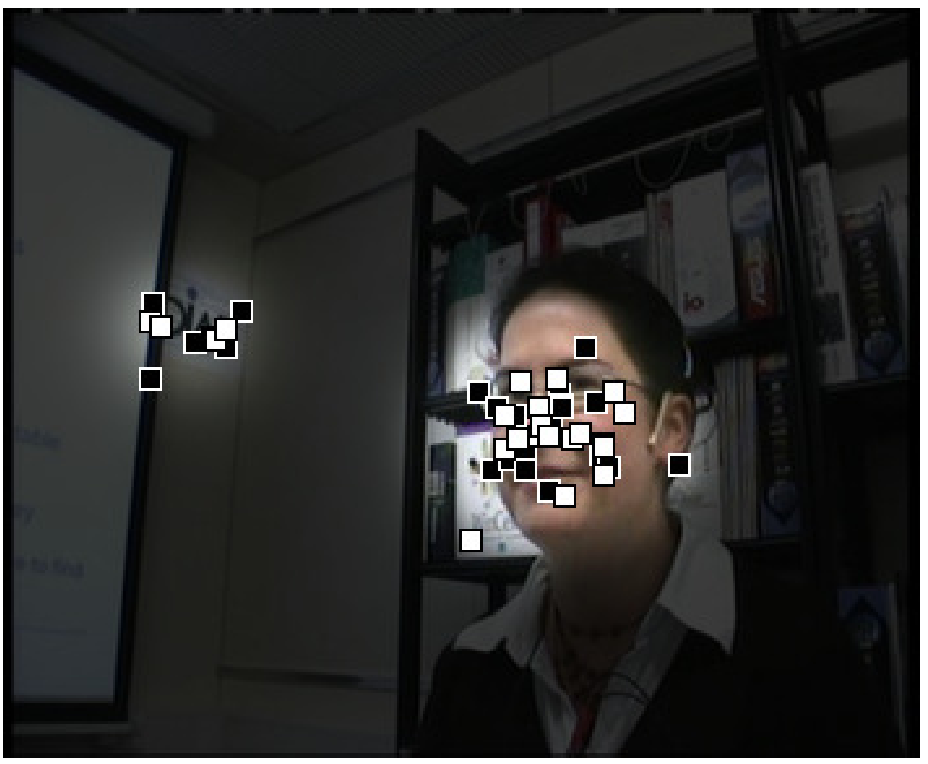}} \hspace{-2em}\hfill \\
\vspace{-2em}
\subfloat{\includegraphics[height=.45in]{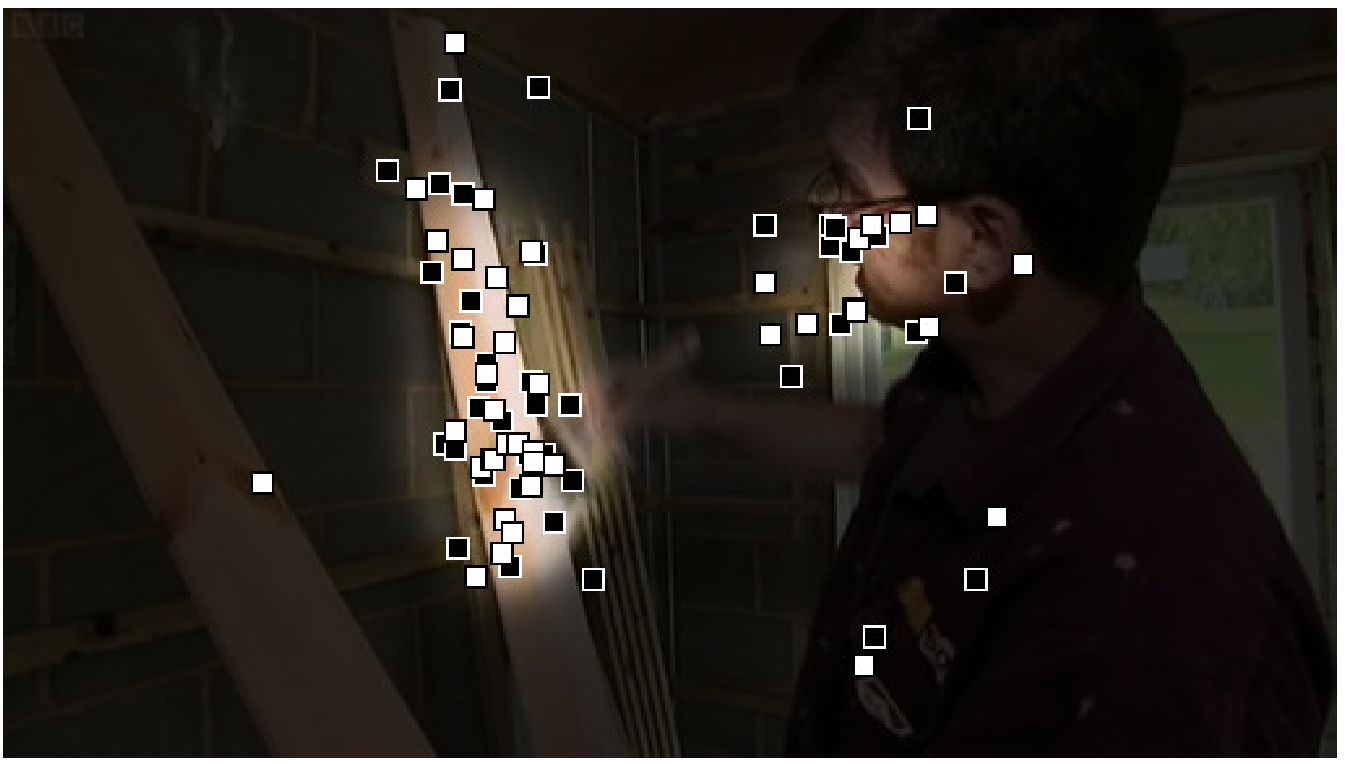}} \hspace{-2em}\hfill &
\subfloat{\includegraphics[height=.45in]{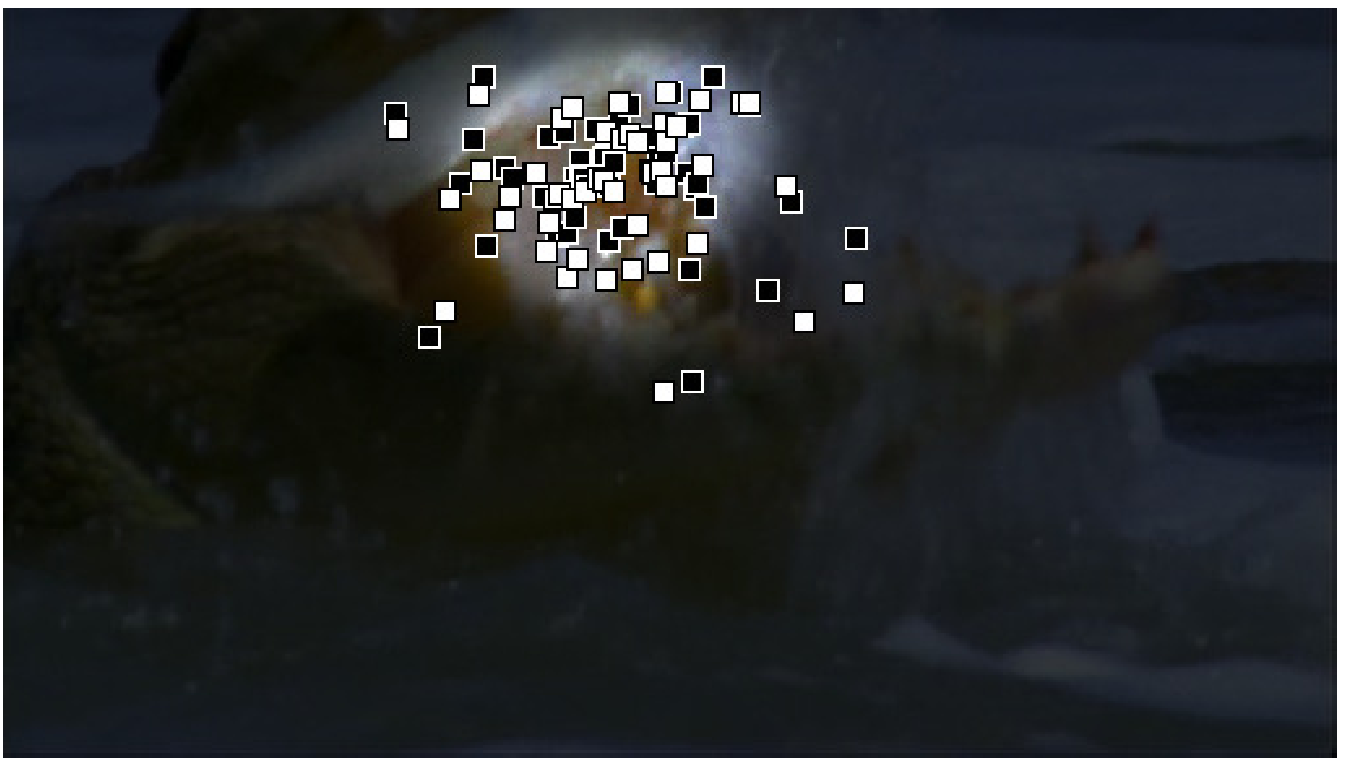}} \hspace{-2em}\hfill & 
\subfloat{\includegraphics[height=.45in]{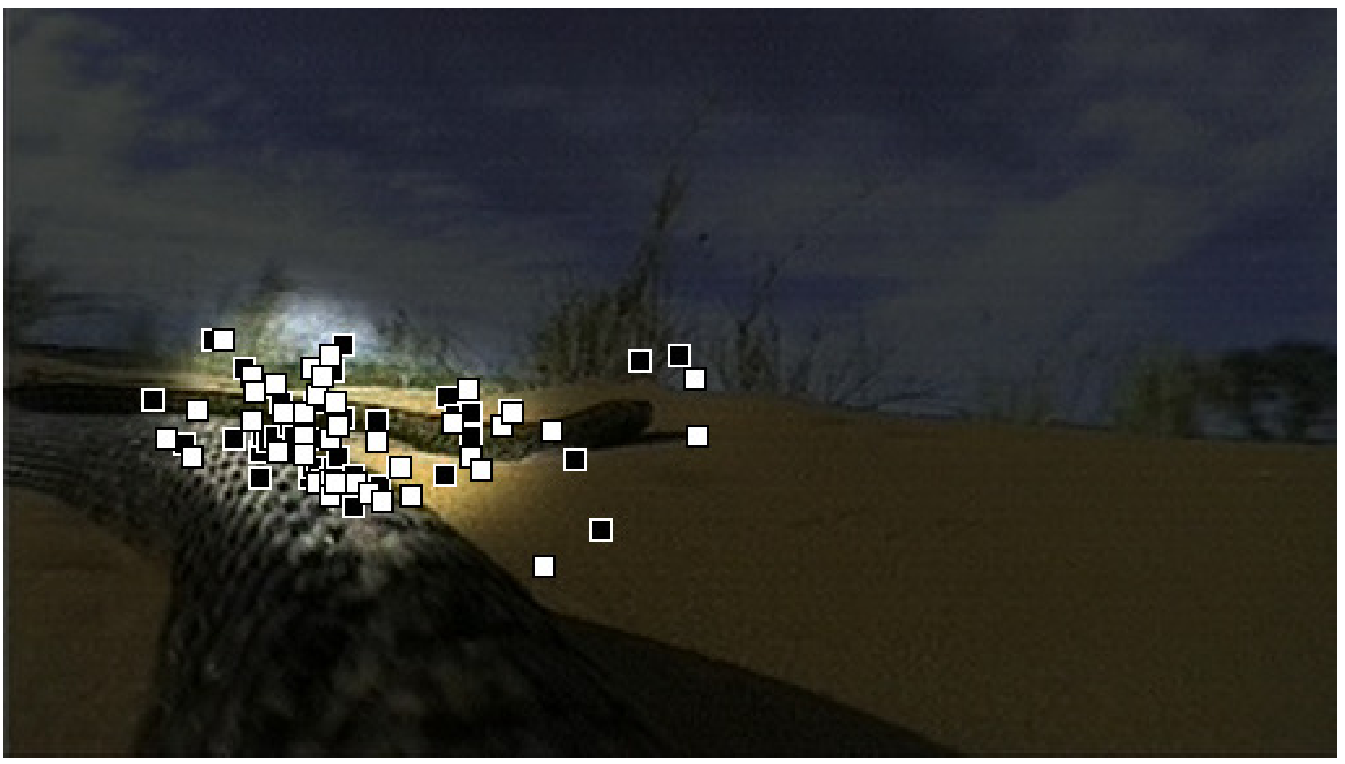}} \hspace{-2em}\hfill & 
\subfloat{\includegraphics[height=.45in]{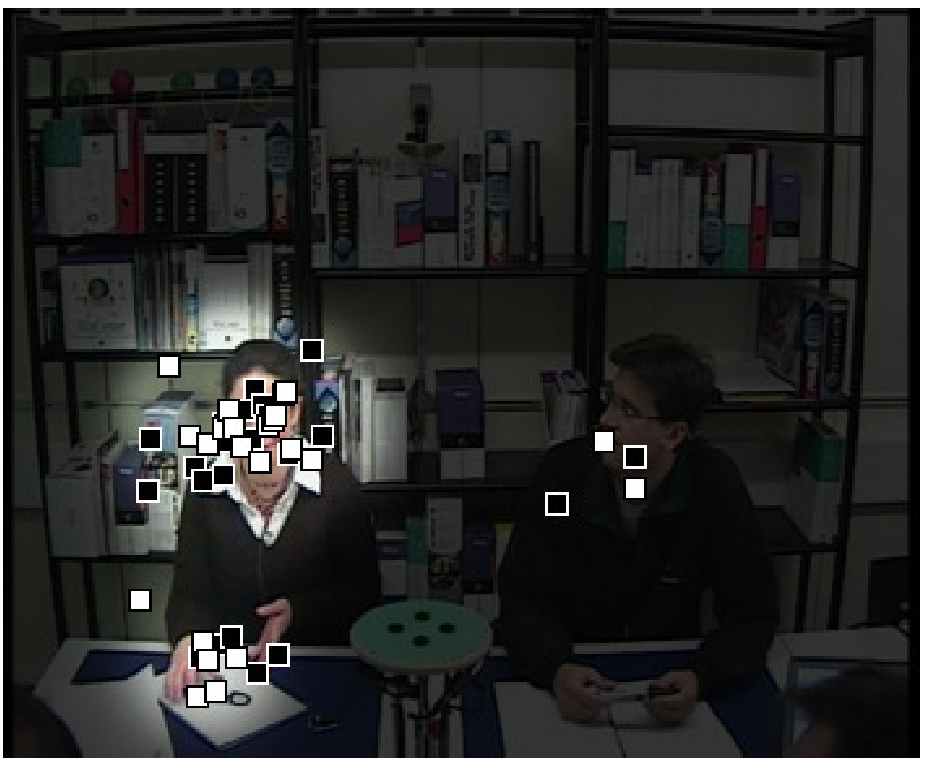}} \hspace{-2em}\hfill \\
\vspace{-2em}
\subfloat{\includegraphics[height=.45in]{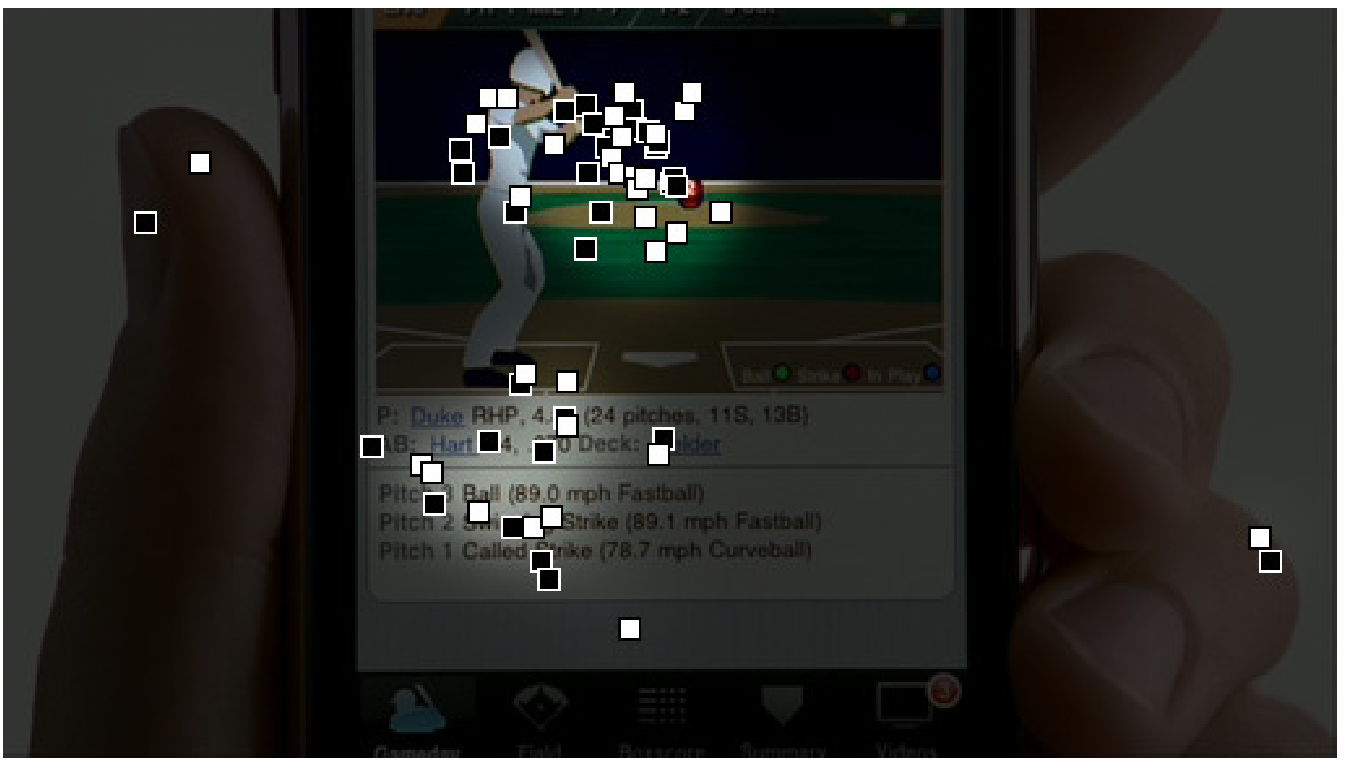}} \hspace{-2em}\hfill &
\subfloat{\includegraphics[height=.45in]{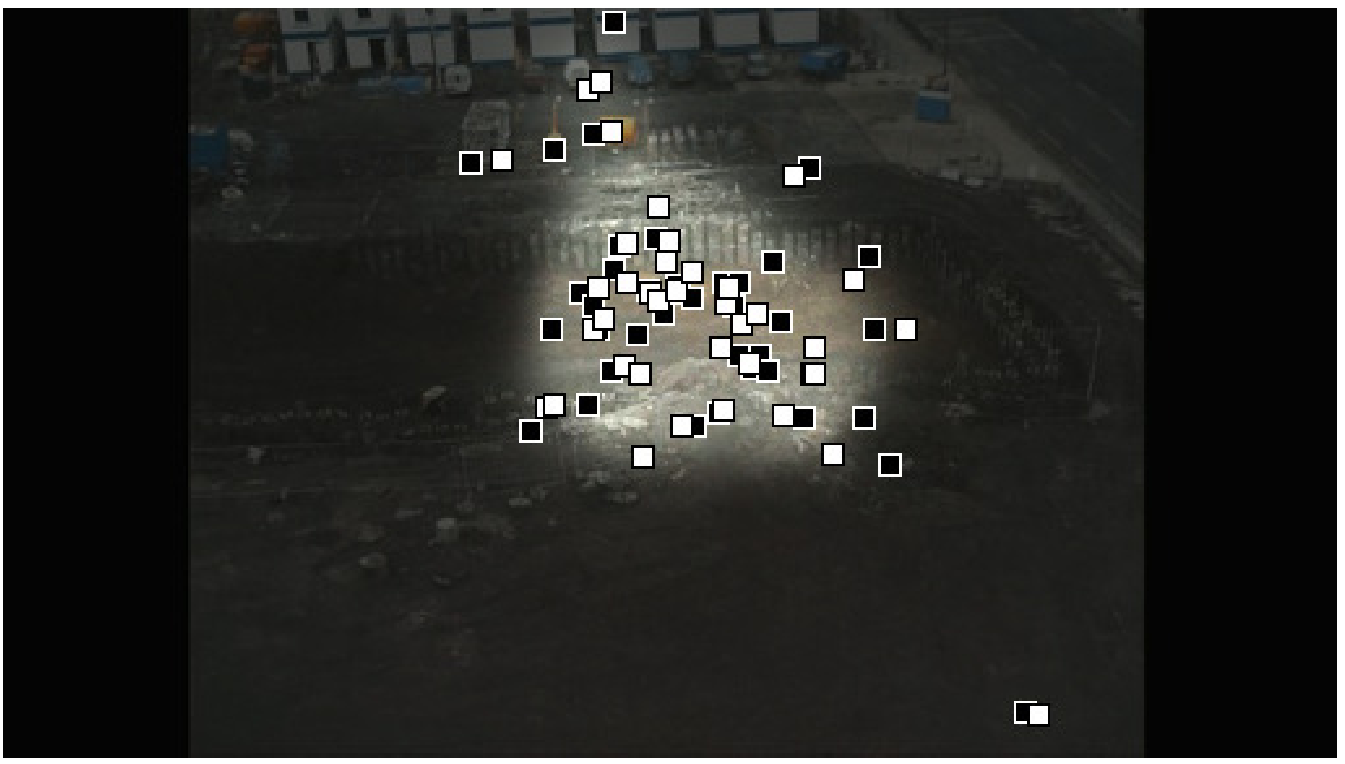}} \hspace{-2em}\hfill &
\subfloat{\includegraphics[height=.45in]{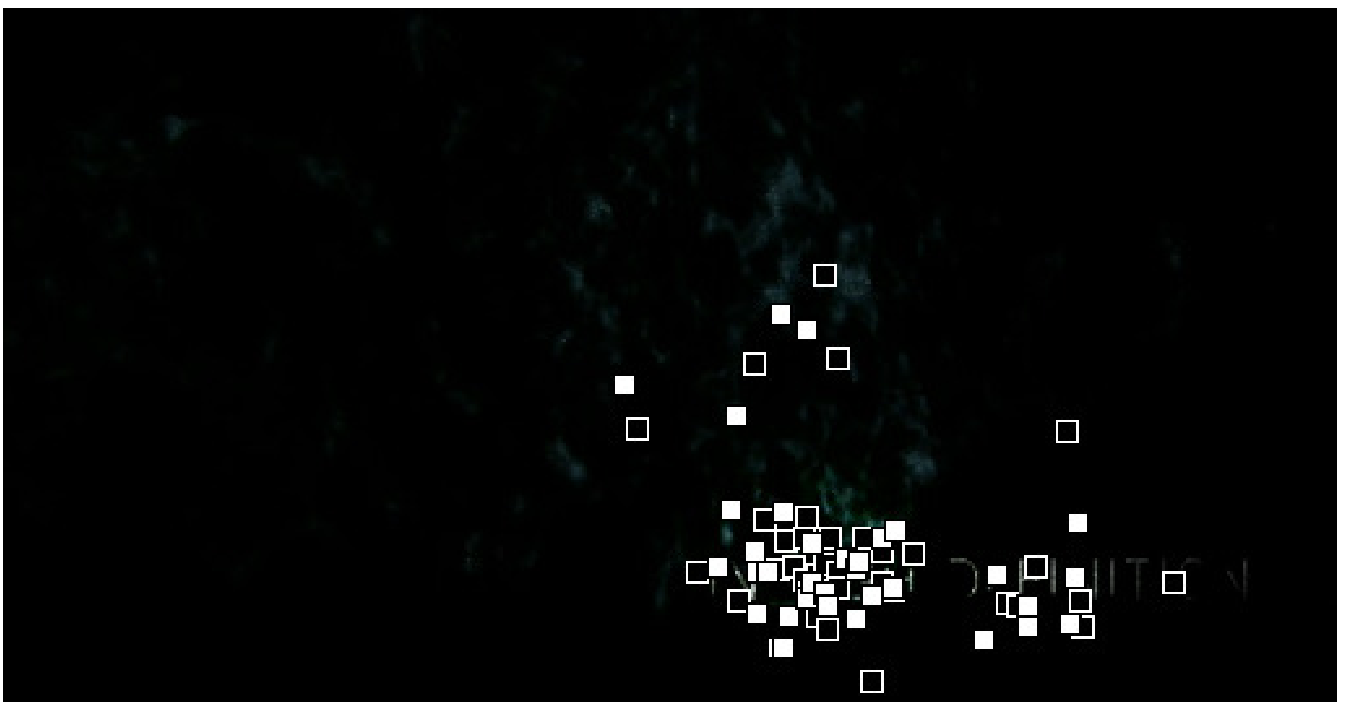}} \hspace{-2em}\hfill &
\subfloat{\includegraphics[height=.45in]{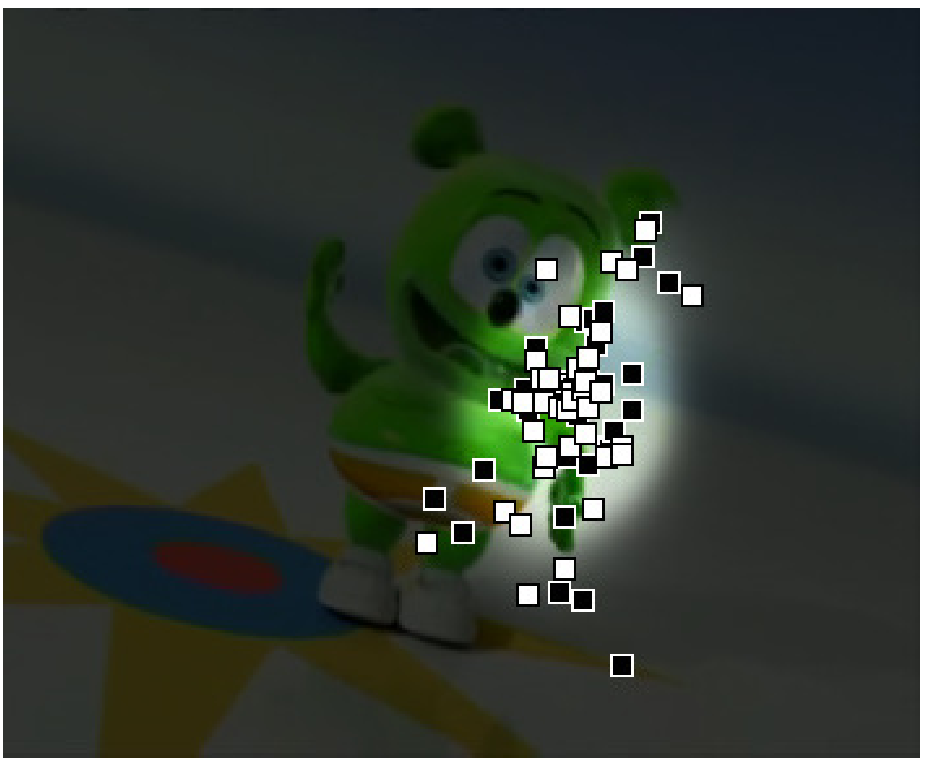}} \hspace{-2em}\hfill \\
\vspace{-2em}
\subfloat{\includegraphics[height=.45in]{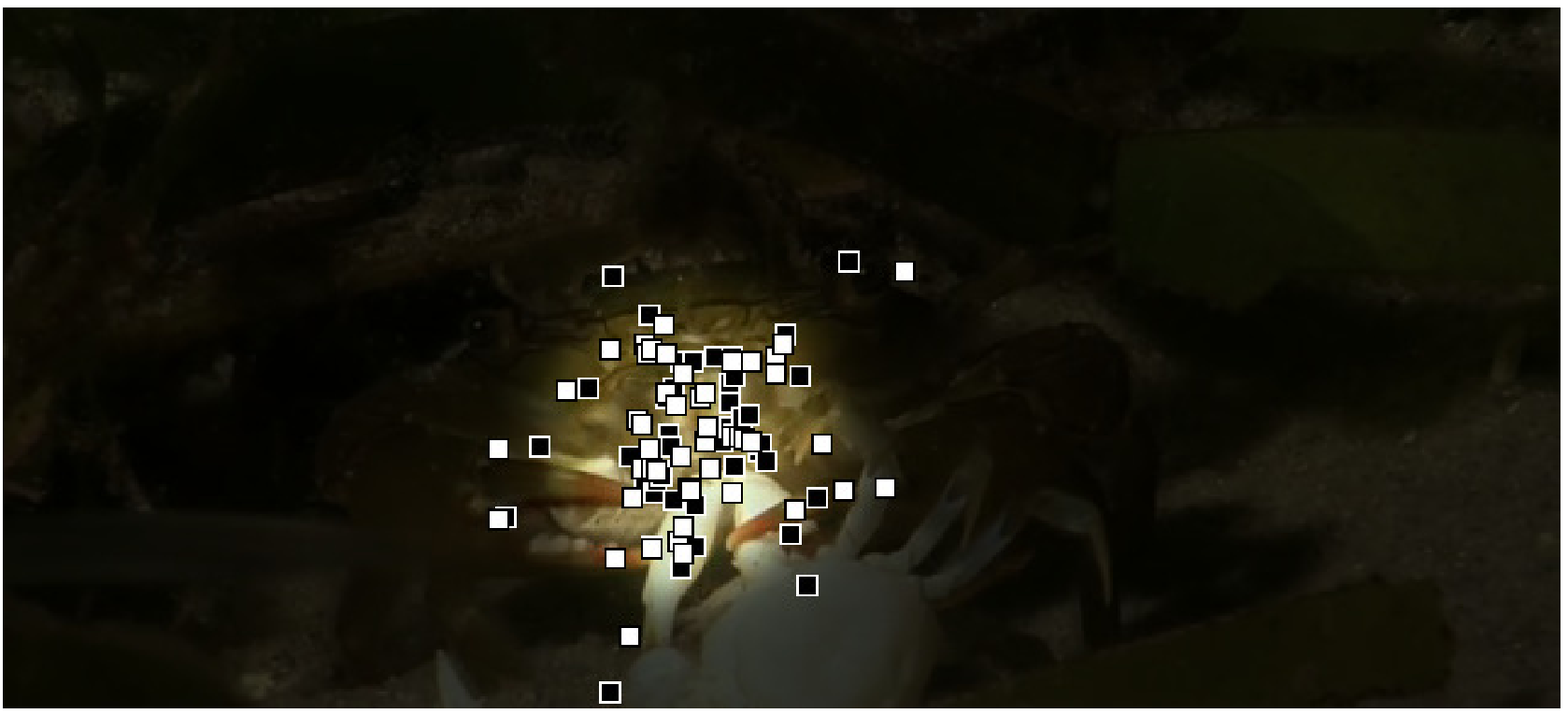}} \hspace{-2em}\hfill &
\subfloat{\includegraphics[height=.45in]{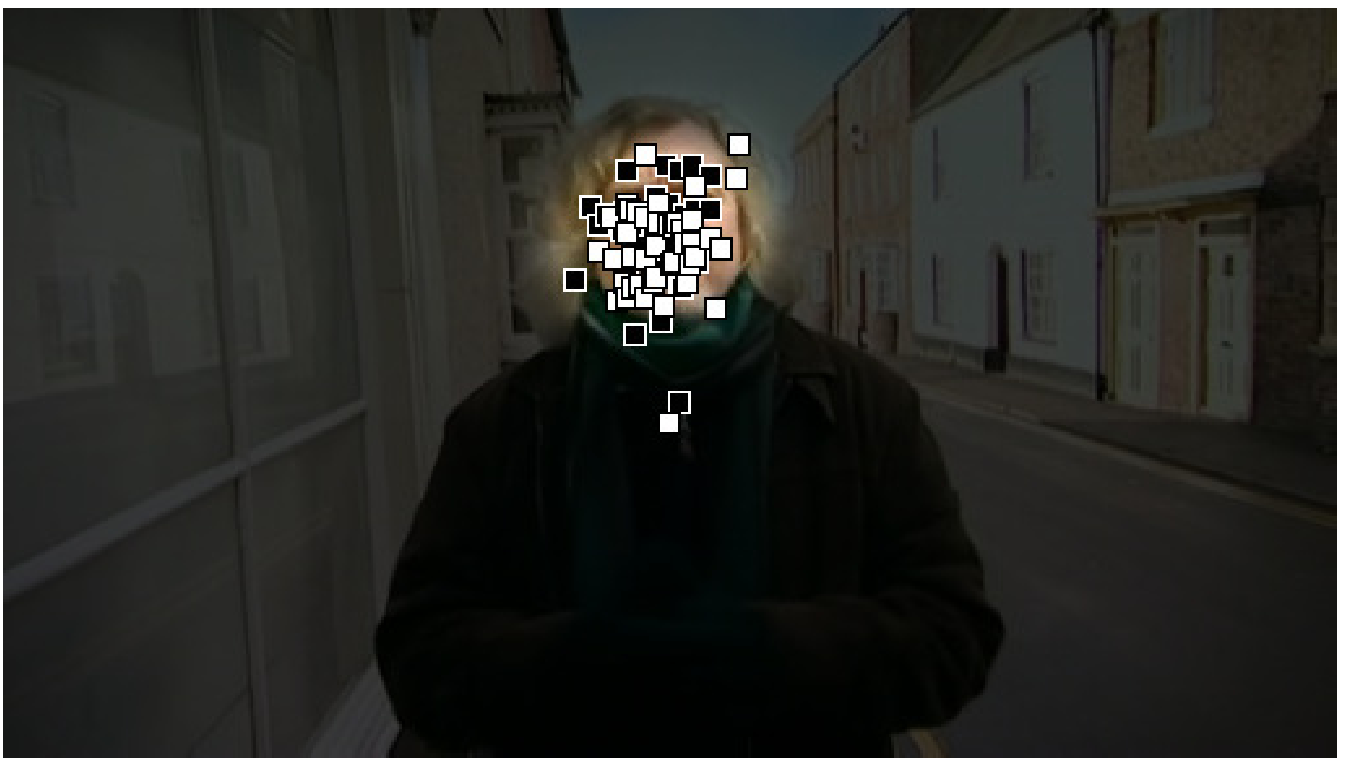}} \hspace{-2em}\hfill &
\subfloat{\includegraphics[height=.45in]{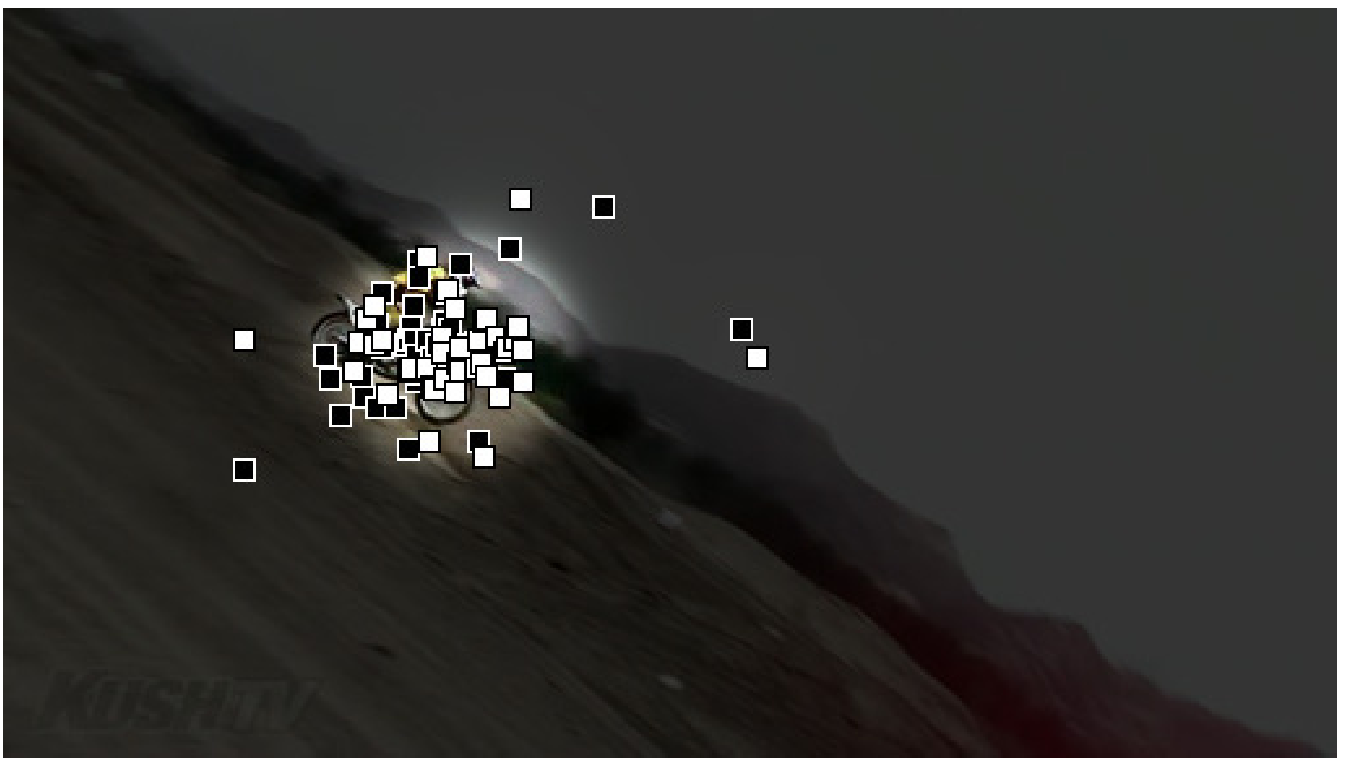}} \hspace{-2em}\hfill &
\subfloat{\includegraphics[height=.45in]{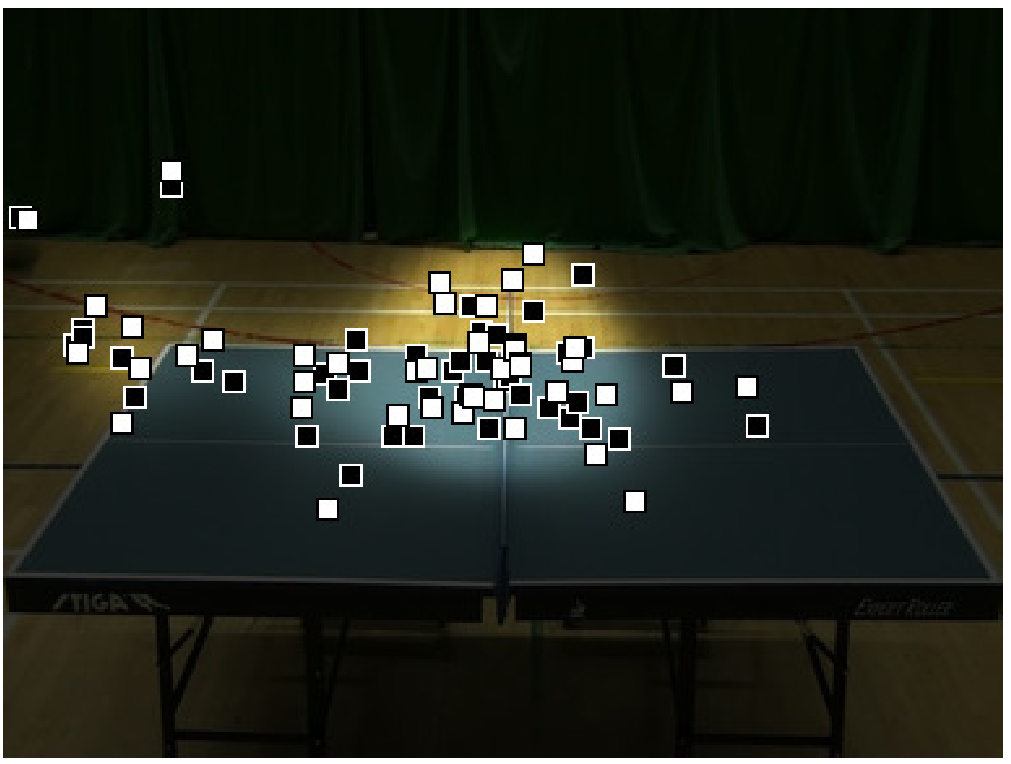}} \hspace{-2em}\hfill \\
\vspace{-2em}
\subfloat{\includegraphics[height=.45in]{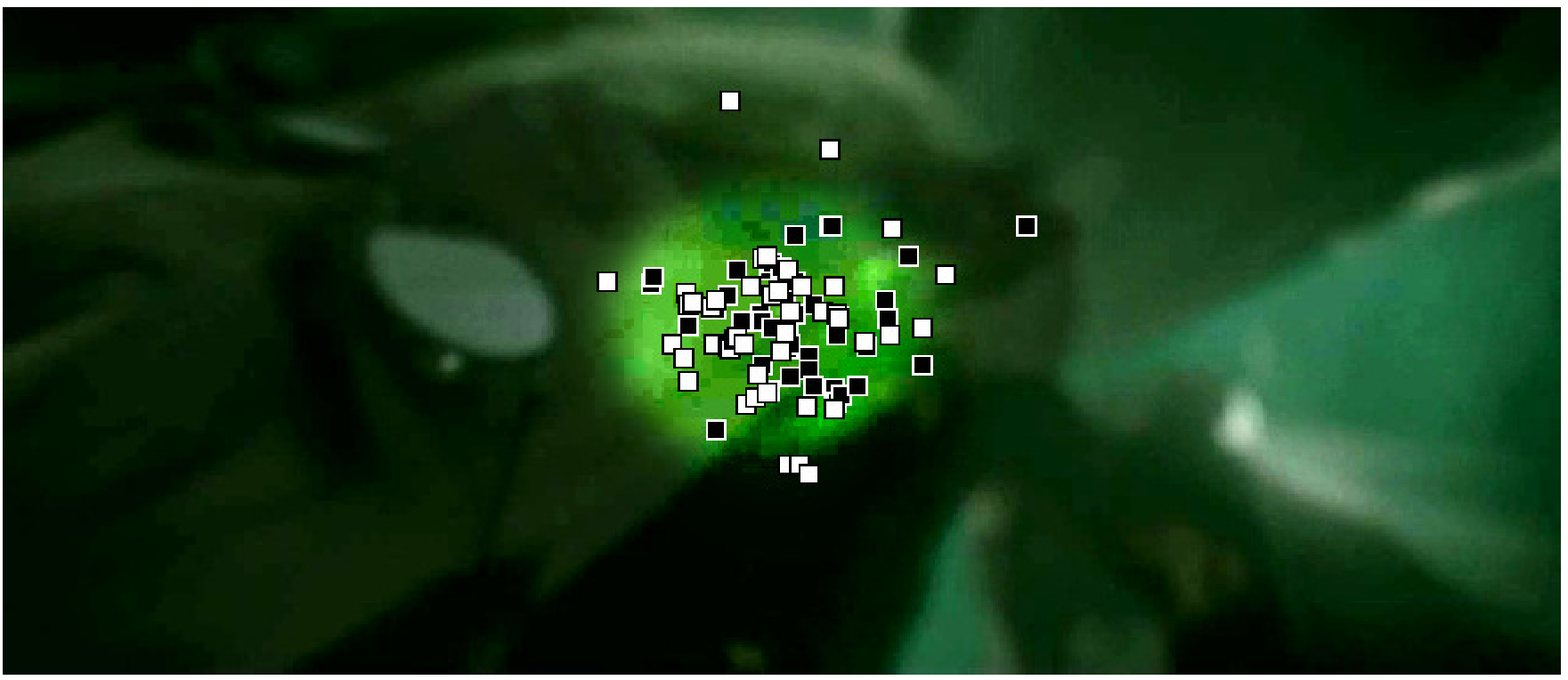}} \hspace{-2em}\hfill &
\subfloat{\includegraphics[height=.45in]{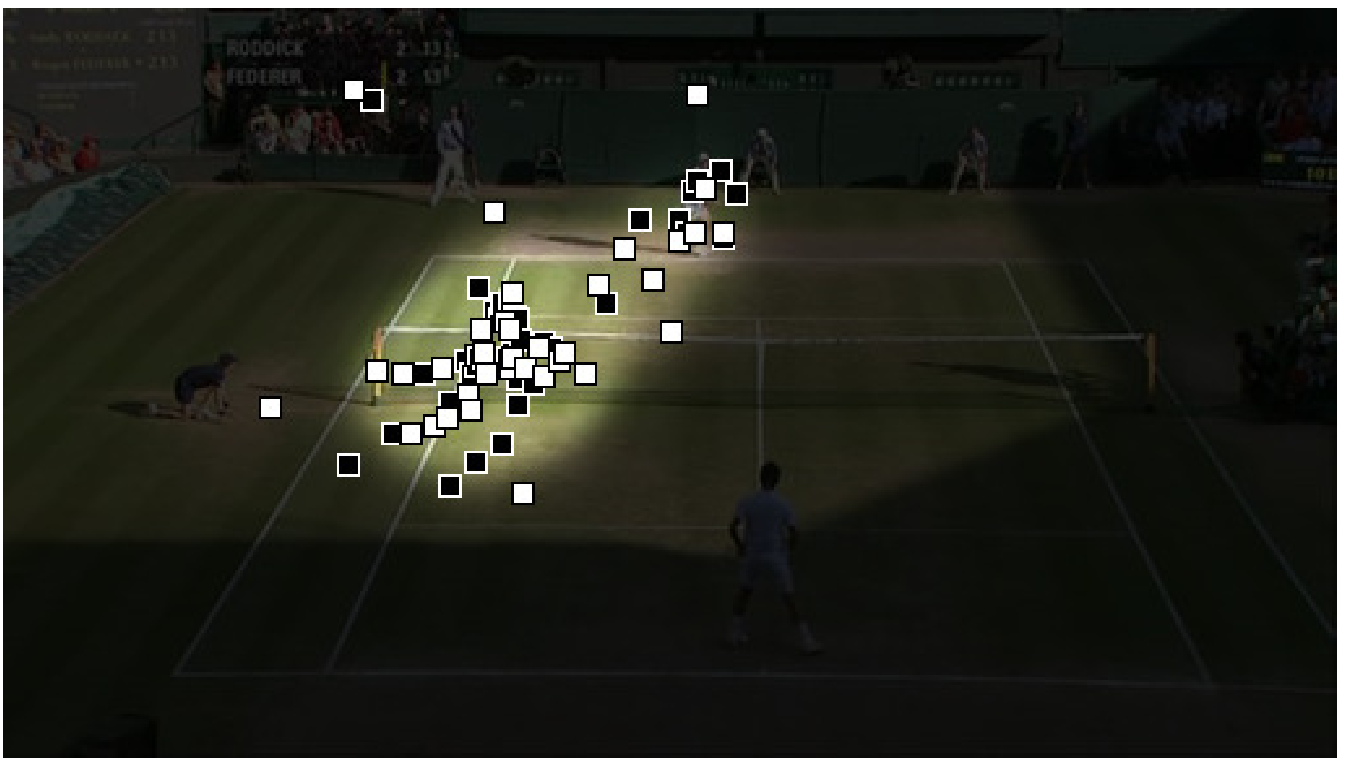}} \hspace{-2em}\hfill &
\subfloat{\includegraphics[height=.45in]{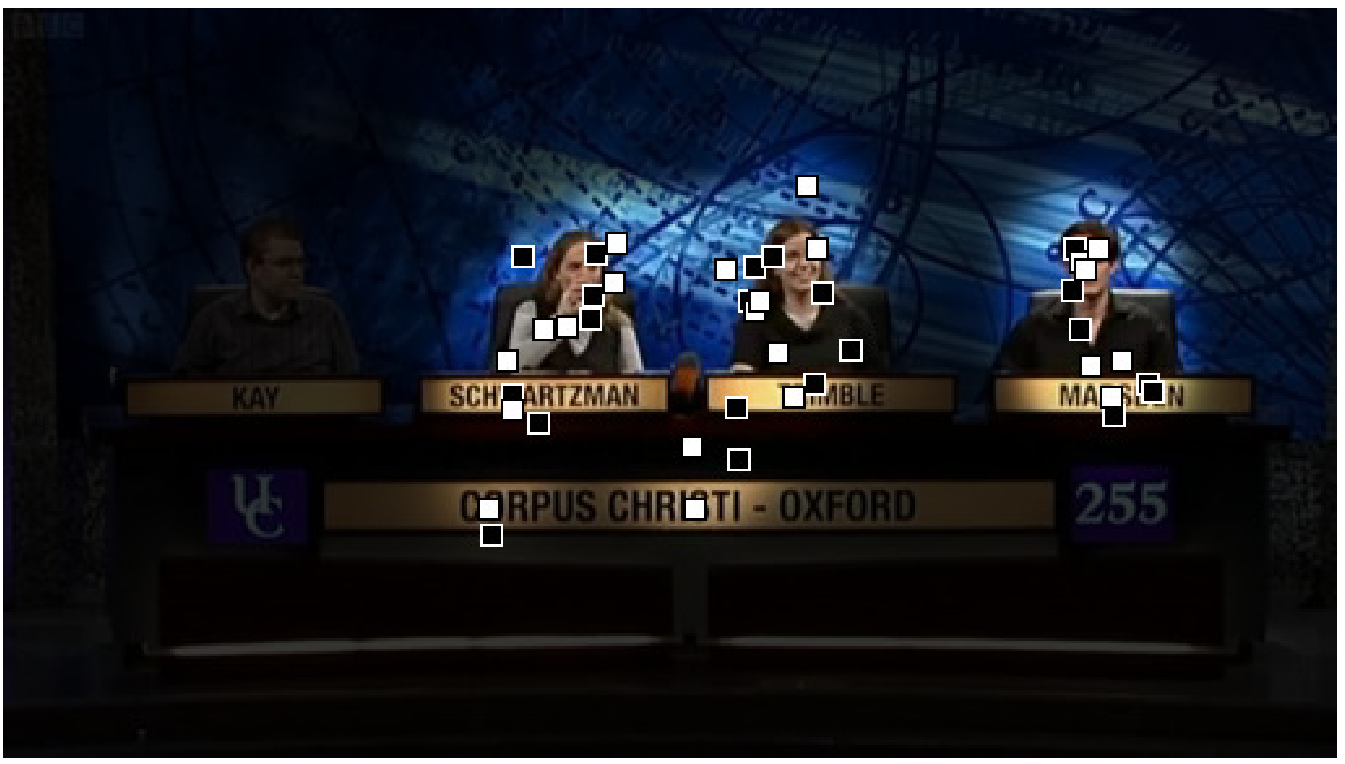}} \hspace{-2em}\hfill &
\subfloat{\includegraphics[height=.45in]{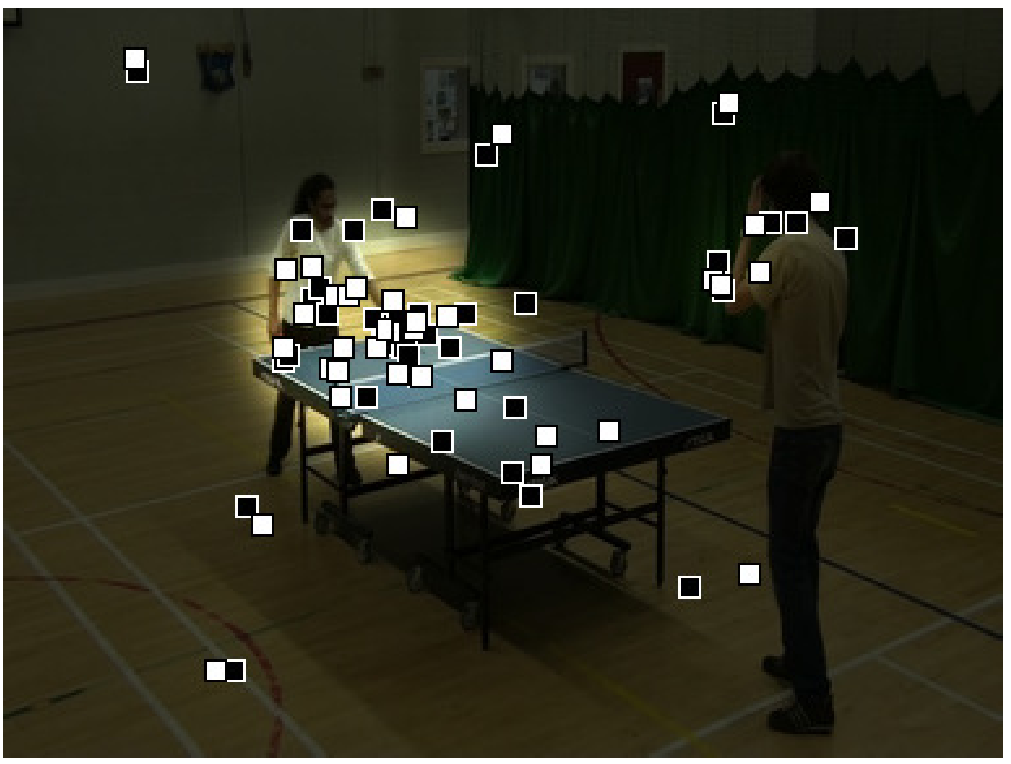}} \hspace{-2em}\hfill \\
\end{tabular}
\vspace{2em}
\caption{Sample gaze visualization from the DIEM Dataset. The gaze points of the right eye are shown as white squares, those of the left eye as black squares.}
\label{fig:gazevisualDIEM}
\end{figure}

%%%%%%%%%%
\subsubsection{The DIEM Dataset} 
Dynamic Images and Eye Movements (DIEM) project~\cite{DIEM} provides tools and data to study how people look at dynamic scenes. So far, DIEM collected gaze data for 85 sequences of 30 fps videos varying in the number of frames and resolution, using the SR Research Eyelink 1000 eye tracker. The videos were taken from various categories including movie trailers, music videos, documentary, news and advertisements. For the purpose of the study, the frames of the sequences from the DIEM dataset were re-sized to 288 pixels height, while securing the original aspect ratio, resulting in five different resolutions: $352 \times 288$, $384 \times 288$, $512 \times 288$, $640 \times 288$ and $672 \times 288$. Among 85 available videos, 20 sequences similar to those used in~\cite{borji13quantitative} were chosen for the study, and, to match the length of the SFU sequences, only the first 300 frames were used in the comparison. In the DIEM dataset, the gaze location of both eyes are available. The gaze locations of the right eye were used as ground truth in the study, while gaze locations of the left eye were used to construct benchmark models, as described in Section~\ref{sec:benchmark_models}. Clearly, the gaze points of the two eyes are very close to each other, closer than the gaze points of the first and second viewing in the SFU dataset. 
A sample frame form each selected sequence, overlaid with gaze locations of both eyes, is illustrated in Fig.~\ref{fig:gazevisualDIEM}. The visualization is such that the less-attended regions (according to the right eye) are indicated by darker colors.

%%%%%%%%%%
\subsection{Benchmark Models}
\label{sec:benchmark_models}
In addition to the computational saliency models, we consider two additional models: Intra-Observer (IO) and Gaussian center-bias (GAUSS). IO saliency map is obtained by the convolution of a 2D Gaussian blob (with standard deviation of $1^{\circ}$ of visual angle) with the second set of gaze points of the same observer within the dataset. Recall that both datasets have two sets of gaze points for each sequence and each observer -- first/second viewing in the SFU dataset, right/left eye in the DIEM dataset. So the IO saliency maps for the sequences in the SFU dataset are obtained using the gaze points from the second viewing, while IO saliency maps for the sequences from the DIEM dataset are obtained using the gaze points of the left eye. These IO saliency maps can be considered as indicators of the best possible performance of a visual saliency model, especially in the DIEM dataset where the right and left eye gaze points are always close to each other. 

\begin{figure}
\centering
%\scriptsize
\begin{tabular}{cc}
\includegraphics[height=1.2in]{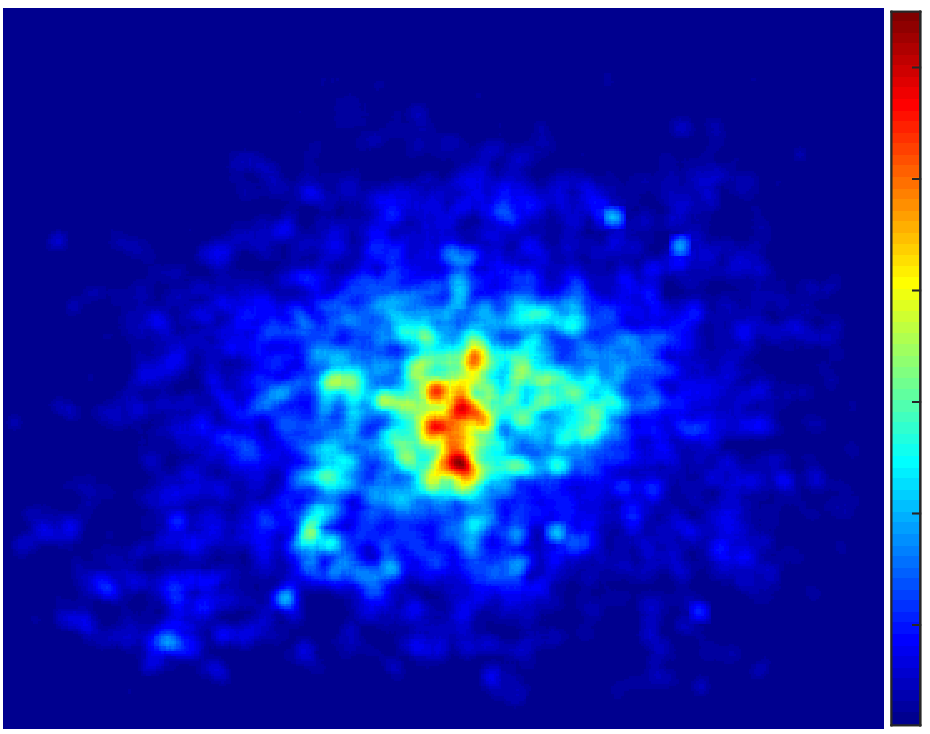}  &
\includegraphics[height=1.2in]{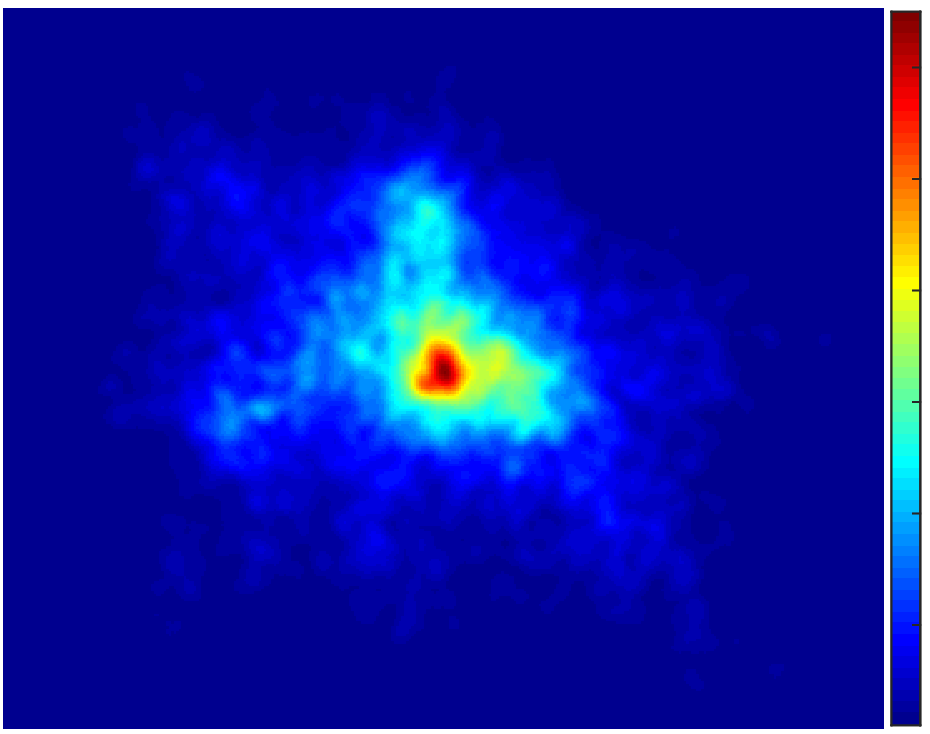} \\
SFU & DIEM
\end{tabular}
\caption{The heatmap visualization of gaze points combined across all frames and all observers, for the first viewing in the SFU dataset and the right eye in the DIEM dataset. Gaze points accumulate near the center of the frame.}
\label{fig:shuffle}
\end{figure}

On the other hand, GAUSS saliency map is just a 2D Gaussian blob with the standard deviation of $1^{\circ}$ located at the center of the frame. This model assumes that the center of a frame is the most salient point. Center bias turns out to be surprisingly powerful and has been used occasionally to boost the performance of saliency models without taking scene content into account. The underlying assumption is that the person recording the image or video will attempt to keep the salient objects at or near the middle of the frame. On average, this assumption is not too bad. Fig.~\ref{fig:shuffle} shows the heatmaps indicating cumulative gaze point locations across all sequences and all participants in the SFU dataset (first viewing) and DIEM dataset (right eye). As seen in the figure, aggregate gaze point locations do indeed cluster around the center of the frame. However, since GAUSS does not take content into account, one could expect a good saliency model to outperform it. 

%%%%%%%%%%%%%%%%%%%
% Section III
%%%%%%%%%%%%%%%%%%%
\section{Evaluation Framework}
\label{sec:evaluation}
%%%%%%%%%%%%%%%%%%%%%%%%%%%%%%%%%%%%%%%
\subsection{Implementation Settings}
\label{sec:implementation}
%%%%%%%%%%%%%%%%%%%%%%%%%%%%%%%%%%%%%%%
In order to have a unified framework for comparison, we have implemented all models in MATLAB 8.5 on the same machine, an Intel (R) Core (TM) i7 CPU at 3.40 GHz and 16 GB RAM running 64-bit Windows 8.1. Where possible, we verified the implementation by comparing the results with those presented in the corresponding papers and/or by contacting the authors. As seen in Table~\ref{tab:Models}, each model assumed a certain video coding standard. However, fundamentally, all models except OBDL-MRF~\cite{khatoonabadi15how} rely on the same type of information -- MVs and DCT of residual blocks (DCT-R) or pixel blocks (DCT-P). OBDL-MRF, on the other hand, directly uses block description length as an indicator of saliency, without the need to decode MVs or prediction residuals. The main difference is in the size of the blocks to which MVs are assigned or to which DCT is applied. In standards up to MPEG-4 ASP, the minimum block size was $8 \times 8$, whereas H.264/AVC allowed block sizes down to $4 \times 4$~\cite{wiegand03overview}. In pursuance of a fair comparison, for which all models should accept the same input data, we chose to encode all videos in two currently most widely used video formats -- MPEG-4 ASP and H.264/AVC. Each choice ensured that seven out of eleven models in the study did not require modification. Minor modification was necessary in order for APPROX~\cite{hadizadeh13visual} to accept compressed input data. Specifically, for MPEG-4 ASP input data, where the spatial saliency map relies on DCT values of $16 \times 16$ pixel blocks, the $16 \times 16$ DCT was computed from the $8 \times 8$ DCTs using a fast algorithm from~\cite{he05bidirectional}. Also, minimum MV block size was set to $8 \times 8$. In case of H.264/AVC input data, only P-frames were considered. In summary, the first group of models that takes MPEG-4 ASP bitstream as an input comprises of models \{1, 2, 3, 4, 6, 7, 8, 9\} from Table~\ref{tab:Models}, while the second group that takes H.264/AVC bitstreams includes models \{1, 2, 3, 4, 5, 8, 10, 11\} in the table.

We considered two configurations to encode the videos used in the evaluation. For the first group, the Group-of-Pictures (GOP) structure was set to IPPP with the GOP size of 12, i.e., the first frame is coded as intra (I), the next 11 frames are coded predictively (P), then the next frame is coded as I, and so on. The MV search range was set to 16 with 1/4-pel motion compensation with QP$\in$\{1, 4, 7, ..., 31\}.
In the decoding stage, the DCT-P values (in I-frames) and DCT-R values (in P-frames), as well as MVs (in P-frames) were extracted from the encoded MPEG-4 ASP bitstream for each $8 \times 8$ block. For the second group, encoding was done using H.264/AVC with QP$\in$\{3, 6, 9, ..., 51\} in the baseline profile. In our setting, for each MB, there exists up to four MVs having 1/4-pixel accuracy with no range restriction. Other settings were set to default. Encoding and partial decoding to extract the required data was accomplished using the FFMPEG library~\cite{ffmpeg}.

%%%%%%%%%%%%%%%%%%%%%%%%%%%%%%%%
\subsection{Accuracy Evaluation}
%%%%%%%%%%%%%%%%%%%%%%%%%%%%%%%%
A number of methods have been used to evaluate the accuracy of visual saliency models with respect to gaze point data~\cite{borji13state,borji13quantitative,einhauser08objects,itti05principled,itti09bayesian,le13methods}. Since each method emphasizes a particular aspect of model's performance, to make the evaluation balanced, a collection of methods and metrics is employed in this study. A model that offers high score across many metrics can be considered to be fairly accurate.

%%%%%%%%%%
\subsubsection{Area Under Curve (AUC)} 
The area under curve or, more precisely, the area under Receiver Operating Characteristic (ROC) curve, is computed from the graph of the True Positive Rate (TPR) versus the False Positive Rate (FPR) at various threshold parameters~\cite{swets96signal}. In the context of saliency maps, the saliency values are first divided into positive and negative sets corresponding to gaze and non-gaze points. Then for any given threshold, TPR and FPR are, respectively, obtained as the fraction of elements in the positive set and in the negative set that are greater than the threshold. Essentially, by varying the threshold, the ROC curve of TPR versus FPR is generated, visualizing the performance of a saliency model across all possible thresholds. The area under this curve quantifies the performance and shows how well the saliency map can predict gaze points. A larger AUC implies a greater correspondence between gaze locations and saliency predictions. A small AUC indicates weaker correspondence. The AUC is in the range $[0,1]$: the value of $1$ indicates the saliency algorithm performs well, the value of $0.5$ represents pure chance performance, and the value of less than $0.5$ represents worse than pure chance performance. This metric is also invariant to monotonic scaling of saliency maps~\cite{brown06receiver}.

It is worth mentioning that instead of using all non-gaze saliency values, these are usually sampled~\cite{schisterman01statistical,einhauser08objects}. The idea behind this approach is that an effective saliency model would have higher values at fixation points than at randomly sampled points. Control points for non-gaze saliency values are obtained with the help of a nonparametric bootstrap technique~\cite{efron93introduction}, and sampled with replacement, with sample size equal to the number of gaze points, from non-gaze parts of the frame multiple times. Finally, the average of the statistic over all bootstrap subsamples is taken as a sample mean. 

%%%%%%%%%%
\subsubsection{Kullback-Leibler Divergence (KLD) and J-Divergence (JD)} 
The KLD is often used to obtain the divergence between two probability distributions. It is given by the relative entropy of one distribution with respect to another~\cite{kullback51information}

\begin{equation}
\label{eq:KLD}
KLD(P\|Q)=\sum_{i=1}^r P(i) \cdot \log_b\left(\frac{P(i)}{Q(i)} \right),
\end{equation}

\noindent where $P$ and $Q$ are discrete probability distributions, $b$ is the logarithmic base, and $r$ indicates the number of bins in each distribution. Note that KLD is asymmetric. The symmetric version of KLD, also called J-Divergence, is~\cite{jeffreys46invariant}

\begin{equation}
\label{eq:JD}
JD(P\|Q)=KLD(P\|Q)+KLD(Q\|P).
\end{equation}

To assess how accurately a saliency model predicts gaze locations based on the symmetric KLD, the distribution of saliency values at the gaze locations is compared against the distribution of saliency values at some random points from non-gaze locations~\cite{itti05principled,itti06bayesian,itti09bayesian}. If these two distributions overlap substantially, i.e., the divergence JD approaches zero, then the saliency model predicts gaze points no better than a random guess. On the other hand, as one distribution diverges from the other and the divergence JD increases, the saliency model is better able to predict gaze points. 

Specifically, let there be $n$ gaze points in a frame. Another $n$ points different from the gaze points are randomly selected from the frame. The saliency values at the gaze points and the randomly selected points constitute the two distributions, $P$ and $Q$. A good saliency model would produce a large JD. The process of choosing random samples and computing the JD is usually repeated many times and the resulting JD values are averaged to minimize the effect of random variations. While JD has certain advantages (please refer to~\cite{itti09bayesian,borji13state} for details), it also faces several problems. One of the problems with KLD and JD is the lack of an upper bound~\cite{kullback97information}. Another problem is that if $P(i)$ or $Q(i)$ is zero for some $i$, one of the terms in (\ref{eq:JD}) is undefined. For these reasons, JD was not used in the present study. 

%%%%%%%%%%
\subsubsection{Jensen-Shannon Divergence (JSD)} 
The Jensen-Shannon divergence (JSD) is a KLD-based metric that avoids some of the problems faced by KLD and JD~\cite{lin91divergence}. For two probability distributions $P$ and $Q$, JSD is defined as~\cite{dagan97similarity}: 

\begin{equation}
JSD(P\|Q)=\frac{KLD(P\|R)+KLD(Q\|R)}{2},
\end{equation}

\noindent where

\begin{equation}
R=\frac{P+Q}{2}.
\end{equation}

\noindent Unlike KLD, JSD is a proper metric, is symmetric in $P$ and $Q$, and is bounded in $[0,1]$ if the logarithmic base is set to $b=2$~\cite{lin91divergence}. The value of the JSD for the saliency map that perfectly predicts gaze points will be equal to $1$. The same sampling strategy employed in AUC computation can also be used for computing JSD.

%%%%%%%%%%
\subsubsection{Normalized Scanpath Saliency (NSS)}
NSS measures the strength of normalized saliency values at gaze locations~\cite{peters05components}. Normalization is affine so that the resulting normalized saliency map has zero mean and unit standard deviation. The NSS is defined as the average of normalized saliency values at gaze points. A positive normalized saliency value at a certain gaze point indicates that the gaze point matches one of the predicted salient regions, zero indicates no link between predictions and the gaze point, while a negative value indicates that the gaze point has fallen into an area predicted to be non-salient.

%%%%%%%%%%
\subsubsection{Pearson Correlation Coefficient (PCC)}
PCC measures the strength of the linear relationship between a predicted saliency map $S$ and the ground truth map $G$. First, the ground truth map $G$ is obtained by convolving the gaze point map with a 2D Gaussian function having the standard deviation of $1^{\circ}$ of visual angle~\cite{le13methods}. Then $S$ and $G$ are treated as random variables whose paired samples are given by values of the two maps at each pixel position in the frame. The Pearson correlation coefficient is defined as
\begin{equation}
corr(G,S)=\frac{cov(G,S)}{\sigma_G\sigma_S},
\end{equation}
where $cov(\cdot,\cdot)$ denotes covariance and $\sigma_G$ and $\sigma_S$ are, respectively, the standard deviations of the ground truth map and the predicted saliency map. The value of PCC is between $-1$ and $1$; the value of $\pm1$ indicates the strongest linear relationship, whereas the value of $0$ indicates no correlation. If the model's saliency values tend to increase as the values in the ground truth map increase, the PCC is positive. Otherwise, if the model's saliency values tend to decrease as the ground truth values increase, the PCC is negative. In this context, a PCC value of $-1$ would mean that the model predicts non-salient regions as salient, and salient regions as non-salient. While this is the opposite of what is needed, such model can still be considered accurate if its saliency map is inverted. While PCC is widely used for studying relationships between random variables, in its default form it has some shortcomings in the context of saliency model evaluation, especially due to center bias, as discussed in the next section.

%%%%%%%%%%
\subsection{Data Analysis Considerations}
Here, we discuss several considerations about the ground truth data, and the methods and metrics used in the evaluation.

\subsubsection{Gaze Point Uncertainty} 
\label{sec:GPUncertainty}
Eye-tracking datasets usually report a single point $(x,y)$ as the gaze point of a given subject in a given frame. However, such data should not be treated as absolute. There are at least two sources of uncertainty in the measurement of gaze points. One is the eye-tracker's measurement error, which is usually on the order of $0.5^{\circ}$ to $1^{\circ}$ of the visual angle~\cite{locarna,sr,mital11clustering}. The other source of uncertainty is the involuntary eye movement during fixations. The human eye does not concentrate on a stationary point during a fixation, but instead constantly makes small rapid movements to make the image more clear~\cite{carlson10psychology}. Depending on the implementation, the eye tracker may filter those rapid movements out, either due to undersampling or to create an impression of a more stable fixation. For at least these two reasons, the gaze point measurement reported by an eye tracker contains some uncertainty. At the current state of technology, the eye tracker measurement errors seem to be larger than the uncertainty caused by involuntary drifts, and so we take them as the dominant source of noise in the ground truth data. To account for this noise, we apply a local maximum operator in a radius of $0.5^{\circ}$ of visual angle. In other words, when computing a saliency value of a given point in a frame, the maximum value within its small neighborhood is used. 

The use of the local maximum operator is meant to counter the effects of measurement noise in the gaze tracking system, which is usually rated at around $0.5^{\circ}$ of visual angle. Hence, the true fixation point may be within $0.5^{\circ}$ of visual angle away from what the gaze measurement system reports. An accurate saliency model produces small saliency values at locations far from fixation points and high saliency values at locations near fixation points. Therefore, for an accurate model, applying a local maximum operator does not change saliency values away from fixations while it ensures that near fixations, the maximum predicted saliency value within the measurement tolerance is considered. So we expect that the accuracy score for an accurate model will increase using this approach. For an inaccurate model (one that produces low saliency values near fixations and large ones away from fixations), we expect little or no change in the score. This is because its predicted saliency values near fixations (which are low) will not increase, while its predicted saliency values away from fixations may get a boost, but they don't matter because they are away from fixations anyway.

\subsubsection{Center Bias and Border Effects}
\label{sec:CBBE}
A person recording a video will generally tend to put regions of interest near the center of the frame~\cite{tatler05visual,parkhurst03scene}. In addition, people also have a tendency to look at the center of the image~\cite{tatler07central}, presumably to maximize the coverage of the displayed image by their field of view. These phenomena are known as center bias. Fig.~\ref{fig:shuffle} illustrates the center bias in the SFU and DIEM datasets by displaying the locations of gaze points accumulated over all sequences and all frames. 

Interestingly, Kanan \textit{et al.}~\cite{kanan09sun} and Borji \textit{et al.}~\cite{borji13quantitative} showed that creating a saliency map merely by placing a Gaussian blob at the center of the frame may result in fairly high scores. Such high scores are partly caused by using a uniform spatial distribution over the image when selecting control samples. Specifically, the computation of ACU, KLD and JSD for a given model involves choosing non-gaze control points randomly in an image. If these are chosen according to a uniform distribution across the image, the process results in many control points near the border, which, empirically, have little chance of being salient. As a result, the saliency values of those control points tend to be small, resulting in an artificially high score for the model under test. At the same time, since gaze points are likely located near the center of the frame, a centered Gaussian blob would tend to match many of the gaze points, which would make its NSS and PCC scores high.

Additionally, Zhang \textit{et al.}~\cite{zhang08sun} thoroughly investigated the effect of dummy zero borders against evaluation metrics. Adding dummy zero saliency values at the border of the image changes the distribution of saliency of the random samples as well as the normalization parameters in NSS, leading to different scores while the saliency prediction is unchanged. To decrease sensitivity to center bias and border effect, Tatler \textit{et al.}~\cite{tatler05visual} and Parkhurst and Niebur~\cite{parkhurst03scene} suggested to distribute random samples according to the measured gaze points. To this end, Tatler \textit{et al.}~\cite{tatler05visual} distributed random samples from human saccades and choose control points for the current image randomly from fixation points in other images in their dataset. Kanan \textit{et al.}~\cite{kanan09sun} also picked saliency values at the gaze points in the current image, while control samples were chosen randomly from the fixations in other images in the dataset. For both techniques, control points are drawn from a non-uniform random distribution according to the measured fixations, decreasing the effect of center bias. Furthermore, this way, dummy zero borders will not affect the distribution of random samples.

In this paper, we use a similar approach for handling center bias and border effects. Instead of directly using the accumulated gaze points over all frames in the dataset (Fig.~\ref{fig:shuffle}), we fit a 2D Gaussian distribution to the accumulated gaze points across both SFU and DIEM datasets. Then, control samples are chosen randomly from the fitted 2D Gaussian distribution. This reduces center bias in AUC and JSD. 

To reduce center bias and border effects in NSS, we modify the normalization as 

\begin{equation}
S'(x,y) = \frac{S(x,y)-\widetilde{\mu}}{\widetilde{\sigma}},
\end{equation}
 \noindent where
\begin{equation}
\widetilde{\mu} = \frac{1}{N}\sum_{(x,y)}{F(x,y) \cdot S(x,y)},
\end{equation}
\begin{equation}
\widetilde{\sigma} = \sqrt{\frac{1}{N-1}\sum_{(x,y)}{\left(F(x,y) \cdot S(x,y)-\widetilde{\mu}\right)^2}}.
\end{equation}

\noindent In the above equations, $(x,y)$ are the pixel coordinates, $N$ is the total number of pixels, and $F(x,y)$ is the fitted 2D Gaussian density evaluated at $(x,y)$ normalized such that it sums up to 1. In the normalization described by the above three equations, the pixels located near the center of the image are given more significance due to $F$. In other words, saliency predictions have the same bias as observers' fixations. These accuracy measures that are modified to reduce the center bias and border effects are indicated by prime ($'$) and referred to as NSS$'$, AUC$'$, and JSD$'$. We summarize all above-mentioned metrics in Table~\ref{tab:Metrics}. Metrics can be divided by symmetry (column 2) or boundedness (column 3). Some metrics favor center-biased saliency models (column 4). Also, some metrics are specific to saliency while others have more general applicability (column 5), e.g., for comparing two distributions. The input data for various metrics comes from three sources (column 6): 1) the locations associated with estimated saliency 2) the distribution of estimated saliency and 3) the values of estimated saliency at fixation points.

\begin{table}[t]
	\centering
	\footnotesize \addtolength{\tabcolsep}{-3pt}
	\caption{\small Summary of evaluation metrics used in the study.}
	\begin{tabular}{cccccc}
	\hline \hline
	\textbf{Metric} & \textbf{Symmetric} & \textbf{Bounded} & \textbf{Center-biased} & \textbf{Applicability} & \textbf{Input} \\ 
	\hline \hline
  \textbf{AUC} & Yes & Yes & Yes & General & Location \\
	\textbf{AUC$'$} & Yes & Yes & No & Saliency & Location \\
	\textbf{KLD} & No & No & Yes & General & Distribution \\
	\textbf{JD} & Yes & No & Yes & General & Distribution \\
	\textbf{JSD} & Yes & Yes & Yes & General & Distribution \\
	\textbf{JSD$'$} & Yes & Yes & No & Saliency & Distribution \\
	\textbf{NSS} & Yes & No & Yes & Saliency & Value \\
	\textbf{NSS$'$} & Yes & No & No & Saliency & Value \\
	\textbf{PCC} & Yes & Yes & Yes & General & Distribution \\
	\hline \hline
	\end{tabular}
	\label{tab:Metrics}
\end{table}

%%%%%%%%%%%%%%%%%%%%%
% Section IV
%%%%%%%%%%%%%%%%%%%%%
\section{Results}
\label{sec:results}

\subsection{Qualitative Comparison}
\label{sec:qualitative_result}

\begin{figure*}
	\centering
%\scriptsize
\begin{tabular}{cccc}
%\textbf{MPEG-4 ASP} & & & \\
\multicolumn{4}{c}{\textbf{MPEG-4 ASP}}\\
%\end{tabular}
%\begin{tabular}{cccc}
IO & \textit{MVF} & PMES & MAM \\
\includegraphics[width=1.0in]{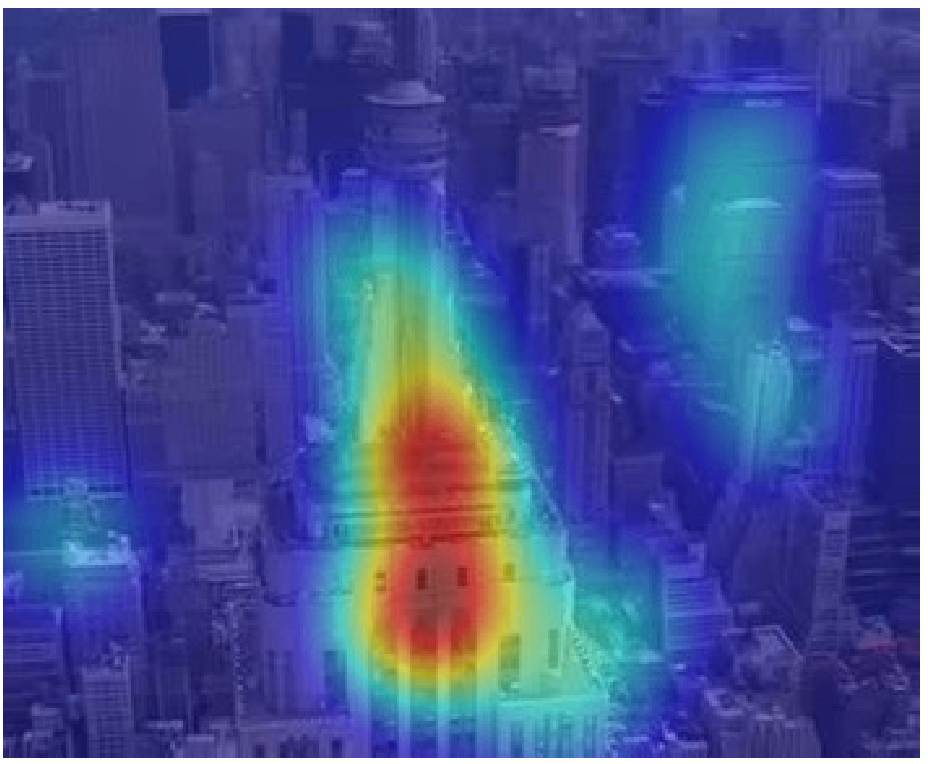}  \hspace*{-.2em}& 
\includegraphics[width=1.0in]{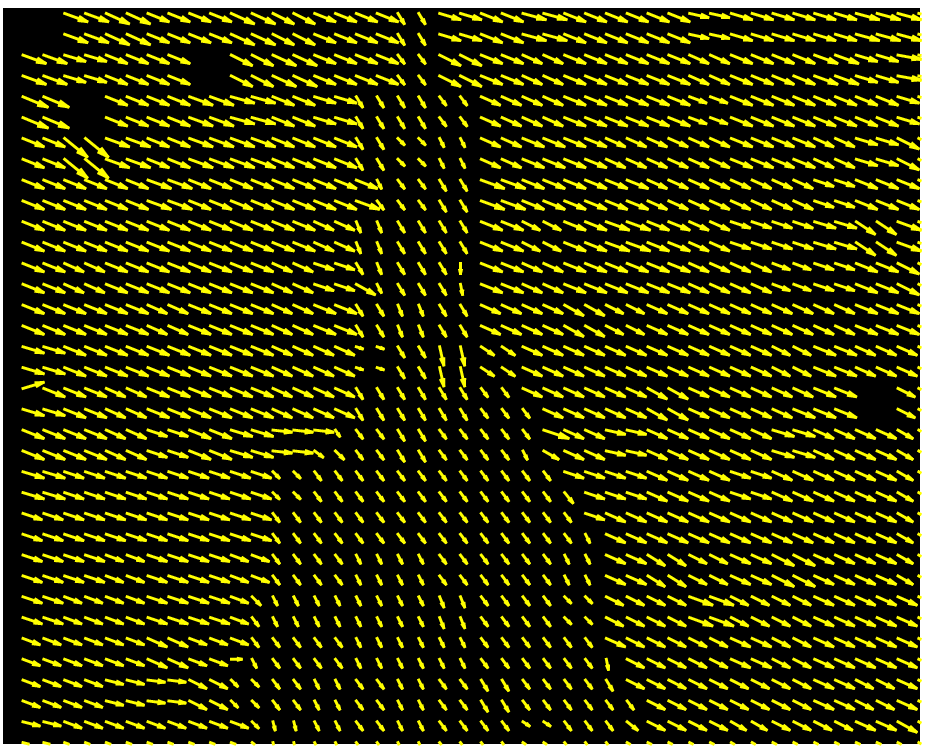} \hspace*{-.2em}& 
\includegraphics[width=1.0in]{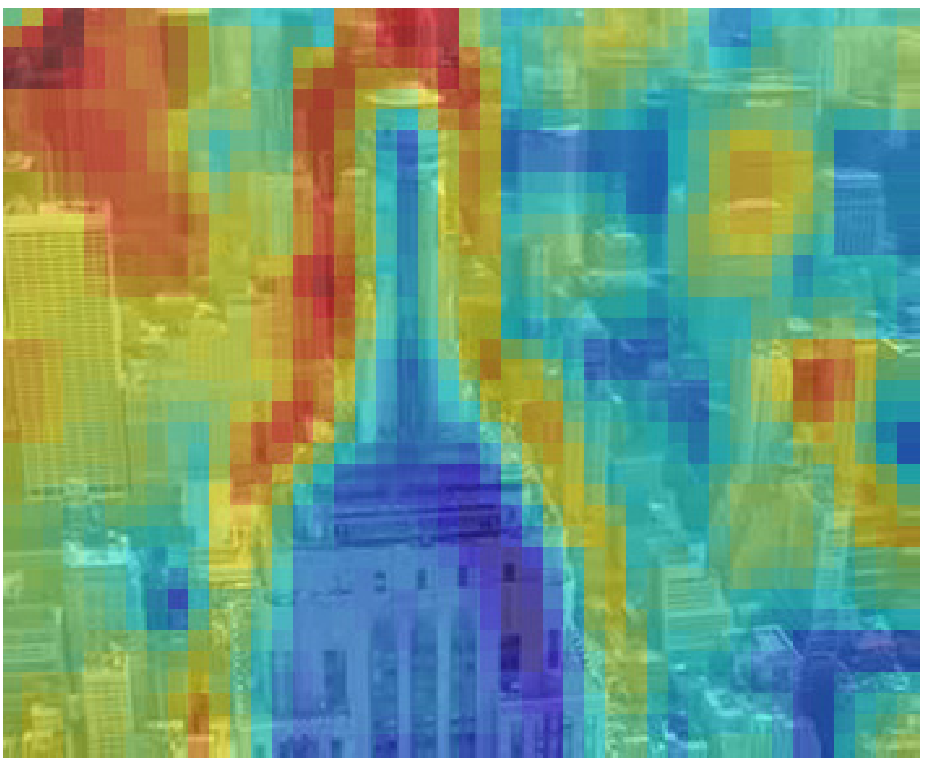}  \hspace*{-.2em}& 
\includegraphics[width=1.0in]{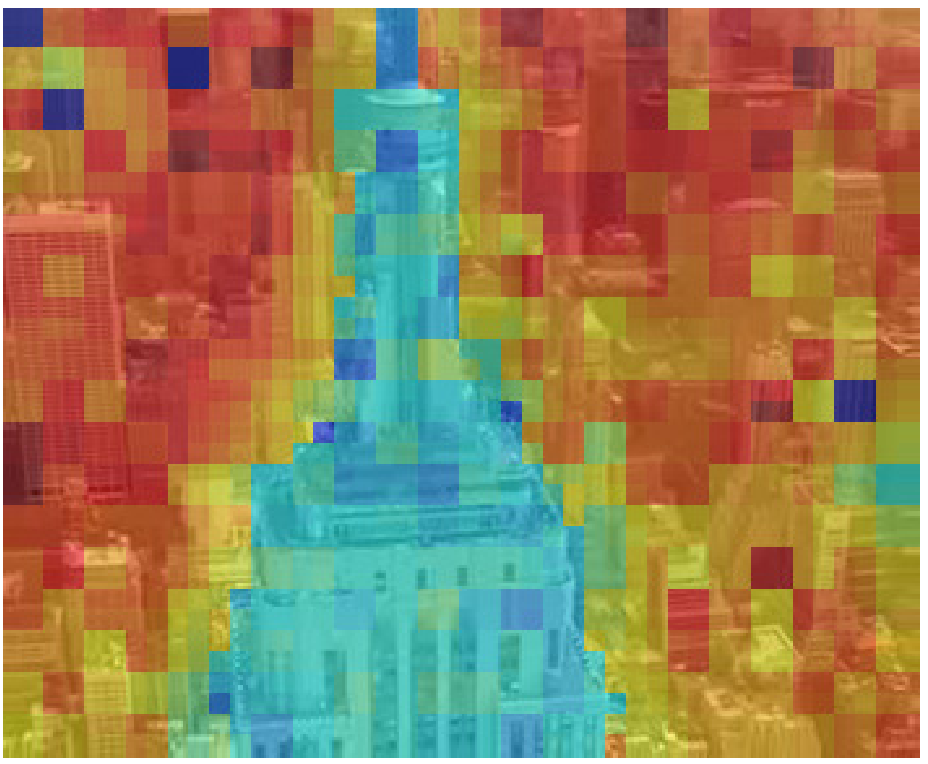}  \hspace*{-.2em}\\

PIM-ZEN & PIM-MCS & MCSDM & APPROX \\
\includegraphics[width=1.0in]{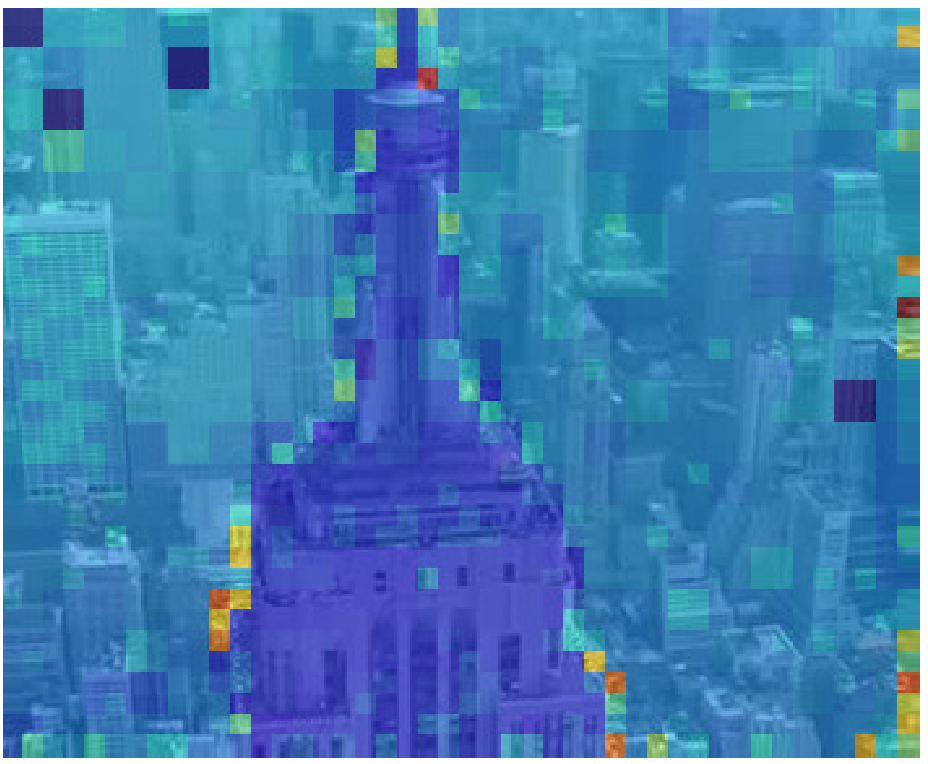}  \hspace*{-.2em}& 
\includegraphics[width=1.0in]{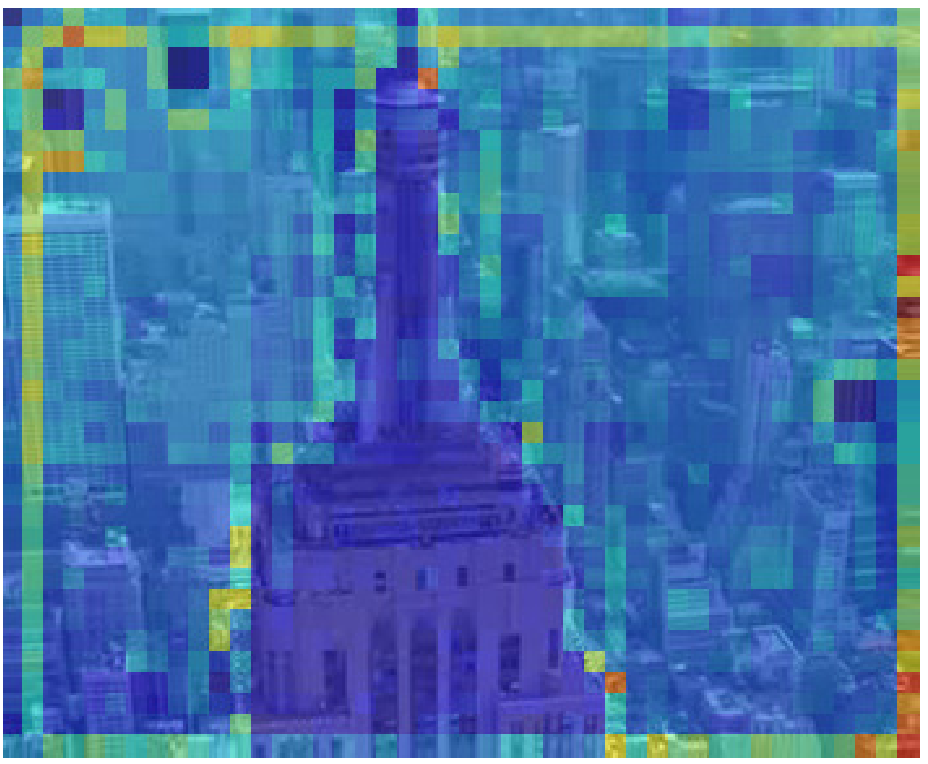}  \hspace*{-.2em} &
\includegraphics[width=1.0in]{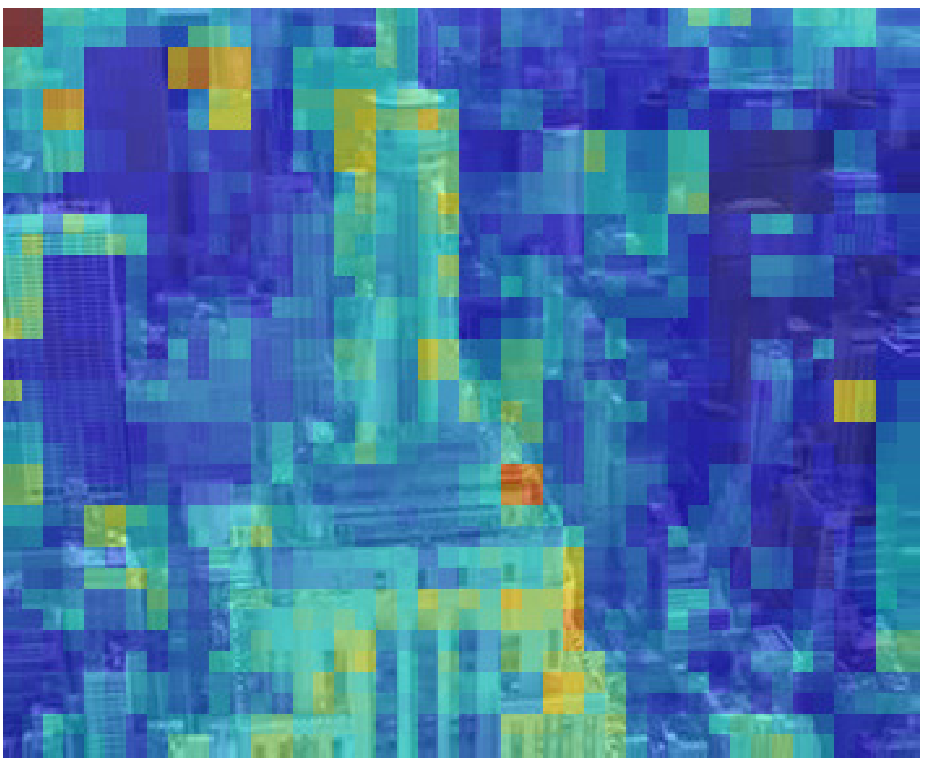}  \hspace*{-.2em} &
\includegraphics[width=1.0in]{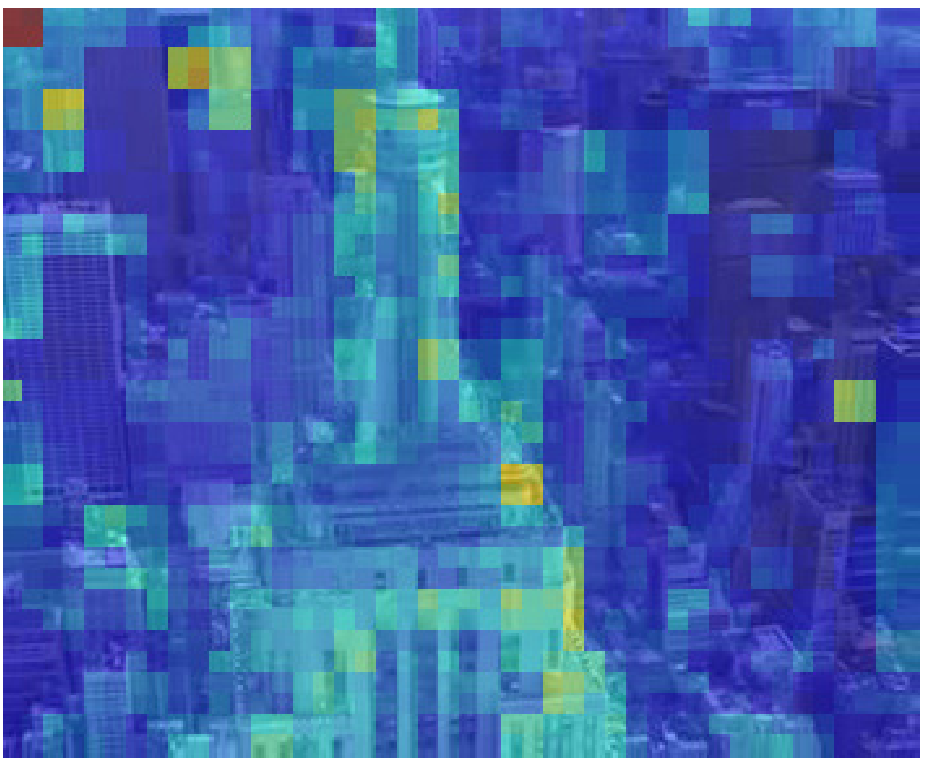}  \hspace*{-.2em}\\

OBDL-MRF & MVE+SRN & AWS & GBVS \\ 
\includegraphics[width=1.0in]{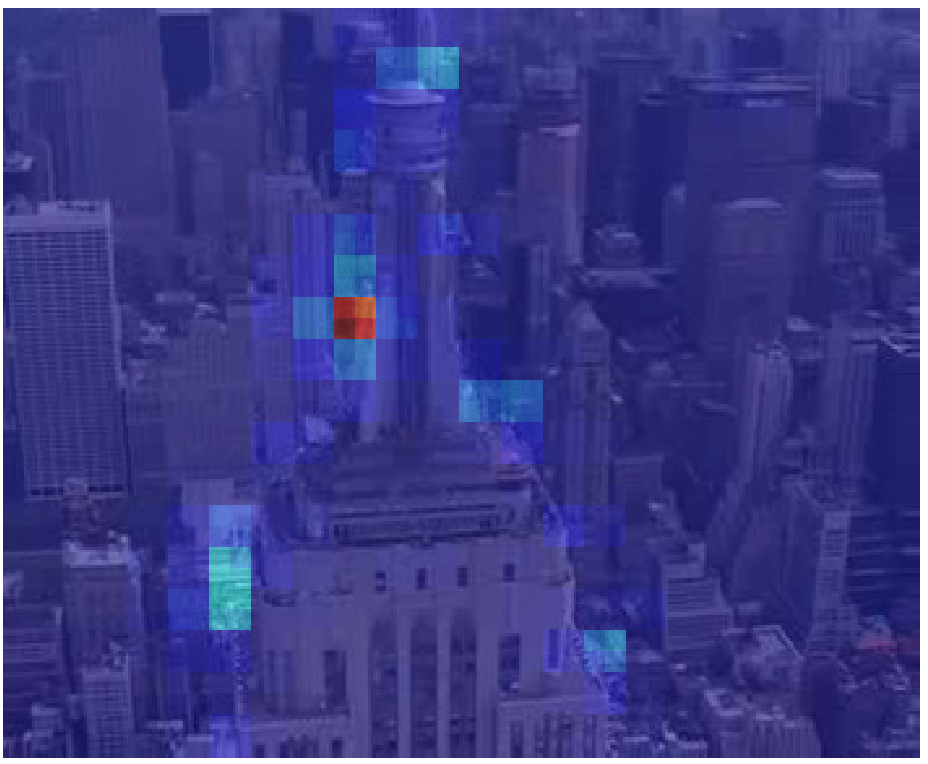}  \hspace*{-.2em}& 
\includegraphics[width=1.0in]{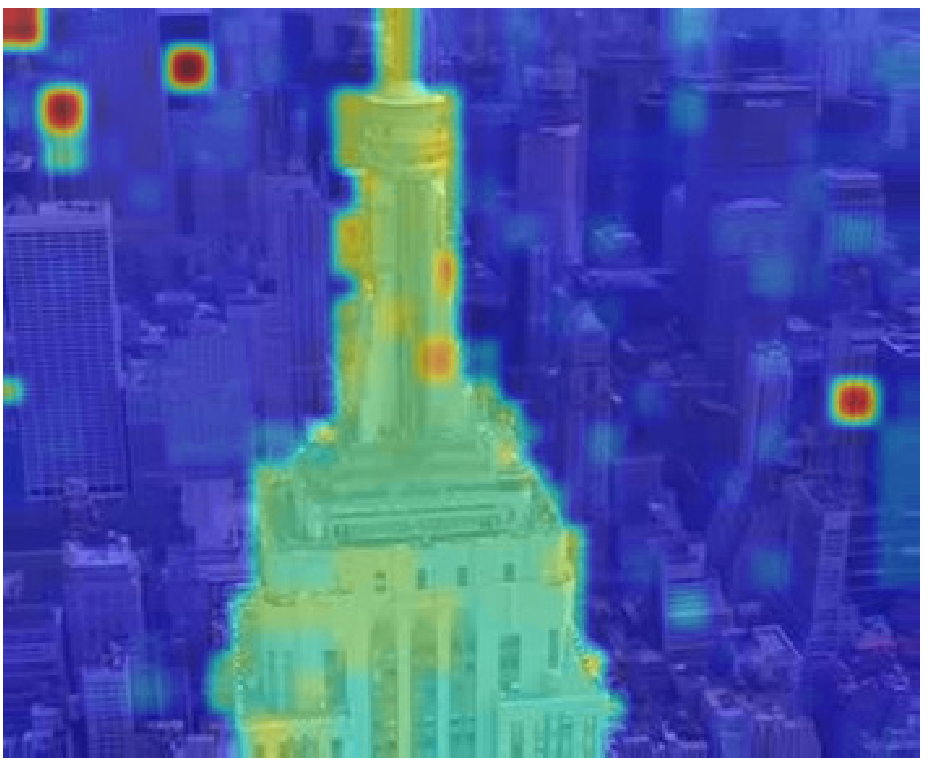}  \hspace*{-.2em}& 
\includegraphics[width=1.0in]{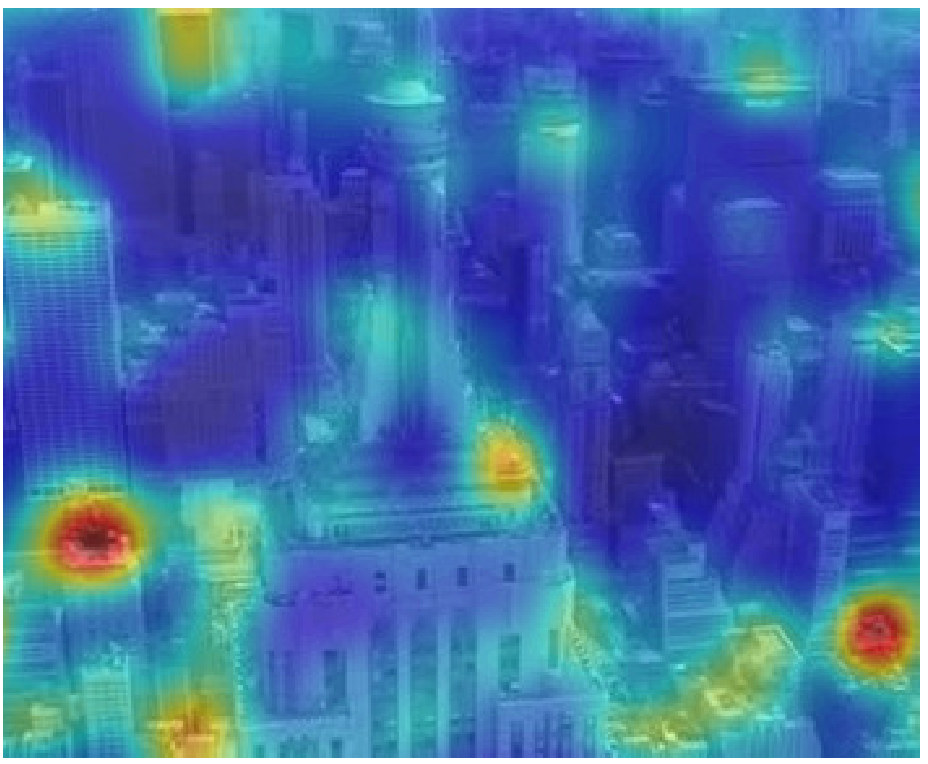}  \hspace*{-.2em}& 
\includegraphics[width=1.0in]{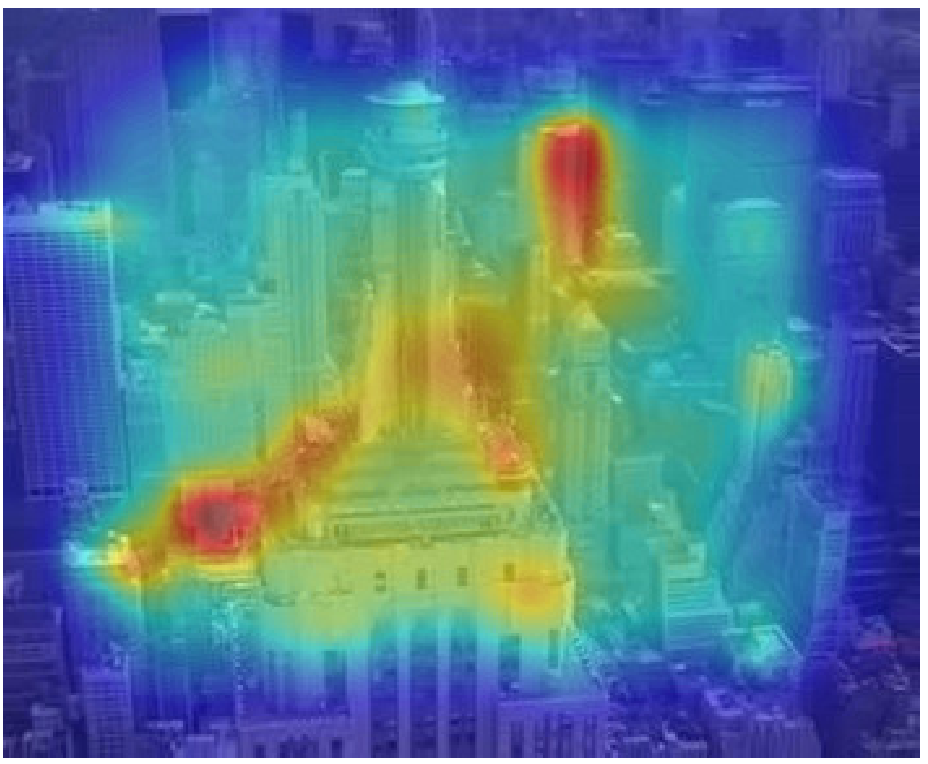} \\
%\end{tabular}	
%\begin{tabular}{c}
\\ \\ 
%\textbf{H.264/AVC} & & & \\
\multicolumn{4}{c}{\textbf{H.264/AVC}}\\
%\end{tabular}
%\begin{tabular}{cccc}
IO & \textit{MVF} & PMES & MAM \\
\includegraphics[width=1.0in]{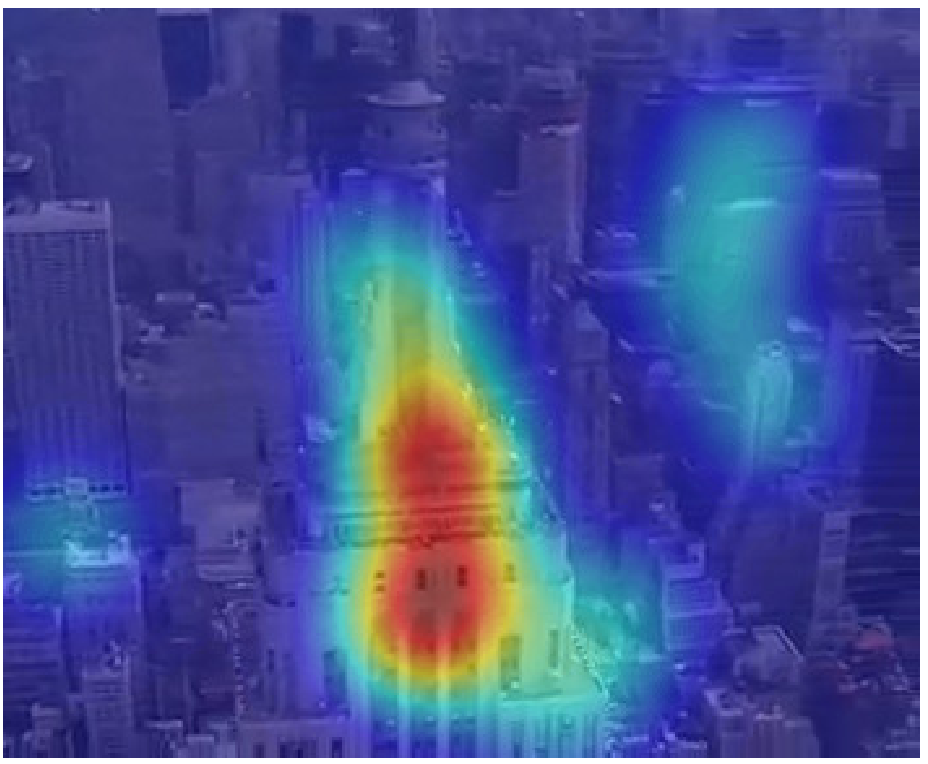}  \hspace*{-.2em}& 
\includegraphics[width=1.0in]{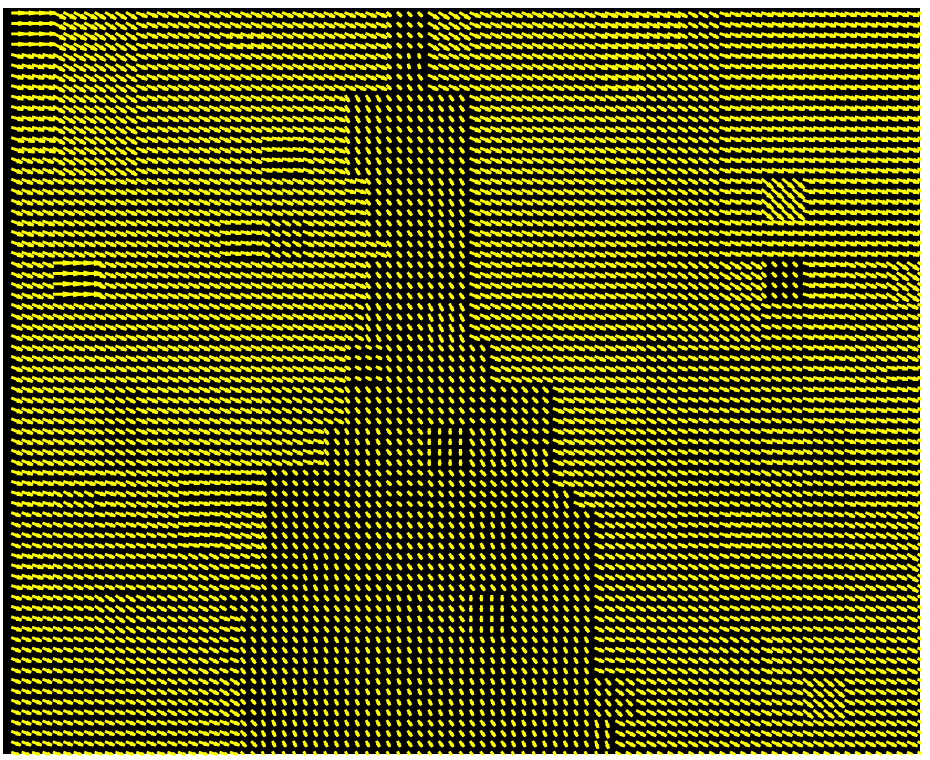} \hspace*{-.2em}& 
\includegraphics[width=1.0in]{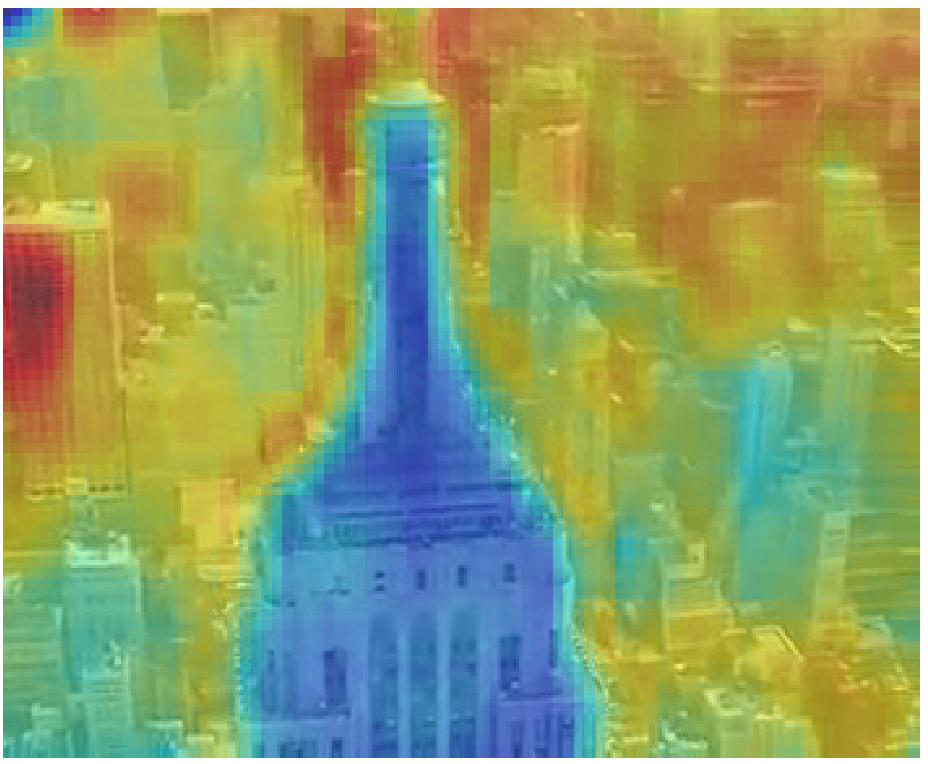}  \hspace*{-.2em}& 
\includegraphics[width=1.0in]{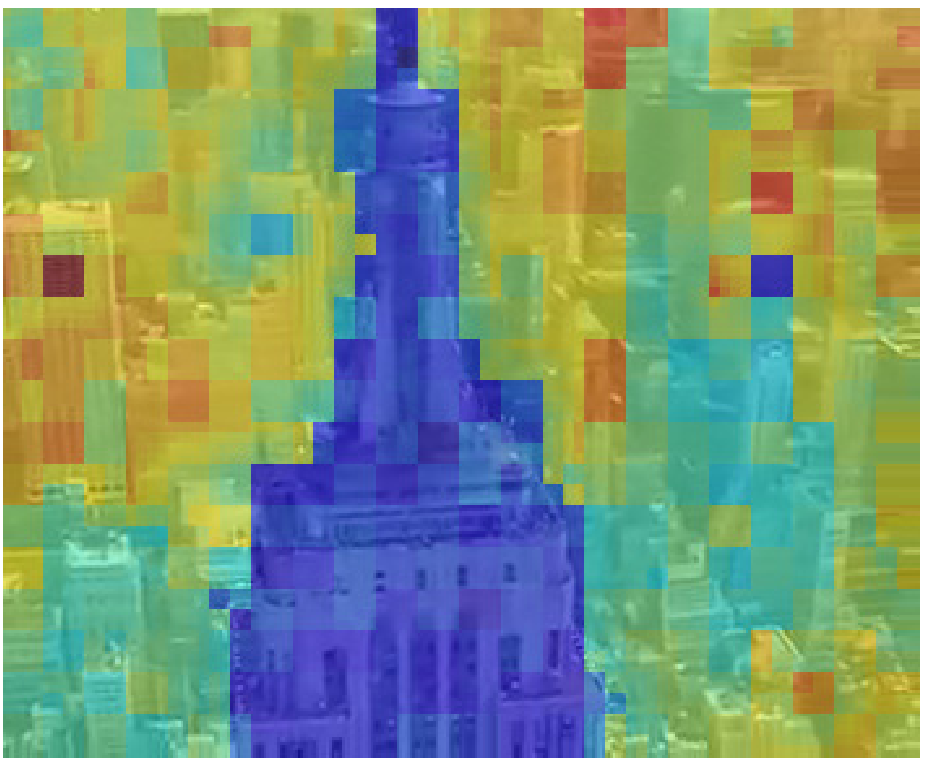}  \hspace*{-.2em}\\

PIM-ZEN & PIM-MCS & MCSDM & APPROX \\
\includegraphics[width=1.0in]{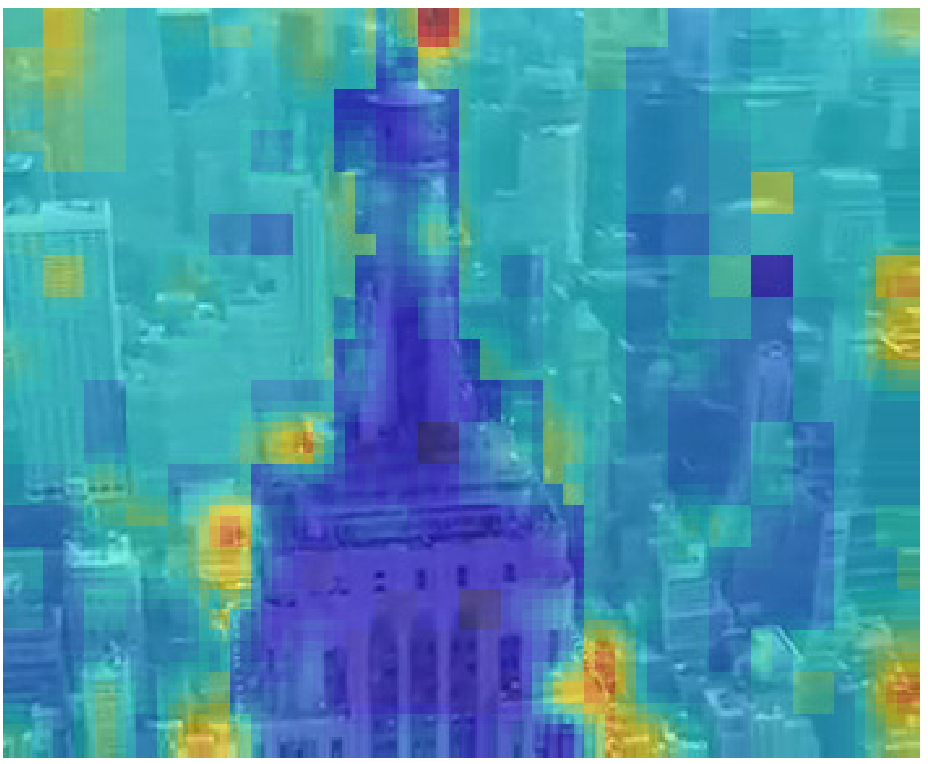}  \hspace*{-.2em}& 
\includegraphics[width=1.0in]{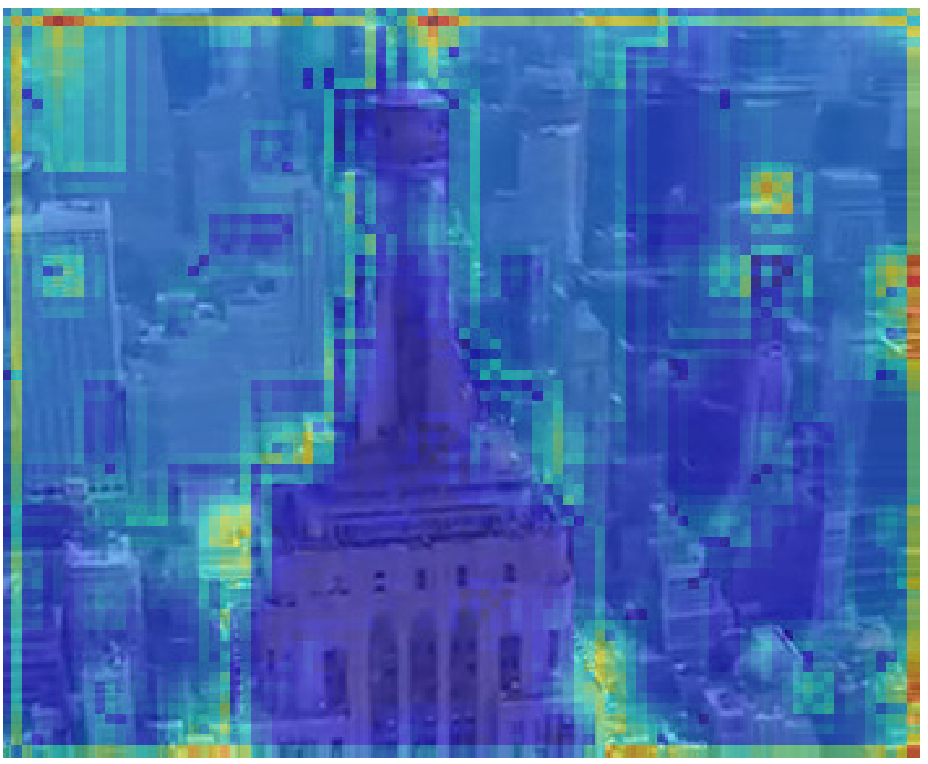}  \hspace*{-.2em} &
\includegraphics[width=1.0in]{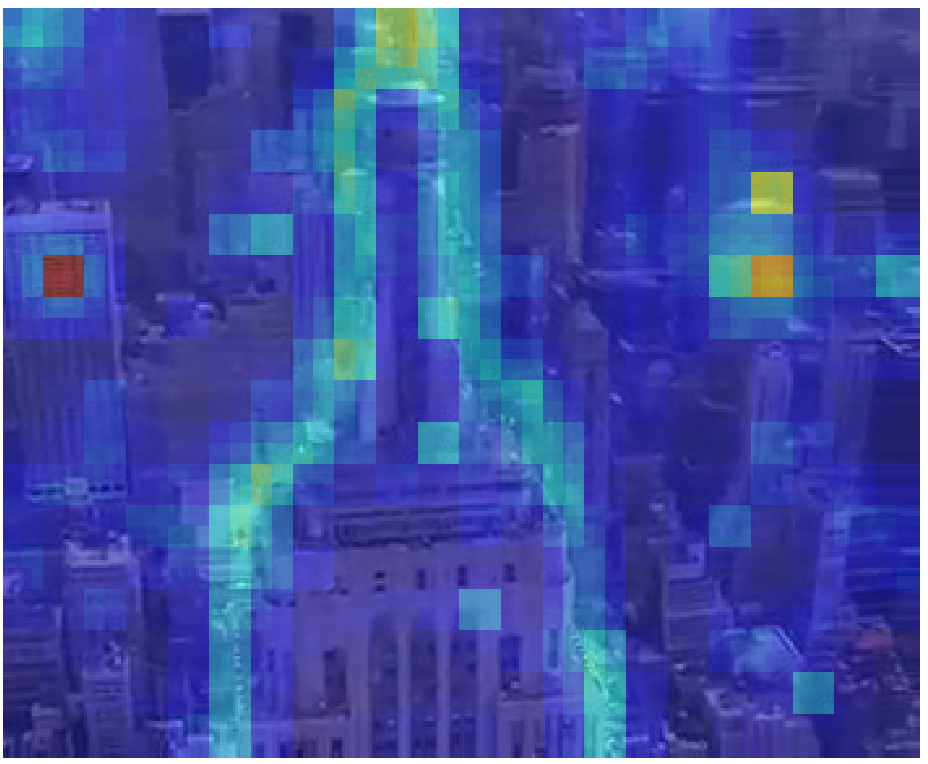}  \hspace*{-.2em} & 
\includegraphics[width=1.0in]{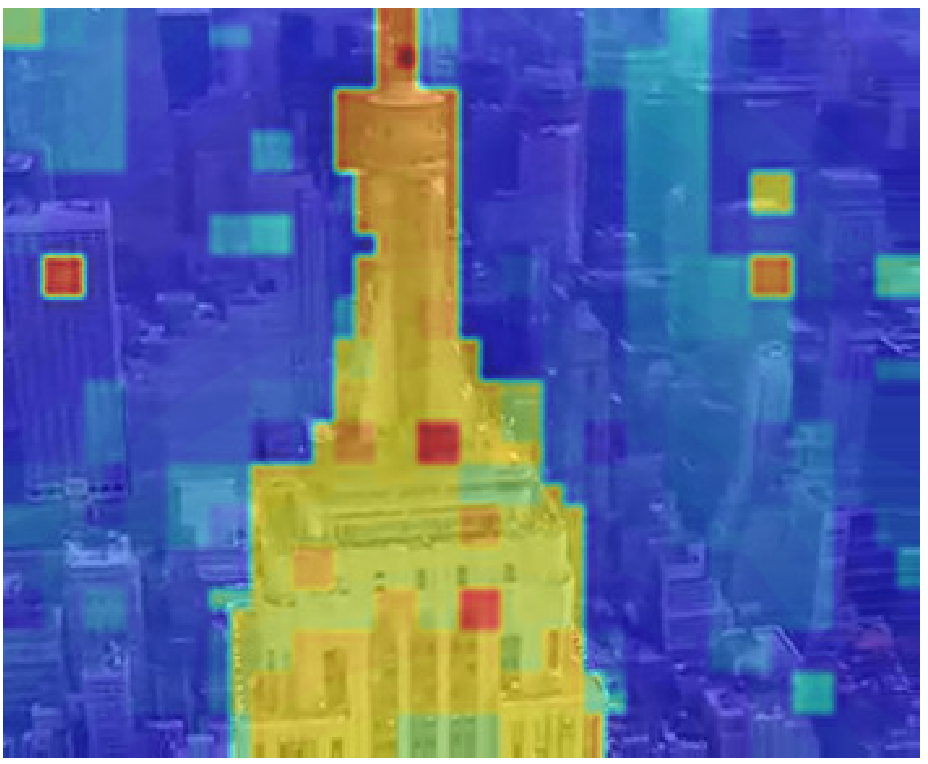}  \hspace*{-.2em}\\

OBDL-MRF & MVE+SRN & AWS & GBVS \\ 
\includegraphics[width=1.0in]{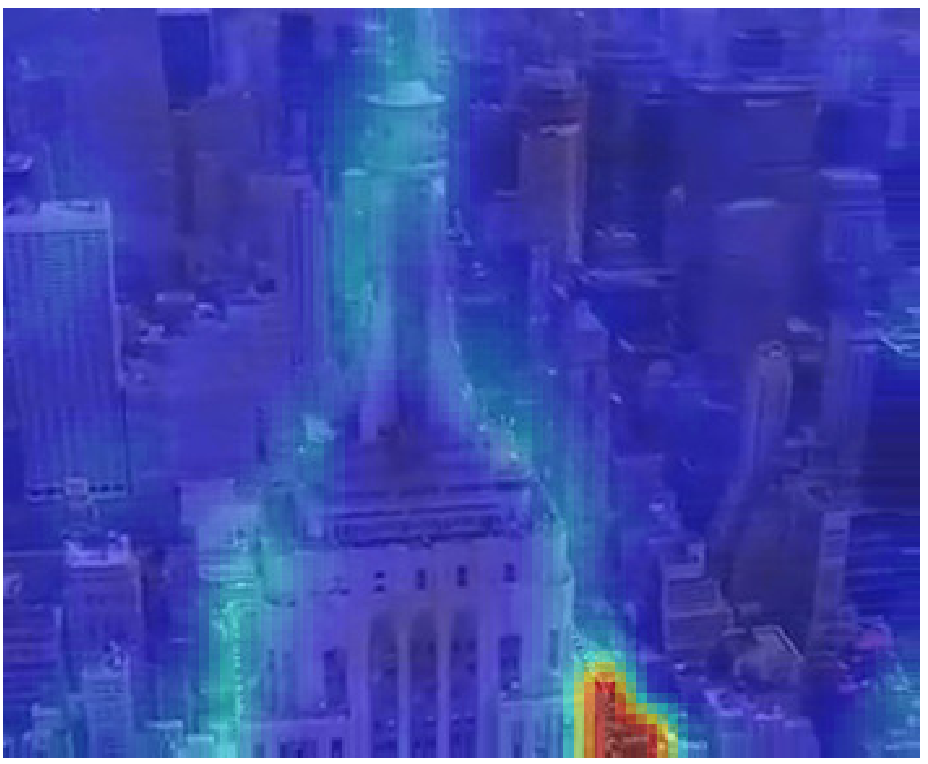}  \hspace*{-.2em}& 
\includegraphics[width=1.0in]{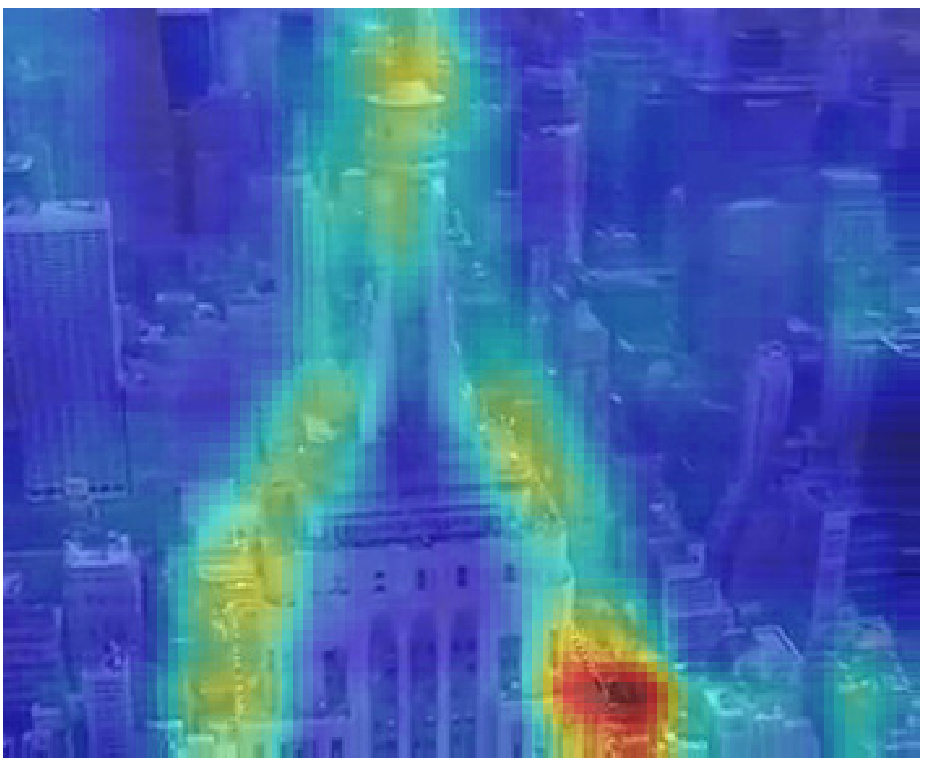}  \hspace*{-.2em}& 
\includegraphics[width=1.0in]{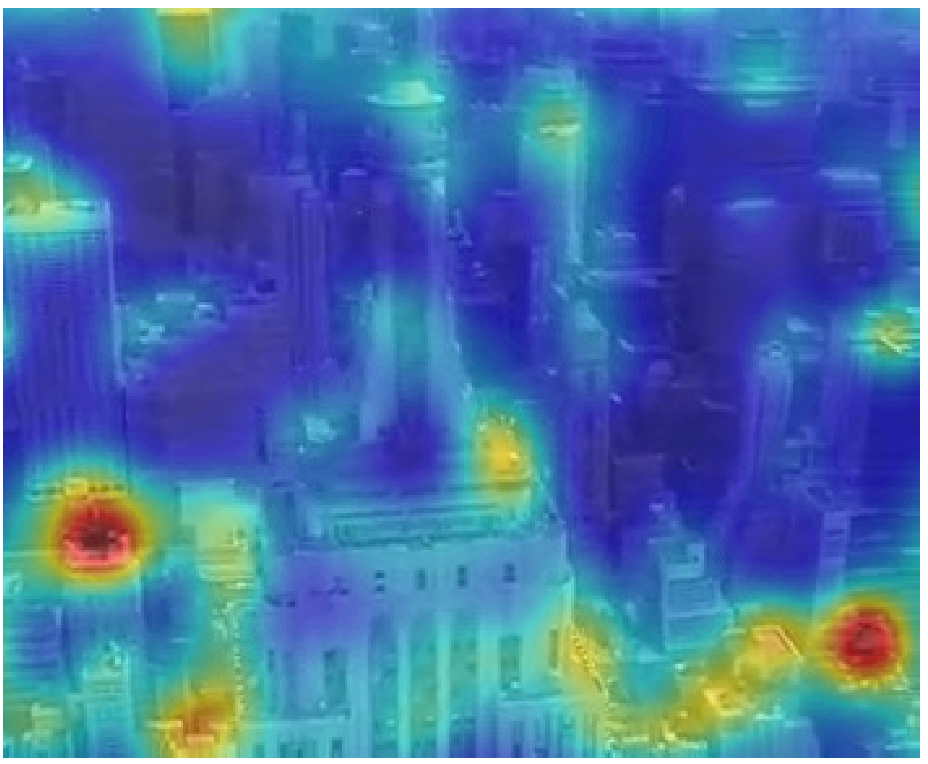}  \hspace*{-.2em}& 
\includegraphics[width=1.0in]{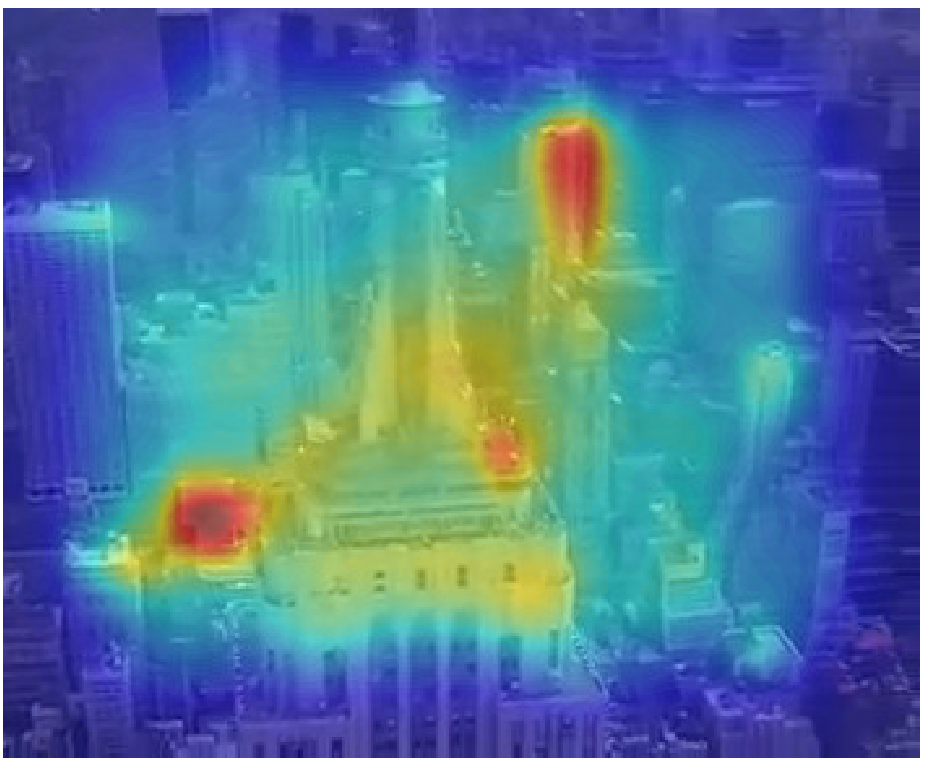} \\
\end{tabular}	
\caption{Sample saliency maps obtained by various models for \textit{City}.}
	\label{fig:SampleSaliency1}
\end{figure*}

\begin{figure*}
	\centering
%\scriptsize
\begin{tabular}{cccc}
%\textbf{MPEG-4 ASP}
\multicolumn{4}{c}{\textbf{MPEG-4 ASP}}\\
%\end{tabular}
%\begin{tabular}{cccc}
IO & \textit{MVF} & PMES & MAM \\
\includegraphics[width=1.0in]{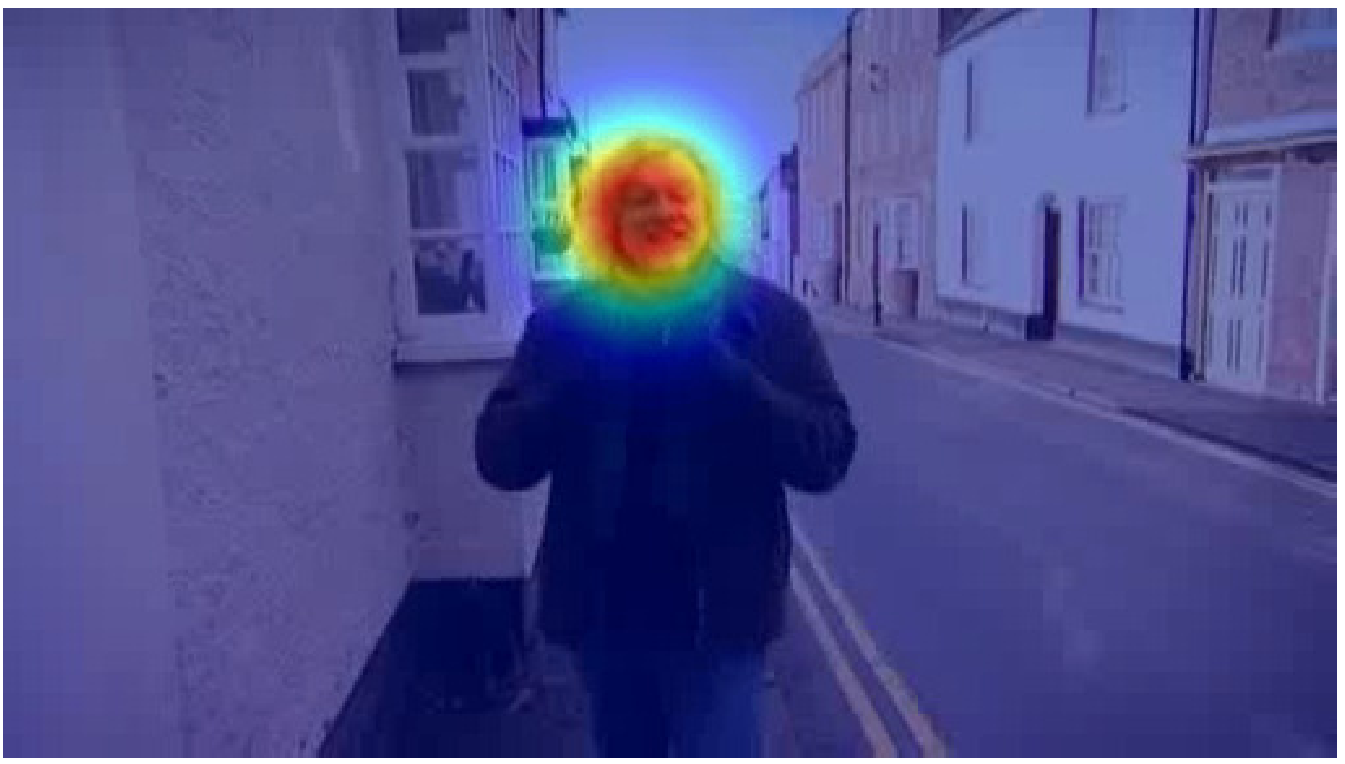}  \hspace*{-.2em}& 
\includegraphics[width=1.0in]{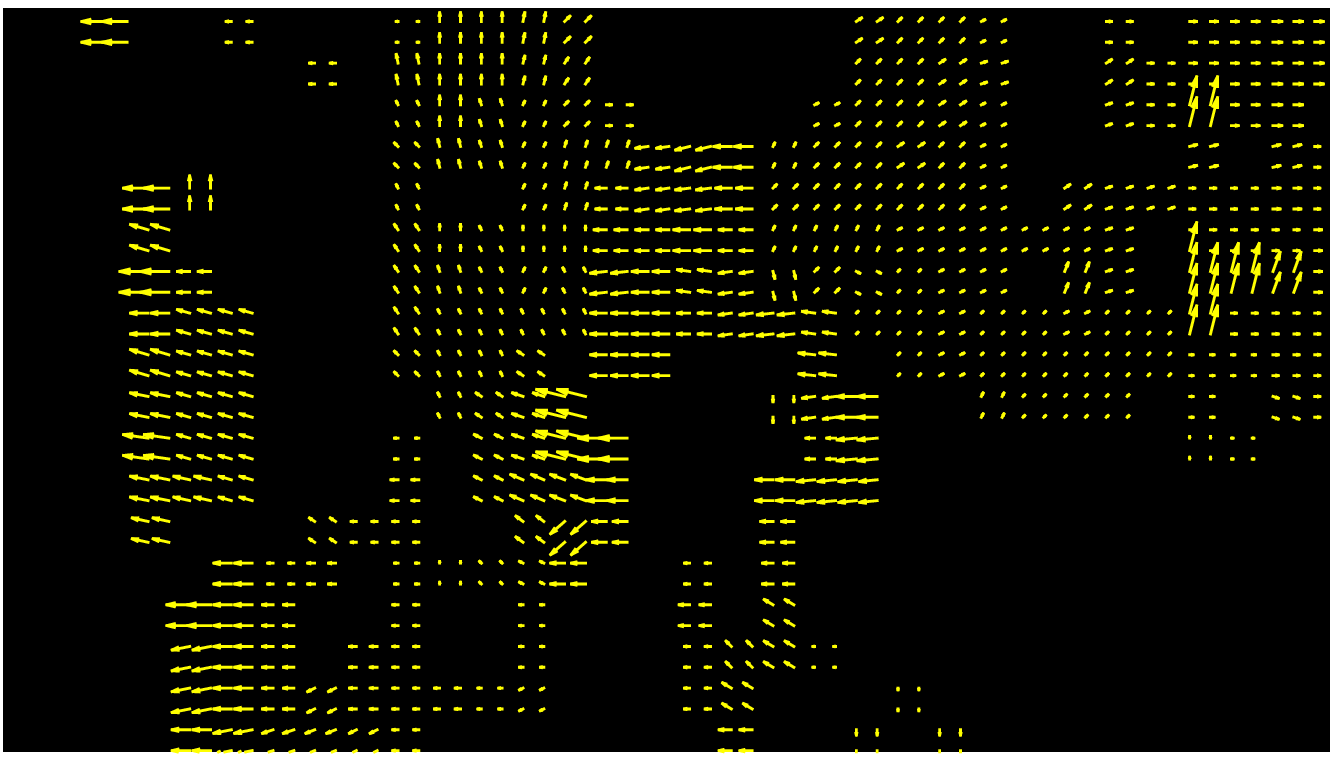} \hspace*{-.2em}& 
\includegraphics[width=1.0in]{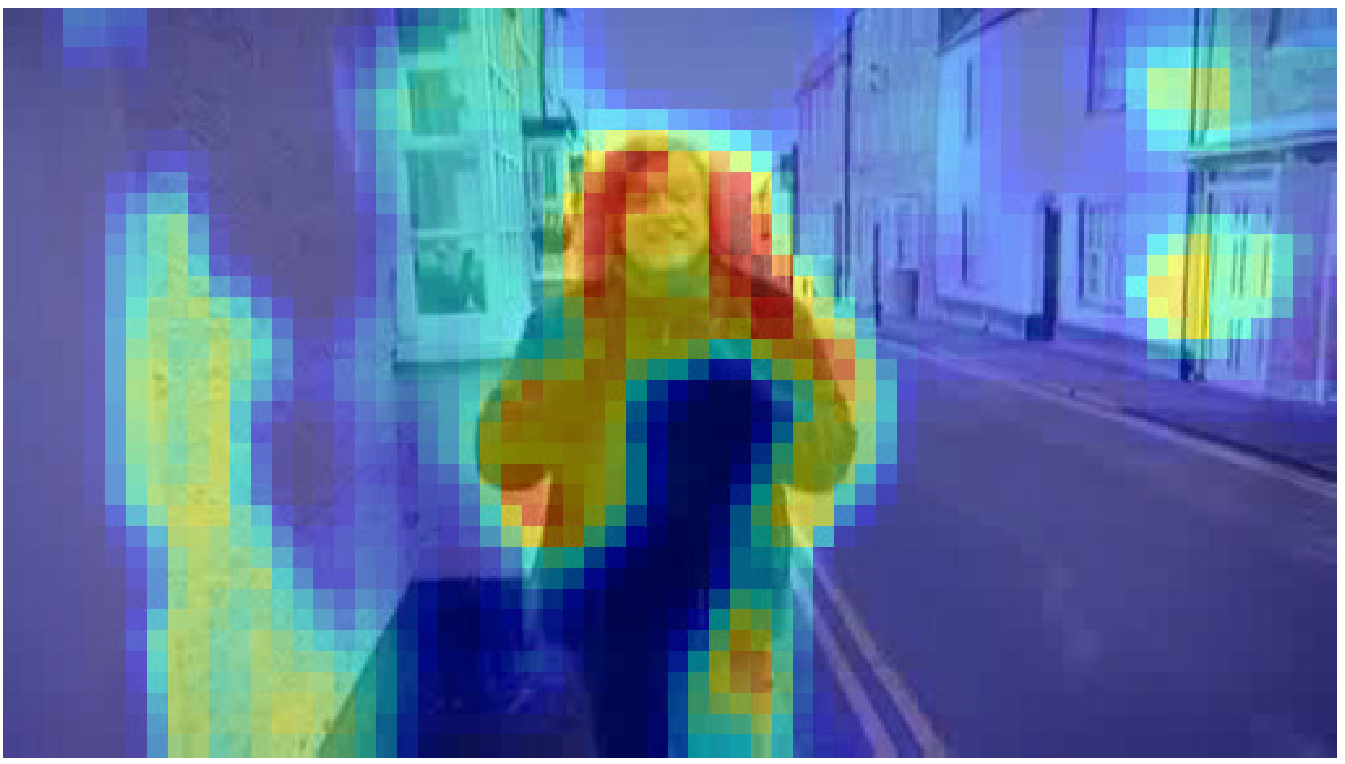}  \hspace*{-.2em}& 
\includegraphics[width=1.0in]{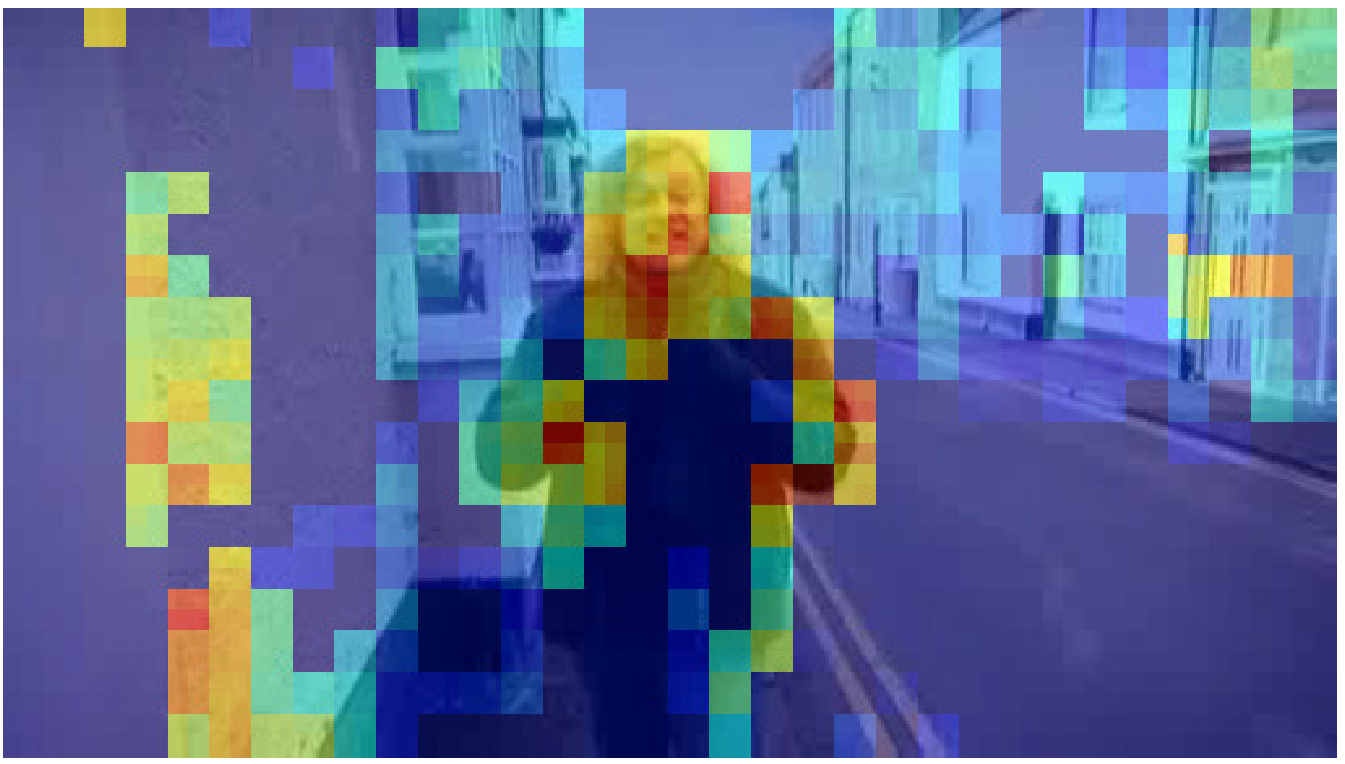}  \hspace*{-.2em}\\

PIM-ZEN & PIM-MCS & MCSDM & APPROX \\
\includegraphics[width=1.0in]{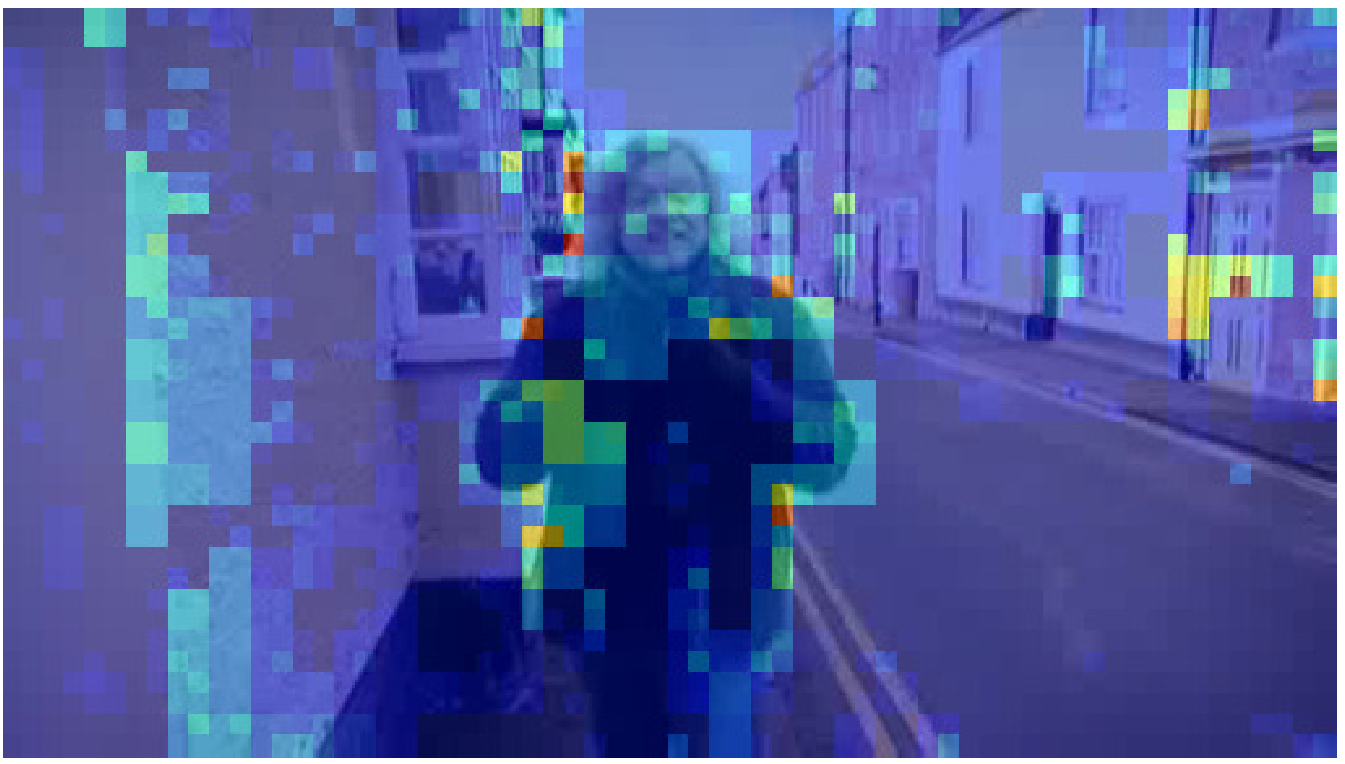}  \hspace*{-.2em}& 
\includegraphics[width=1.0in]{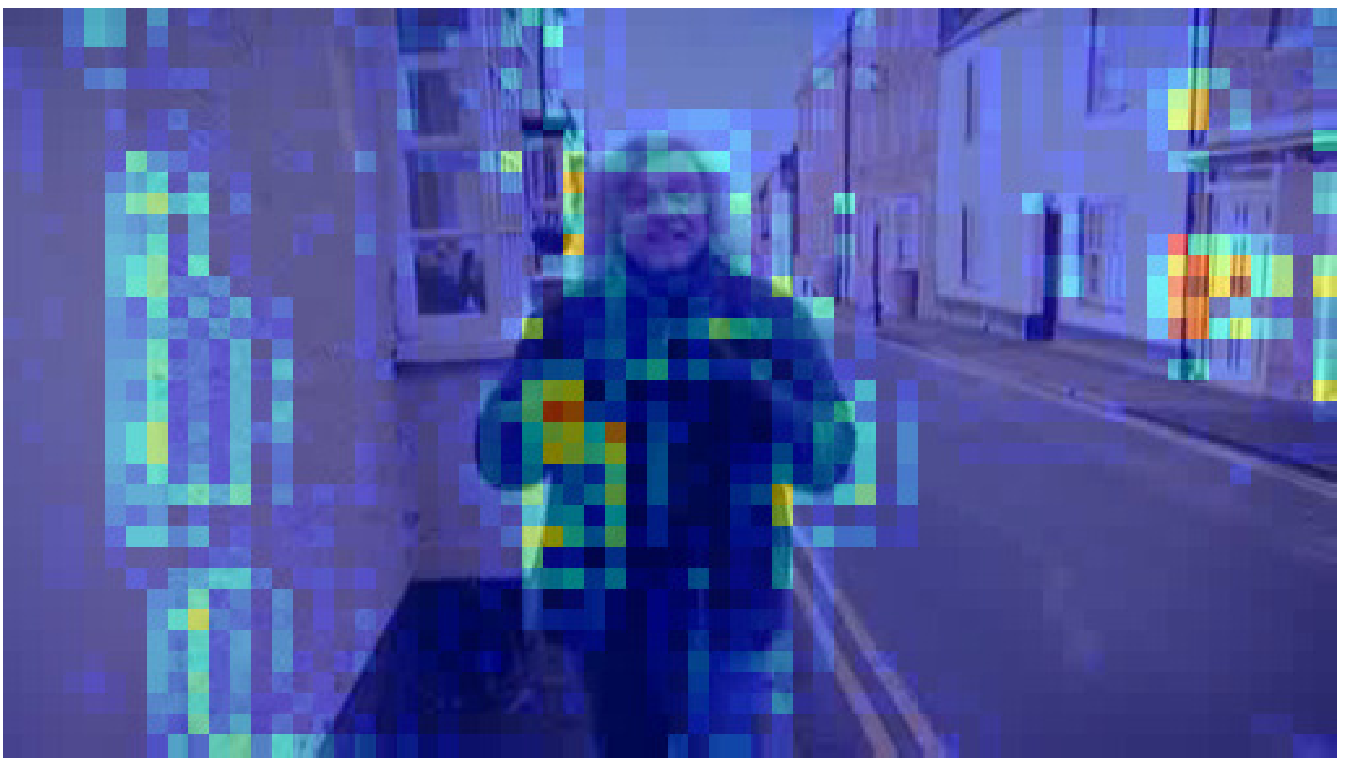}  \hspace*{-.2em} &
\includegraphics[width=1.0in]{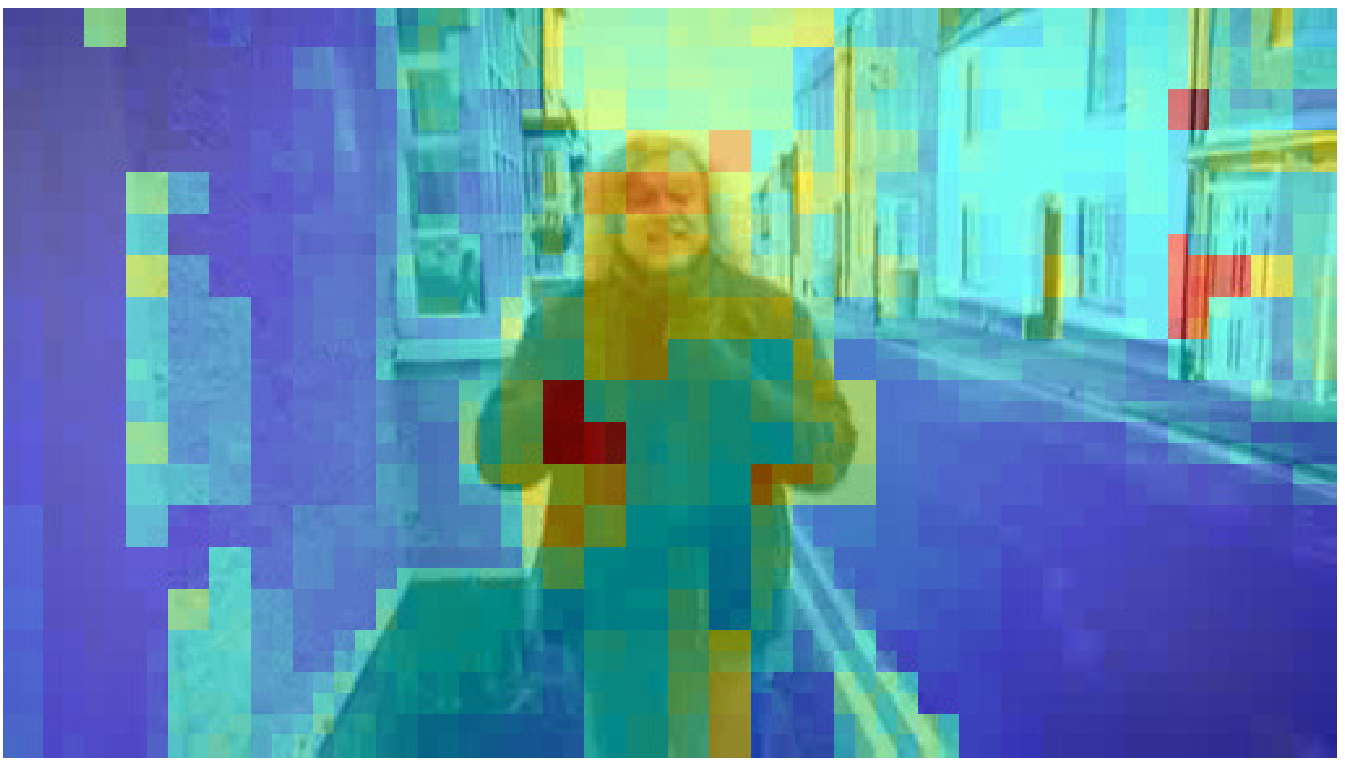}  \hspace*{-.2em} &
\includegraphics[width=1.0in]{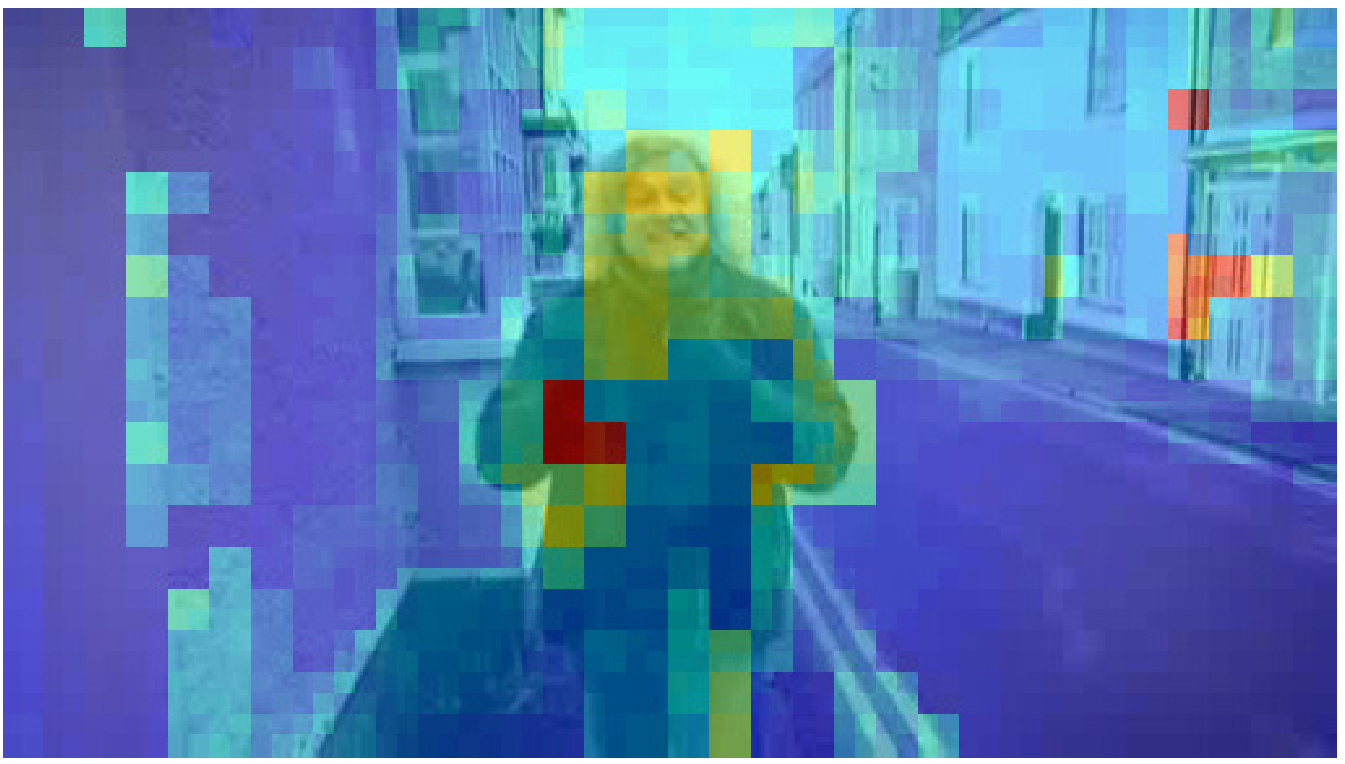}  \hspace*{-.2em}\\

OBDL-MRF & MVE+SRN & AWS & GBVS \\ 
\includegraphics[width=1.0in]{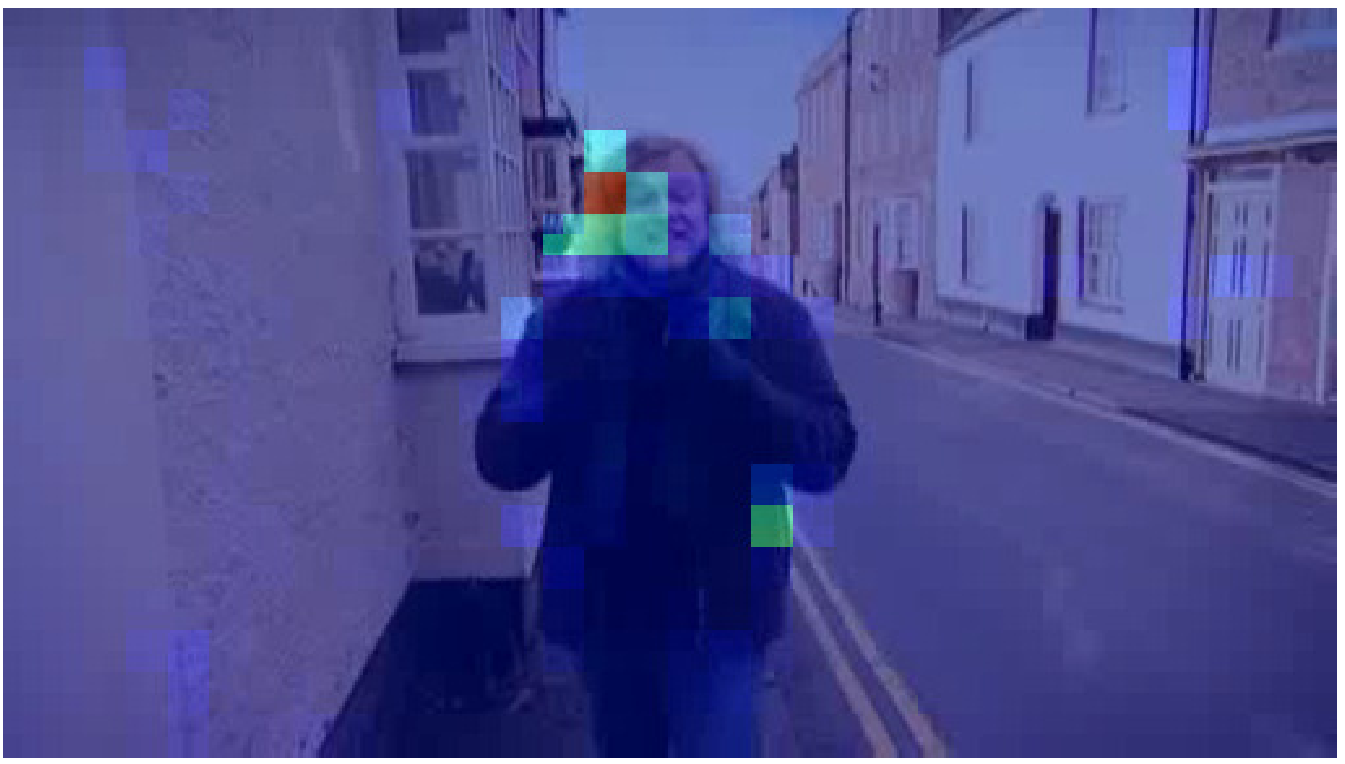}  \hspace*{-.2em}& 
\includegraphics[width=1.0in]{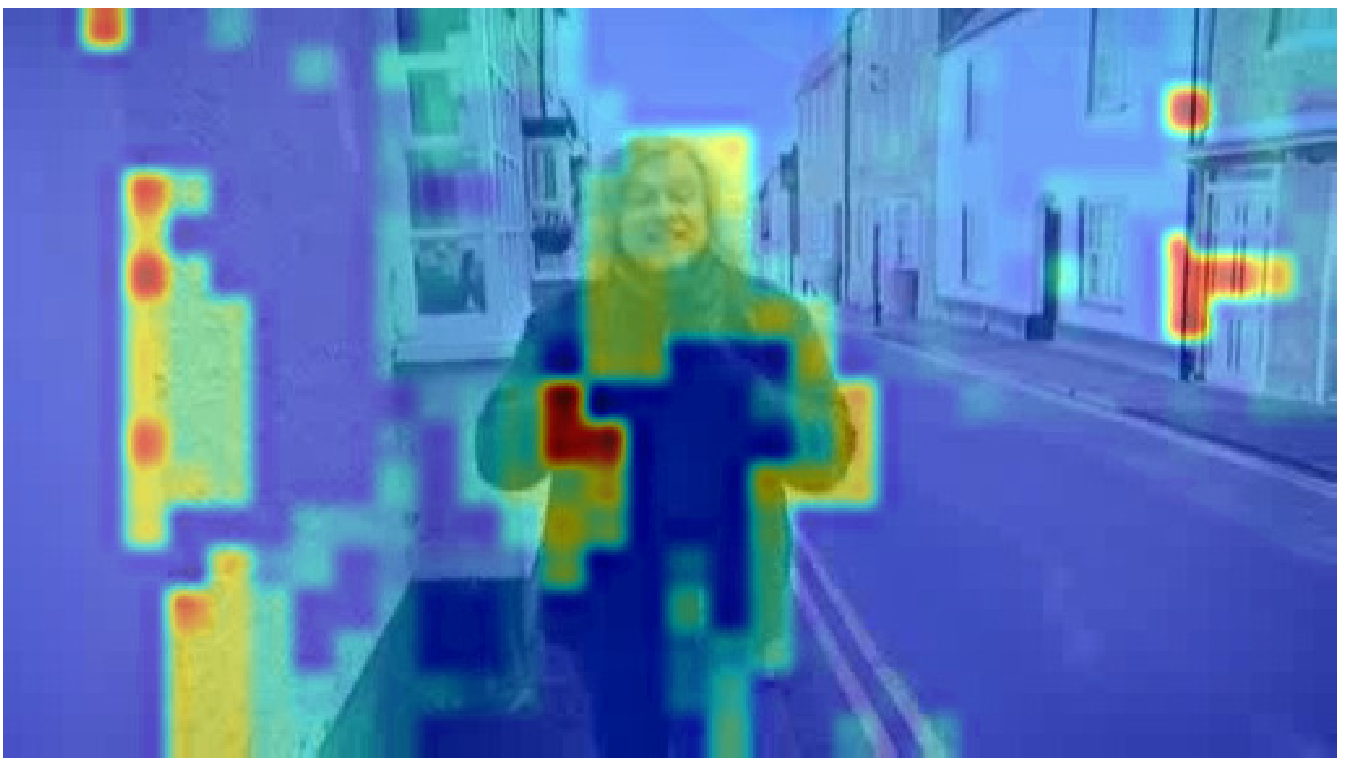}  \hspace*{-.2em}& 
\includegraphics[width=1.0in]{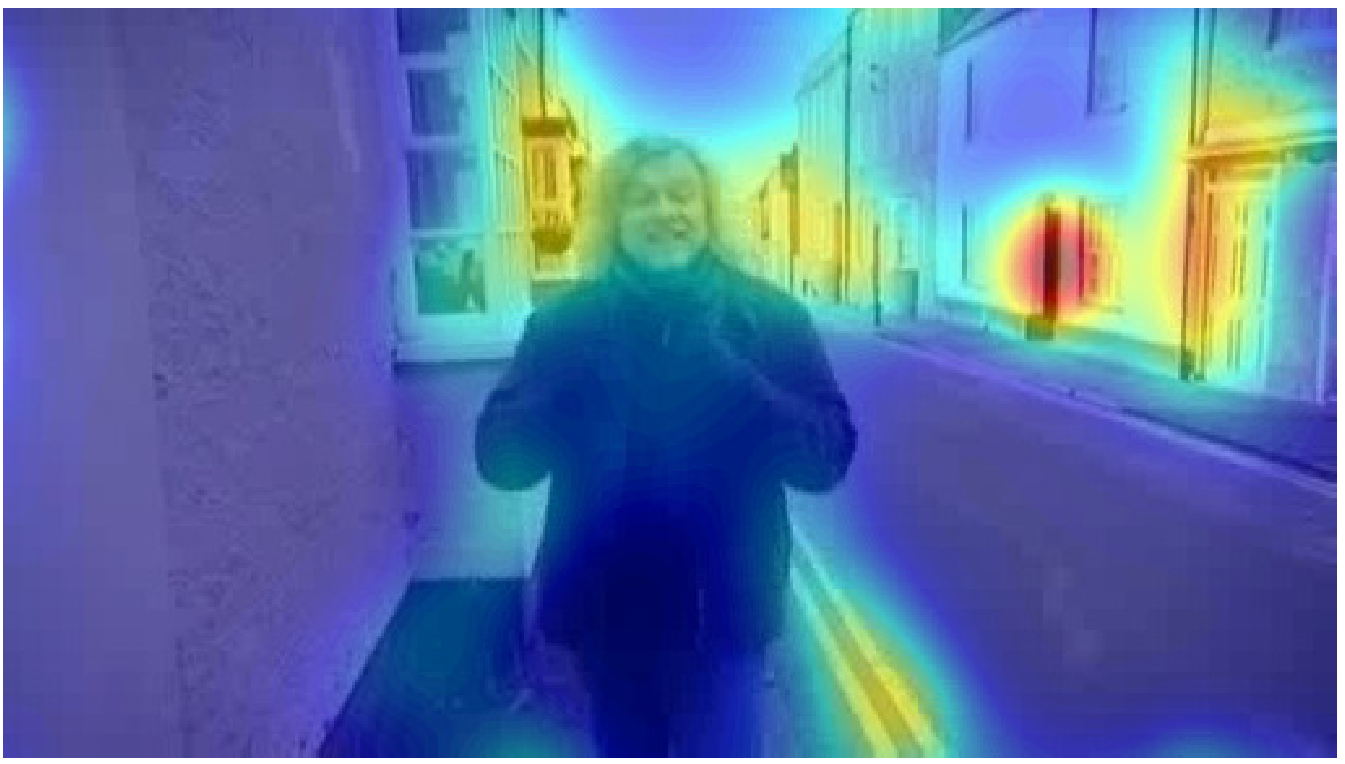}  \hspace*{-.2em}& 
\includegraphics[width=1.0in]{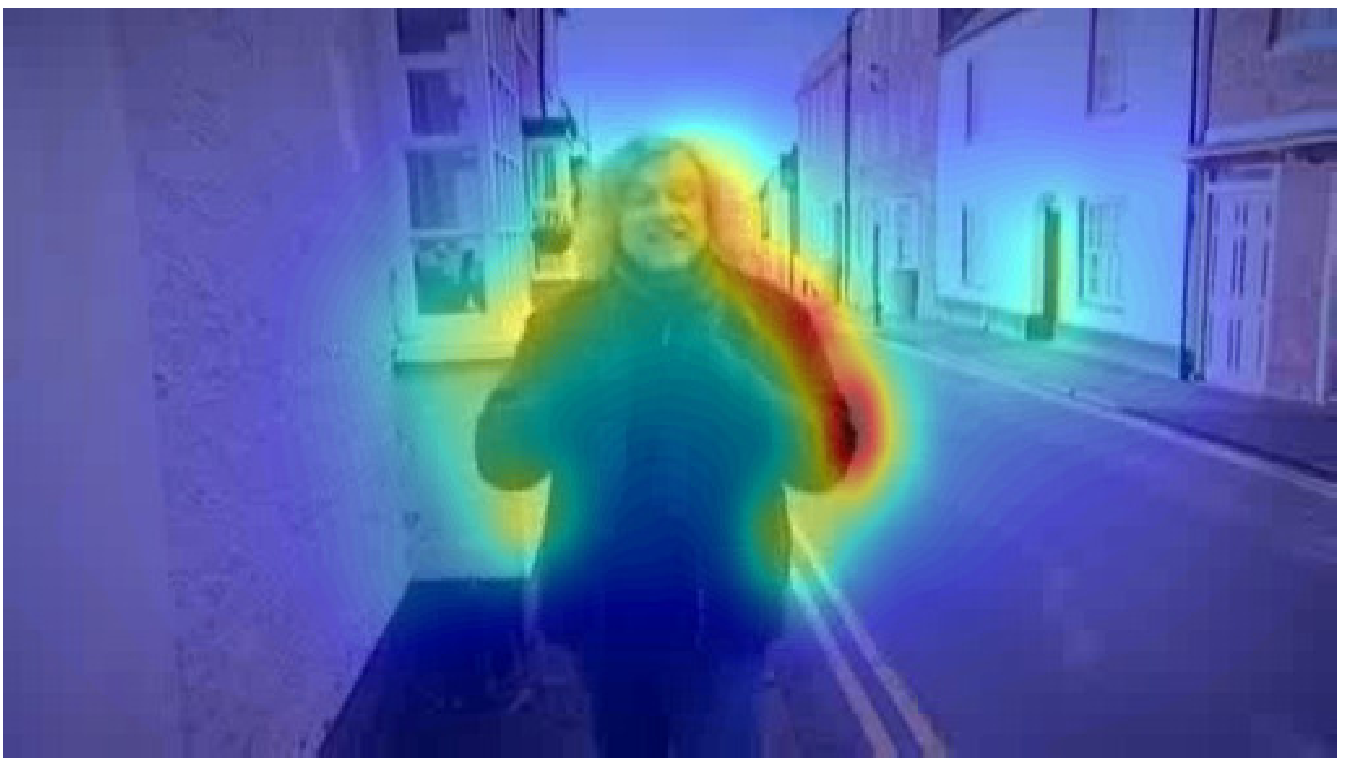} \\
%\end{tabular}	
%\begin{tabular}{c}
\\ \\ 
%\textbf{H.264/AVC}
\multicolumn{4}{c}{\textbf{H.264/AVC}}\\
%\end{tabular}
%\begin{tabular}{cccc}
IO & \textit{MVF} & PMES & MAM \\
\includegraphics[width=1.0in]{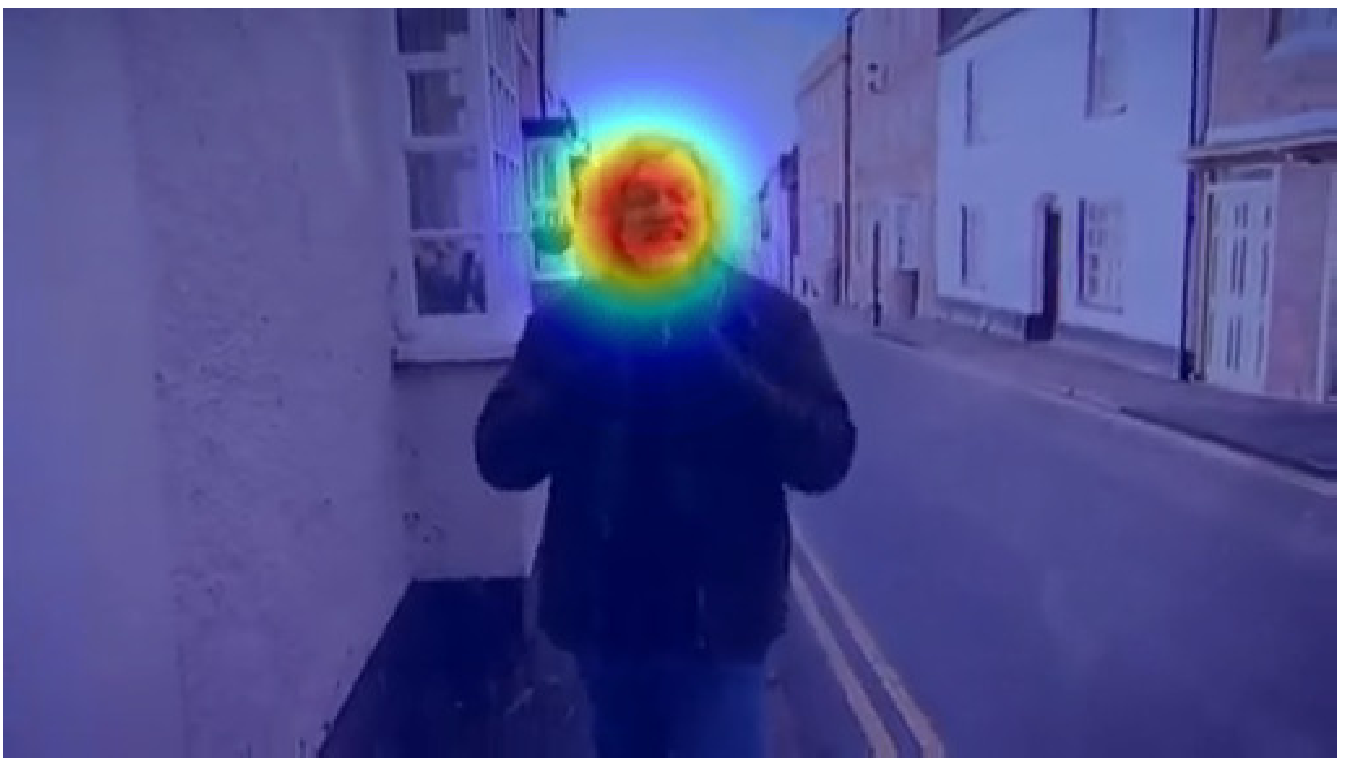}  \hspace*{-.2em}& 
\includegraphics[width=1.0in]{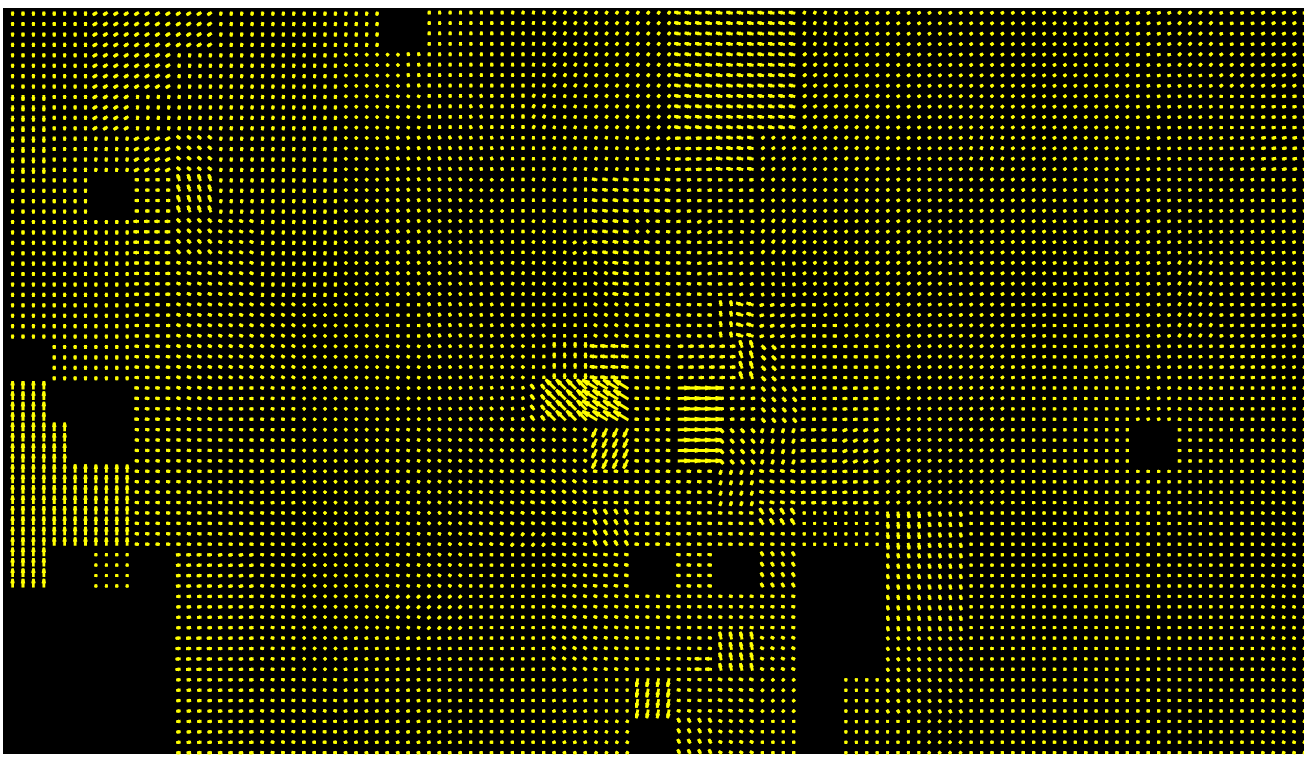} \hspace*{-.2em}& 
\includegraphics[width=1.0in]{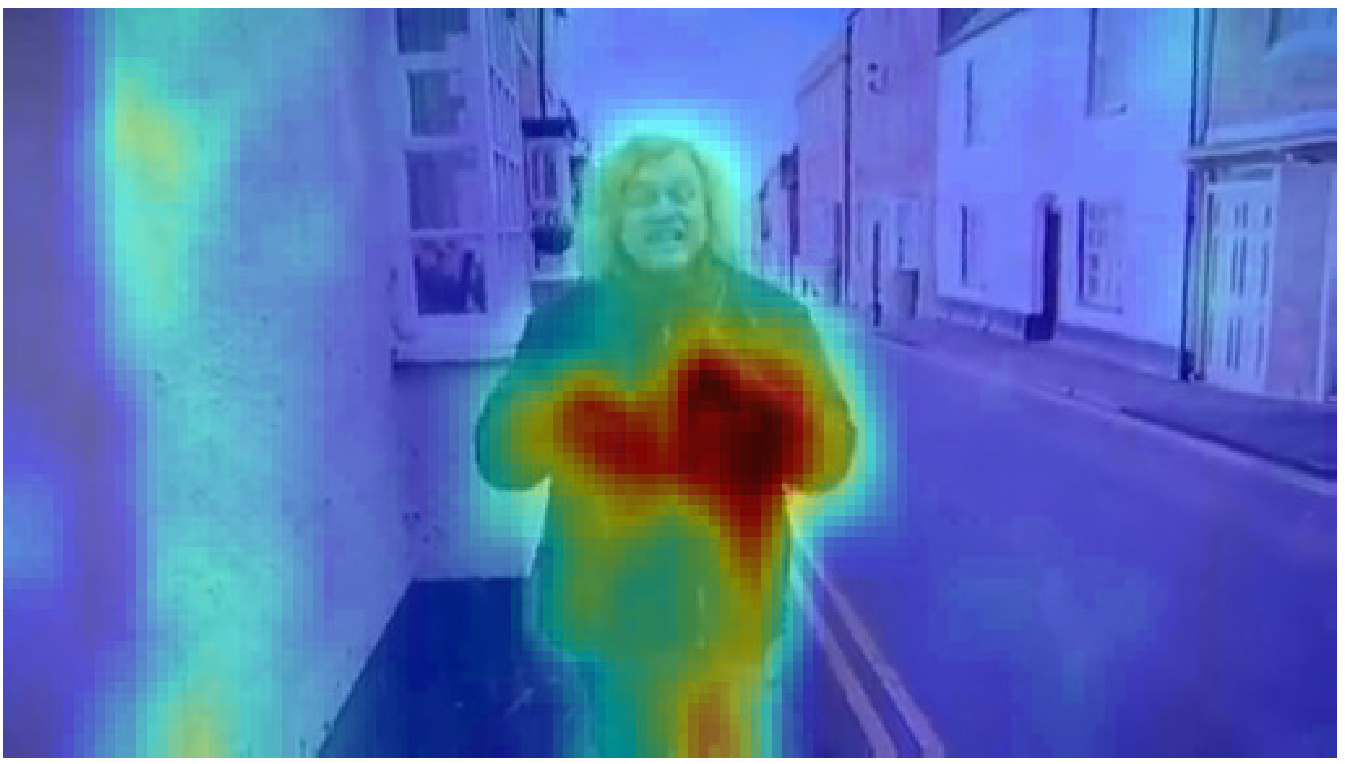}  \hspace*{-.2em}& 
\includegraphics[width=1.0in]{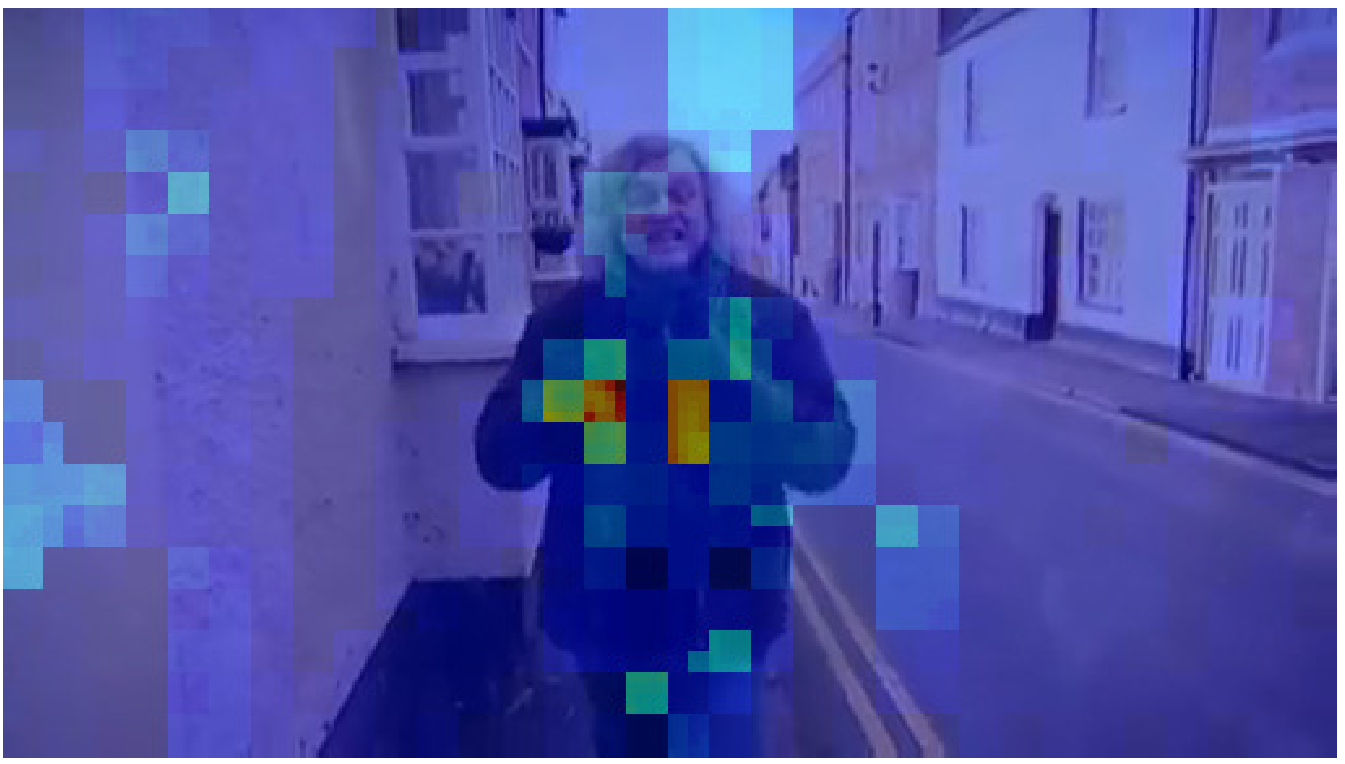}  \hspace*{-.2em}\\

PIM-ZEN & PIM-MCS & MCSDM & APPROX \\
\includegraphics[width=1.0in]{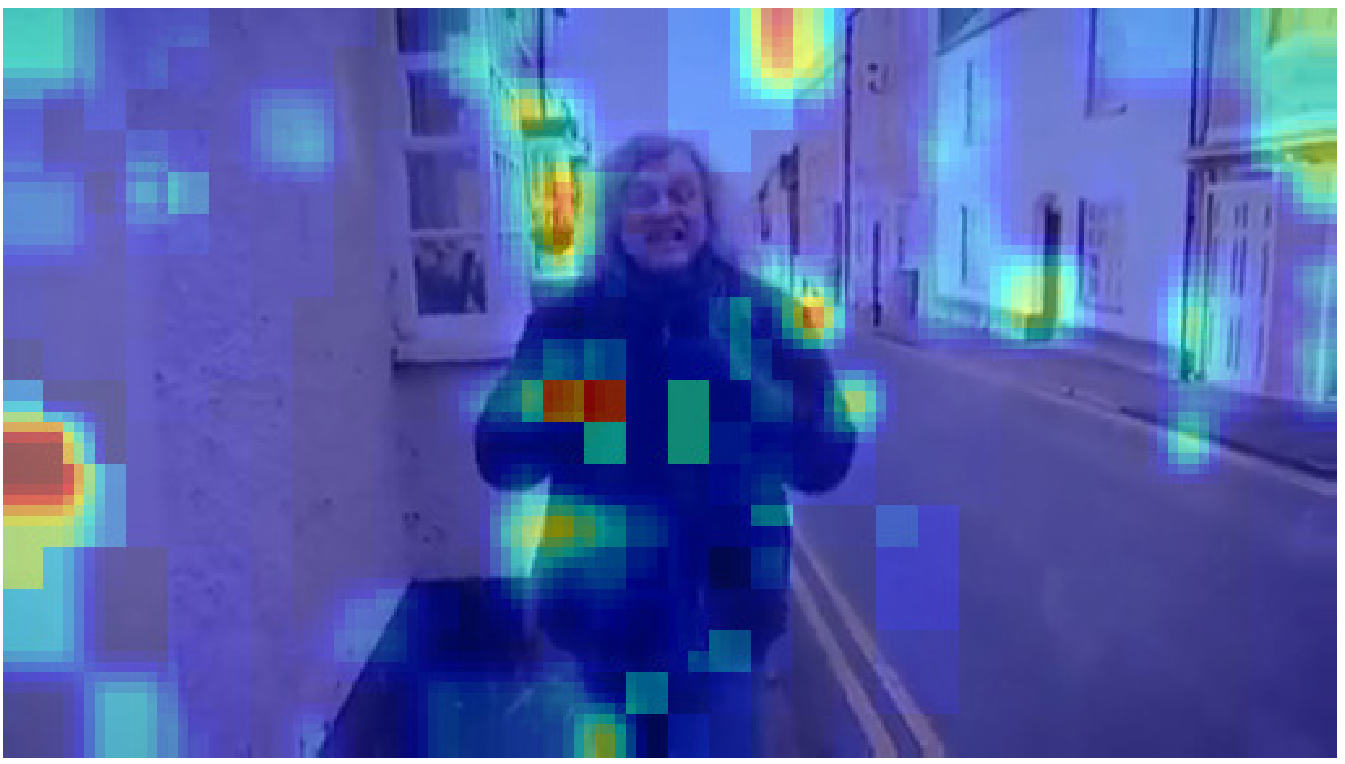}  \hspace*{-.2em}& 
\includegraphics[width=1.0in]{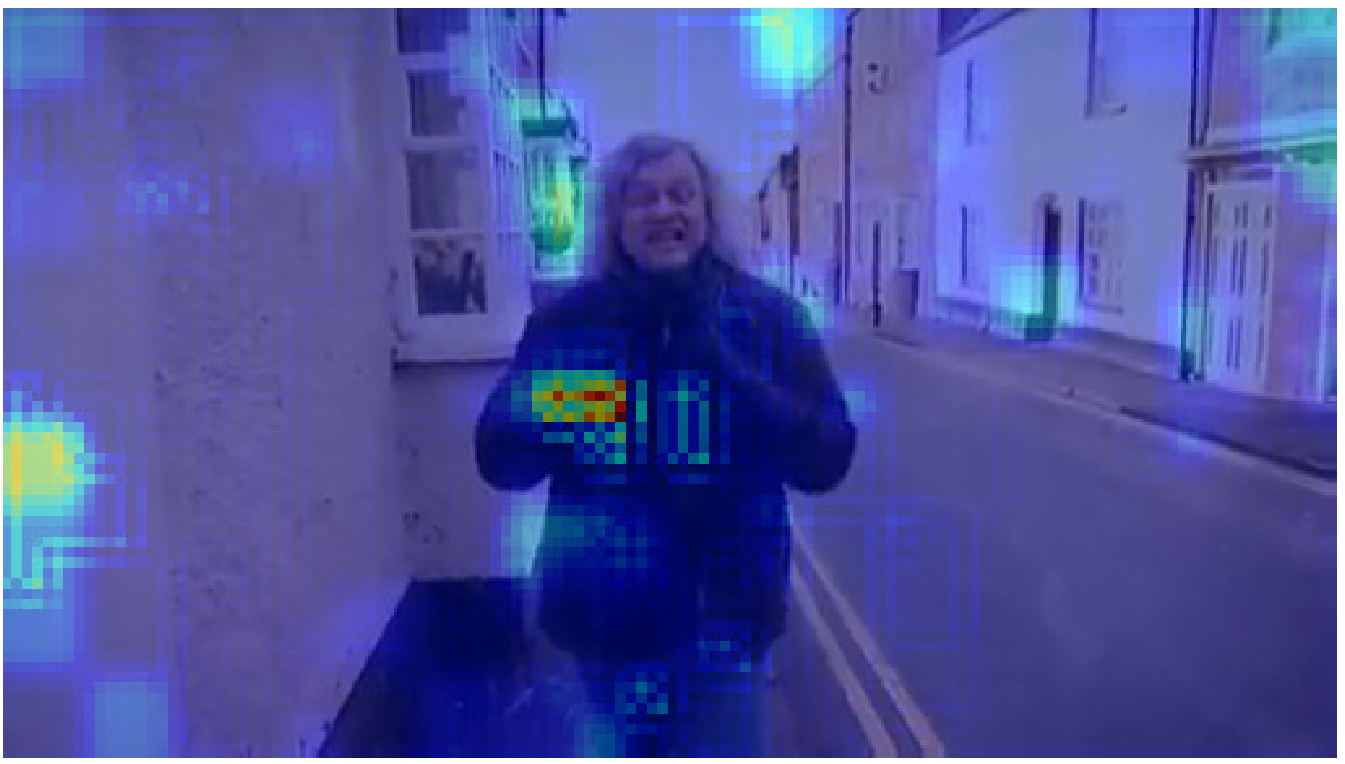}  \hspace*{-.2em} &
\includegraphics[width=1.0in]{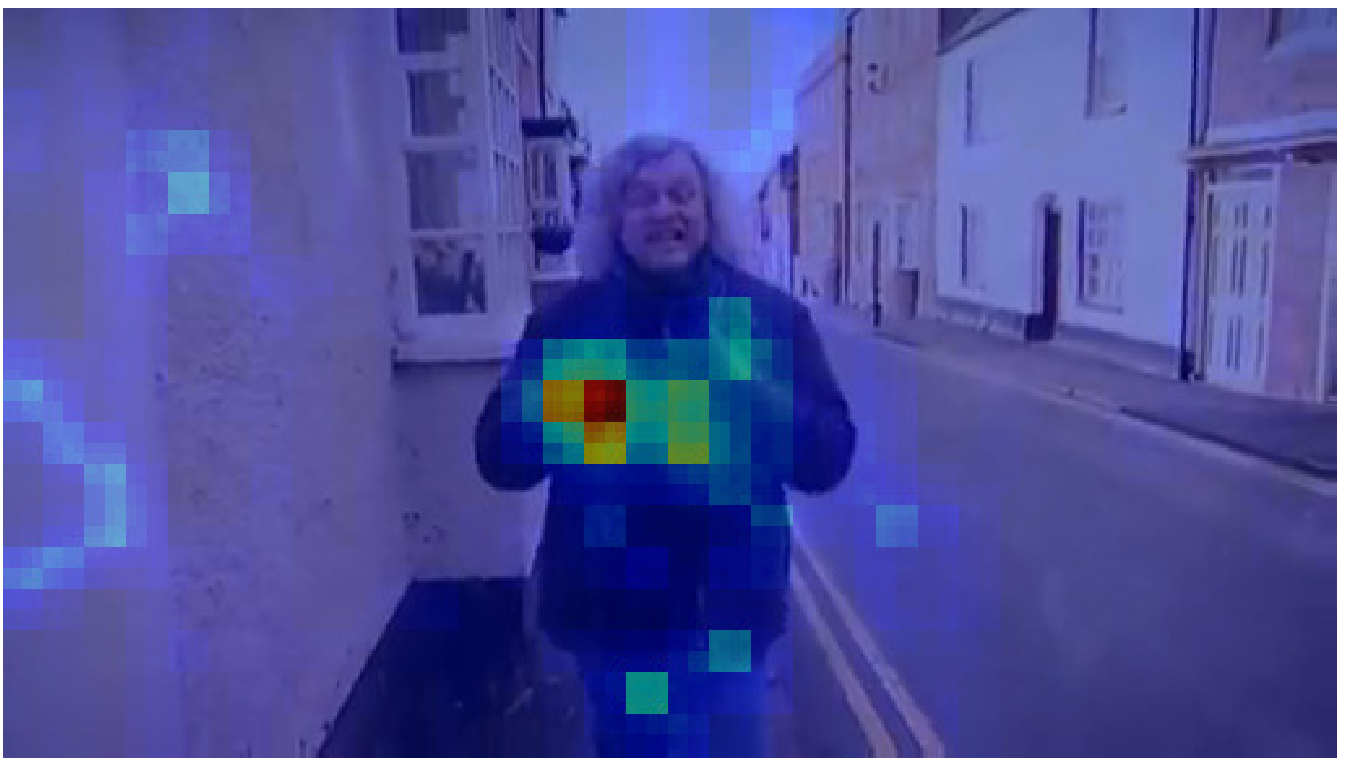}  \hspace*{-.2em} &
\includegraphics[width=1.0in]{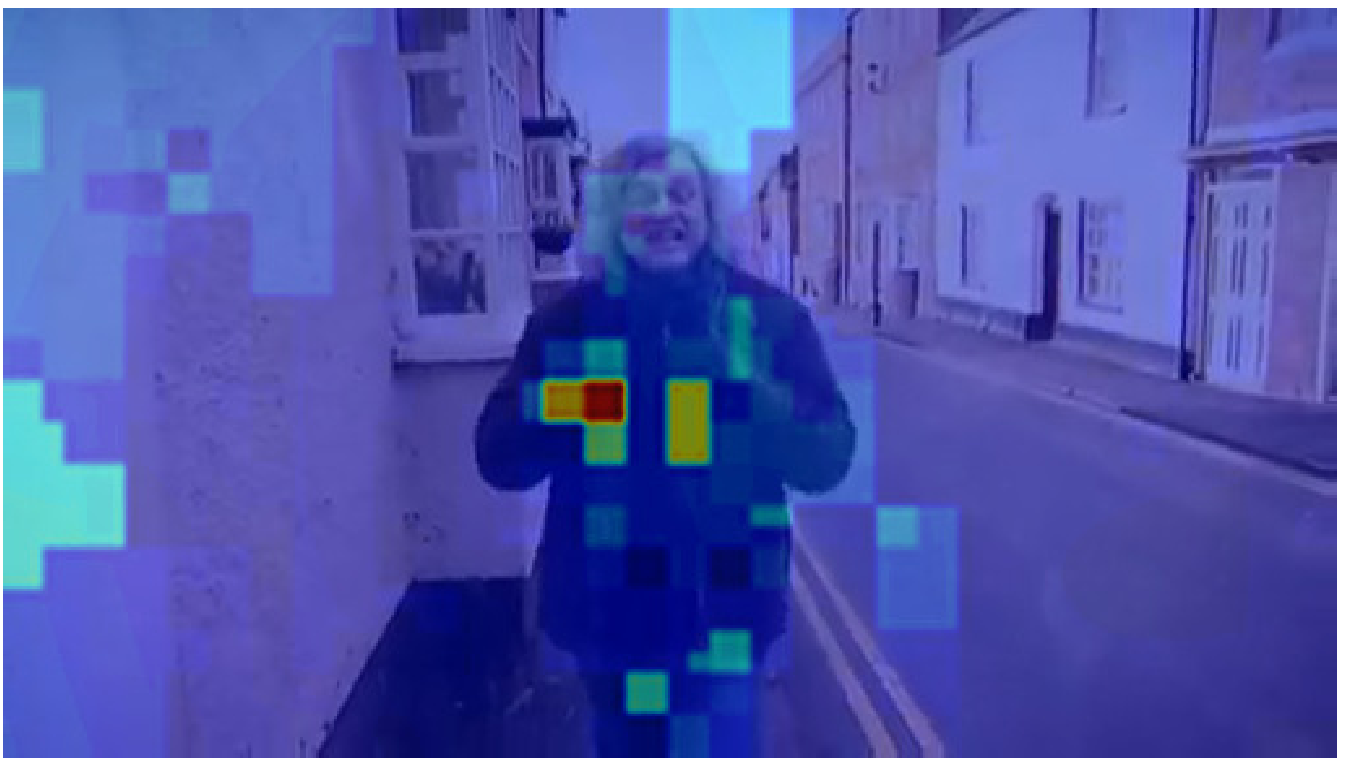}  \hspace*{-.2em}\\ 

OBDL-MRF & MVE+SRN & AWS & GBVS \\ 
\includegraphics[width=1.0in]{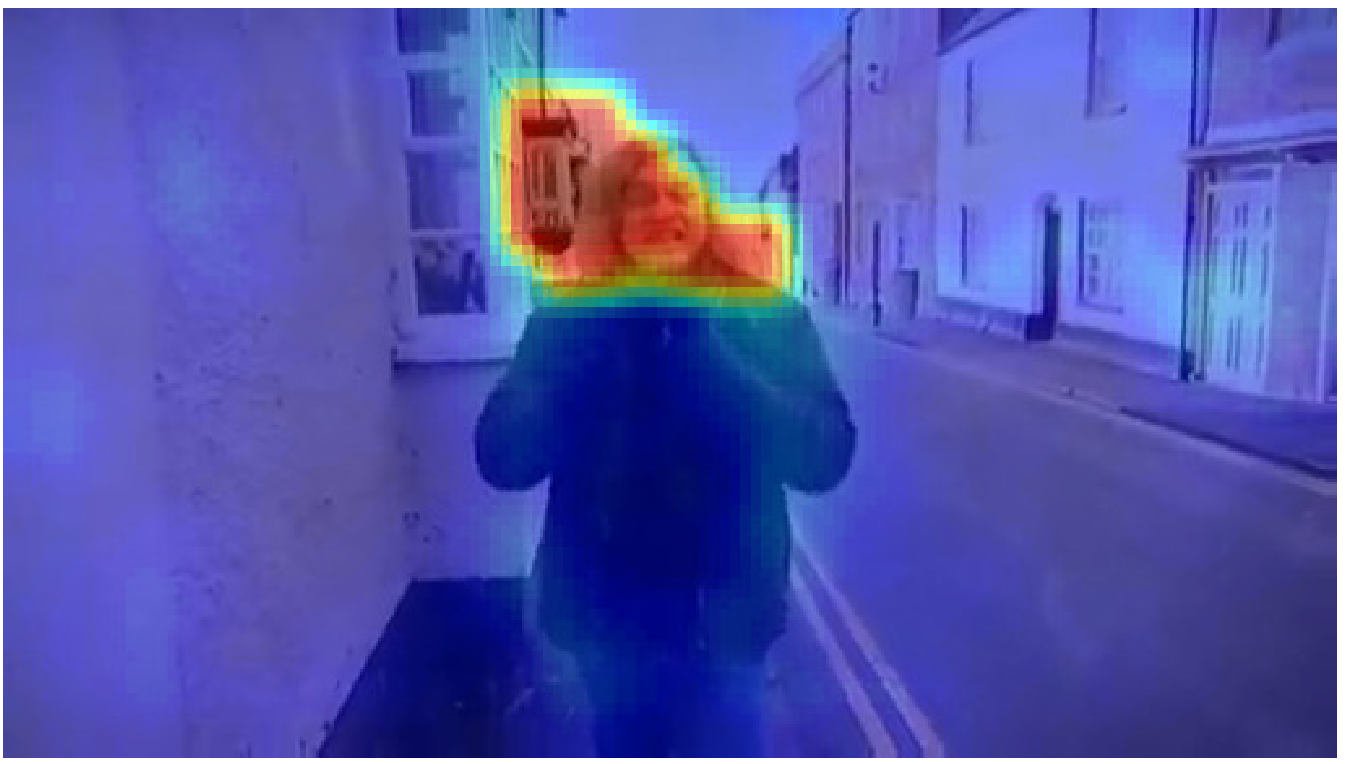}  \hspace*{-.2em}& 
\includegraphics[width=1.0in]{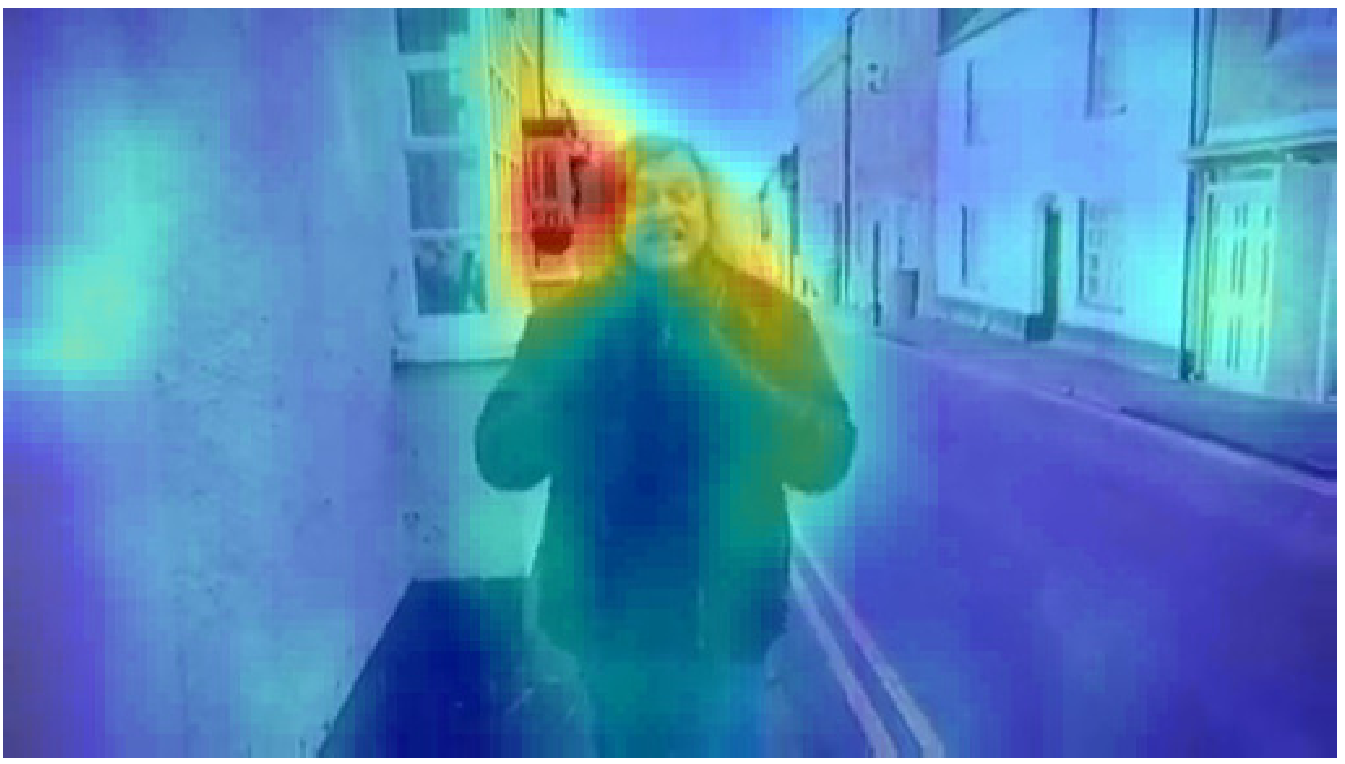}  \hspace*{-.2em}& 
\includegraphics[width=1.0in]{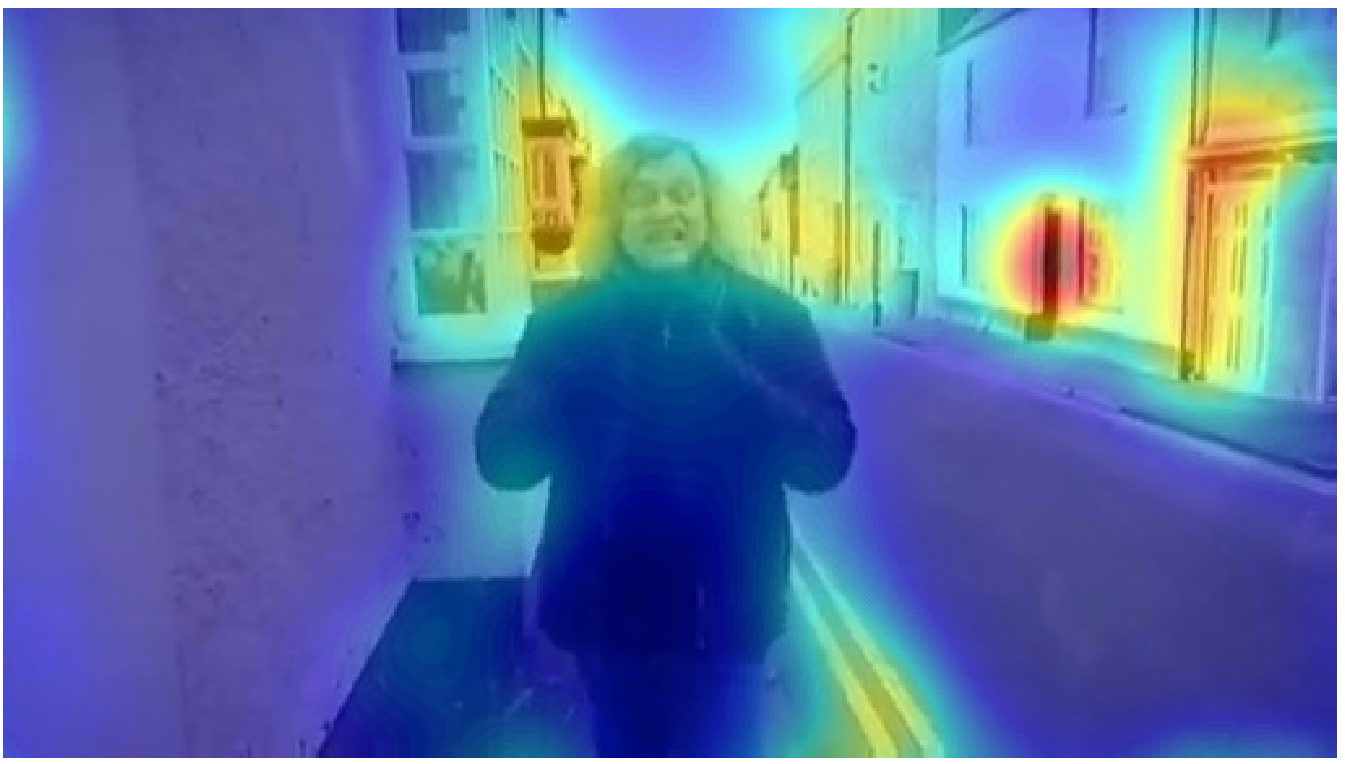}  \hspace*{-.2em}& 
\includegraphics[width=1.0in]{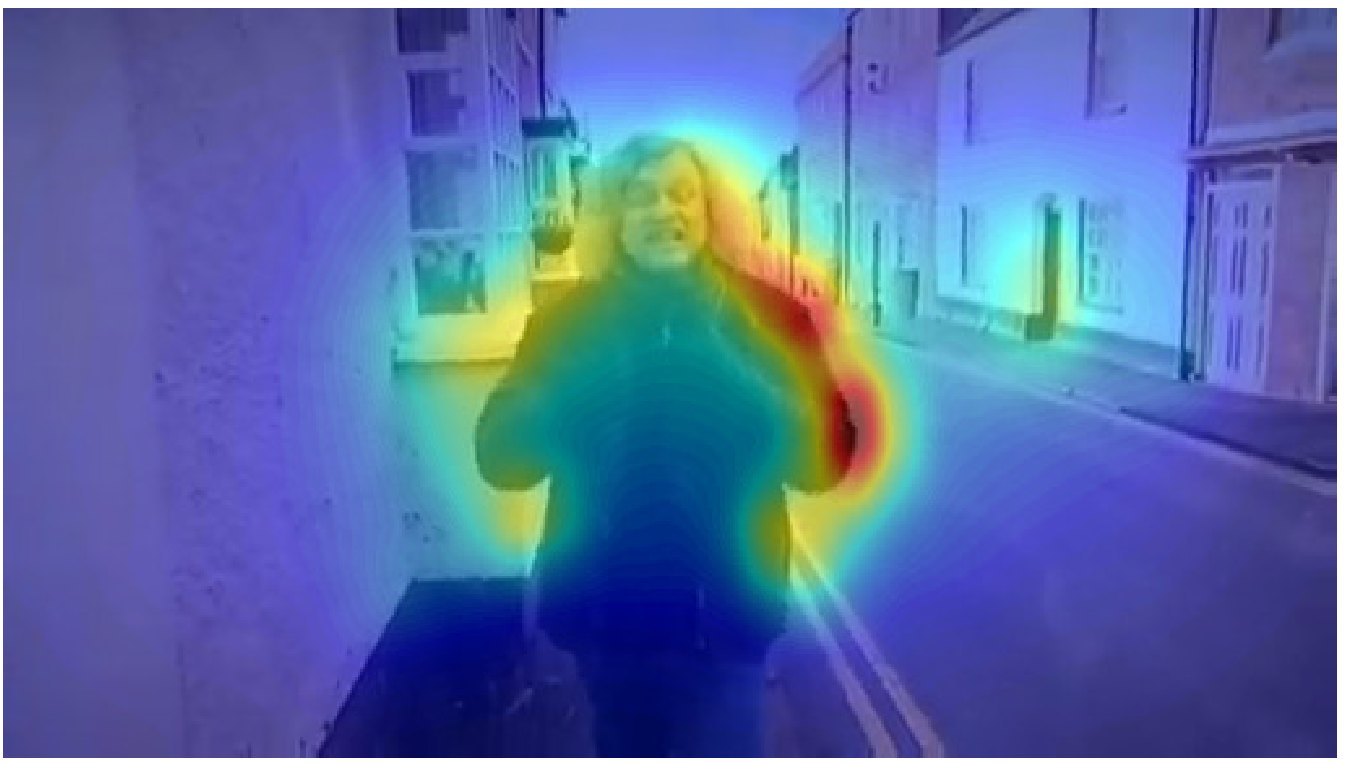} \\
\end{tabular}	
\caption{Sample saliency maps obtained by various models for \textit{one-show}.}
	\label{fig:SampleSaliency2}
\end{figure*}

In this section, we show a qualitative comparison of saliency maps produced by various models on two specific examples.\footnote{Additional examples are provided in the supplementary material~\cite{CSEsupplementary}.}
Figs.~\ref{fig:SampleSaliency1} and \ref{fig:SampleSaliency2} show the saliency maps for frame~\#150 of \textit{City} and frame~\#150 of \textit{one-show} produced from MPEG-4 ASP and H.264/AVC bitstreams. The QP values for MPEG-4 ASP and H.264/AVC were set to 16 and 36, respectively. This selection brings about the same average PSNR ($\approx 30.0dB$) over the whole sequence. In the figure, the MVF of each frame is also shown. Note that due to the differences in MVs and residuals of MPEG-4 ASP and H.264/AVC, the resulting saliency maps for those models that are able to accept both formats can be different. In the figures, these would be PMES, MAM, PIM-ZEN, PIM-MCS and APPROX. On \textit{City}, saliency maps produced by the same model from two different bitstreams do resemble each other, but on \textit{one-show} they could be quite different. This is mainly due to the fact that the MVF on \textit{City} is more consistent between MPEG-4 ASP and H.264/AVC, whereas on \textit{one-show}, the two encoders produce fairly different MVFs.

In \textit{City}, all the motion is due to camera movement. While observers typically look at the building in the center of the frame (see IO in Figs.~\ref{fig:SampleSaliency1} and~\ref{fig:SampleSaliency2}), all models, no matter which encoding is used, declare the boundary of the building as salient, where local motion is different from the global motion. Meanwhile, APPROX is also able to detect the central building as salient. Note that APPROX is the only model in the study that employs global motion compensation (GMC) and its high scores on \textit{City} are an indication that other models could benefit from incorporating GMC. 

In \textit{one-show}, large noisy MVs in low-texture areas cause all compressed-domain models except MSM-SM and OBDL-MRF to mistakenly declare them as salient regions. Note that MSM-SM does not directly use motion magnitude but rather uses processed MVs in the form of a motion binary map. Meanwhile, OBDL-MRF uses the number of bits per block, rather than any direct measure of motion magnitude, to predict saliency. In this sequence, observers mostly focus on the face (see IO in Figs.~\ref{fig:SampleSaliency1} and \ref{fig:SampleSaliency2}) so a model that was able to perform face detection would have done well in this example. Unfortunately, none of the models is currently able to do face detection in the compressed domain - this seems like a rather challenging problem. AWS and GBVS also declare some part of non-salient regions as salient.

\subsection{Quantitative Comparison}
\label{sec:quantitative_result}
\begin{figure}
\centering
\includegraphics[width=.8\linewidth]{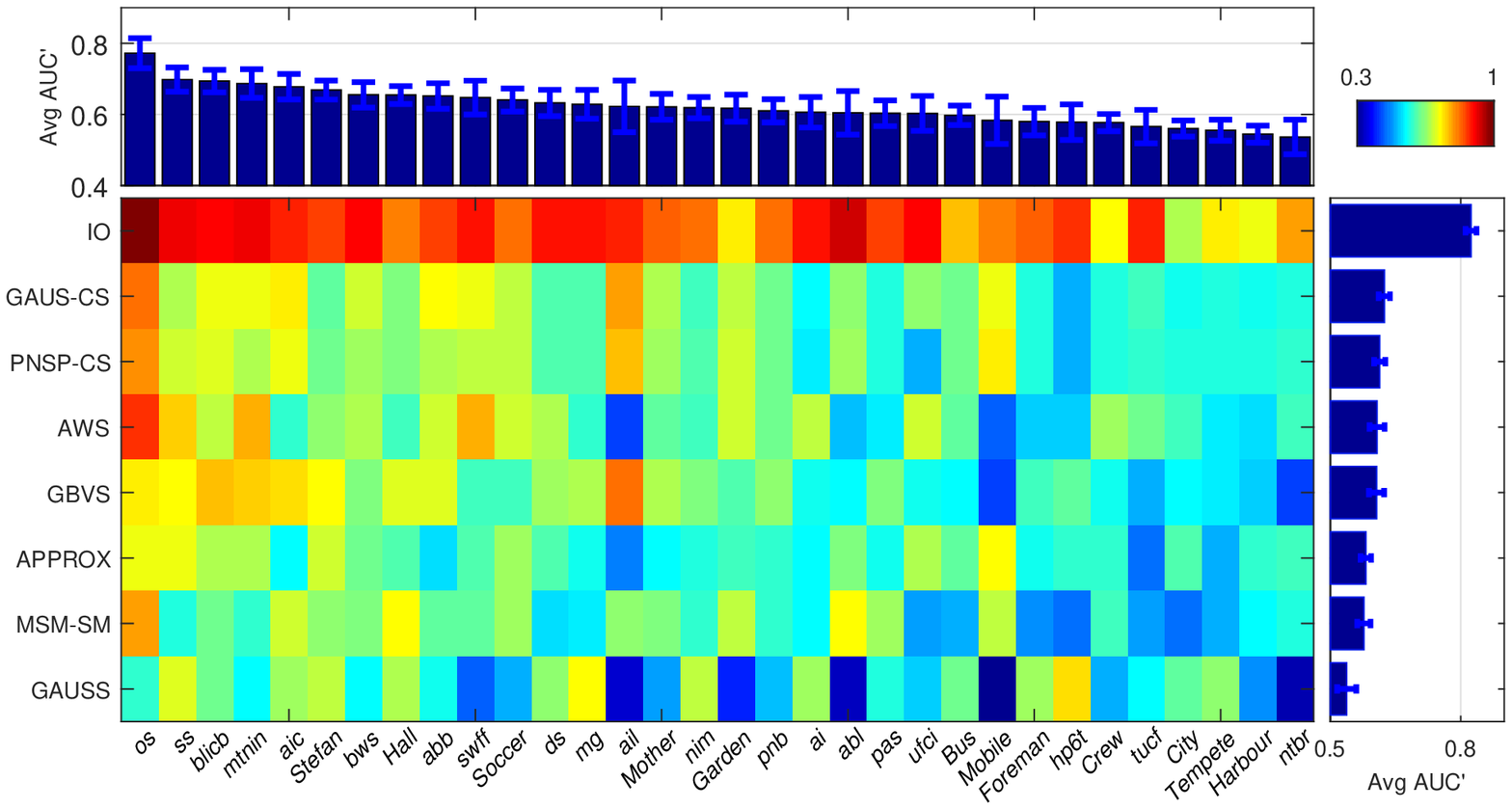} \\
\includegraphics[width=.8\linewidth]{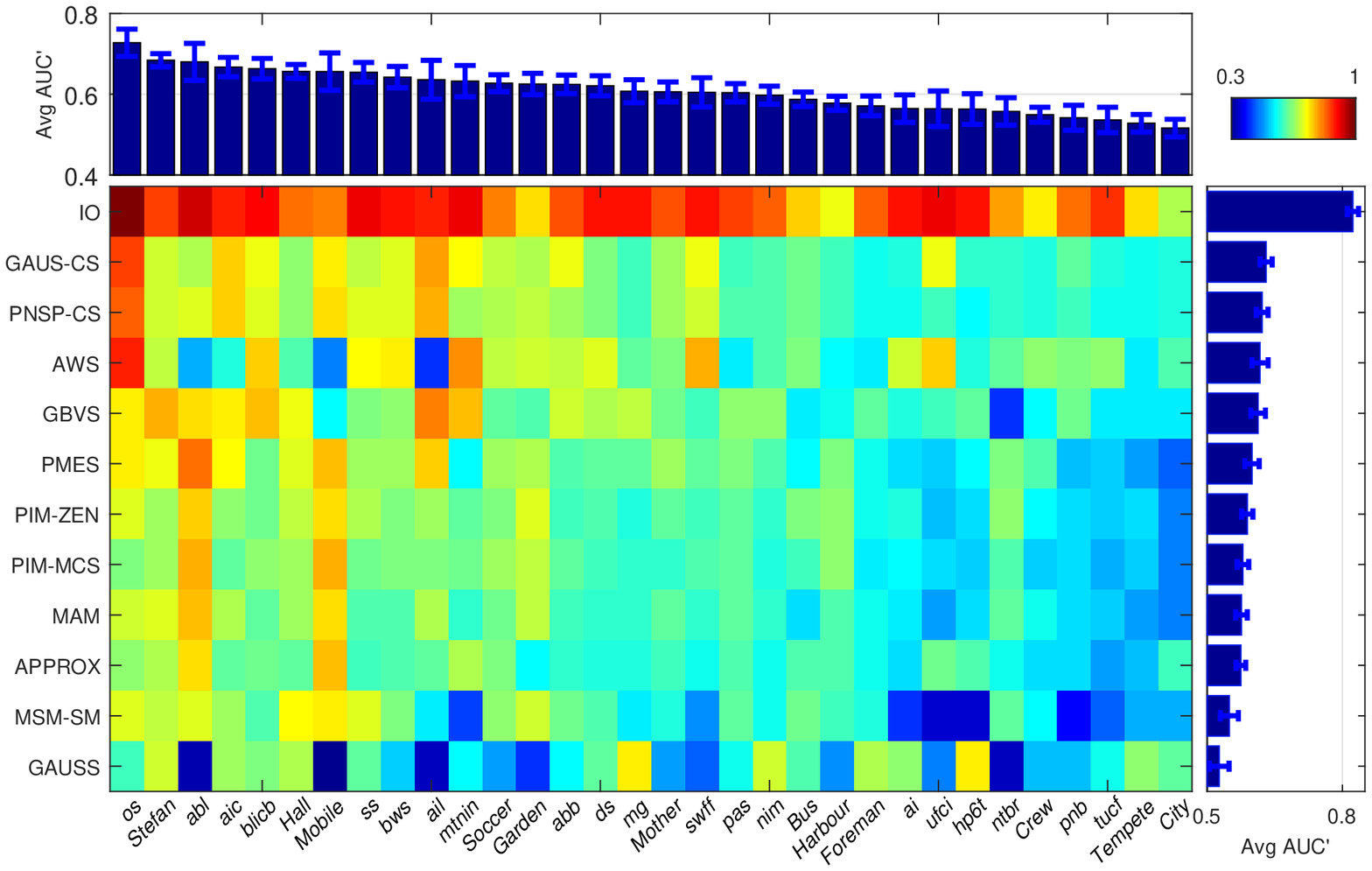} \\
\caption{Accuracy of various saliency models over MPEG-4 ASP encoded sequences according to AUC$'$ score for (top) I-frames and (bottom) P-frames. The 2D color map shows the average AUC$'$ score of each model on each sequence. \textit{Top}: Average AUC$'$ score for each sequence, across all models. \textit{Right}: Average AUC$'$ scores each model across all sequences. Error bars represent standard error of the mean (SEM), $\sigma/\sqrt{n}$, where $\sigma$ is the sample standard deviation of $n$ samples.}
\label{fig:AUC}
\end{figure}

First, we present quantitative assessment of the saliency models using the MPEG-4 ASP encoded data from the SFU and DIEM datasets. We start with the assessment based on AUC$'$. Fig.~\ref{fig:AUC} shows the average AUC$'$ scores of various models across the test sequences. Note that all models are able to produce saliency maps for P-frames, while only some of them are able to produce a saliency map for I-frames. Hence, Fig.~\ref{fig:AUC} (top) shows the average AUC$'$ scores on I-frames for those models able to handle I-frames, while Fig.~\ref{fig:AUC} (bottom) shows the average AUC$'$ scores for all models on P-frames. Sequences from the SFU dataset are indicated with capital first letter.

As seen in the figure, all models achieved average AUC$'$ scores between those of IO, which represents a kind of an upper bound (especially on the DIEM dataset), and GAUSS, which represents center-biased, content-independent static saliency map. Note that GAUSS itself has a slightly better AUC$'$ score than the pure chance score of $0.5$. Recall that AUC$'$ corrects for center bias by random sampling of control points based on empirical gaze distribution across all frames and all sequences. It is encouraging that all models are able to surpass GAUSS and achieve average AUC$'$ scores around $0.6$.  

Another interesting point in Fig.~\ref{fig:AUC} is an indication of how difficult or easy is saliency prediction in a given sequence according to AUC$'$. In the figure, the sequences are sorted along the horizontal axis in decreasing order of average AUC$'$ score across all models. Although the order is not the same for I- and P-frames, overall, it seems that \textit{one-show} is the one for which saliency prediction is easiest, whereas \textit{City} is the one for which saliency prediction is hardest. We will return to this issue shortly. Note that IO has better performance on the sequences from the DIEM dataset. Here, IO saliency maps are formed by the left eye gaze points and represent an excellent indicator of the ground-truth right eye gaze points. In the sequences from the SFU dataset, where IO saliency map is formed from the gaze points of the second viewing, the IO scores are not as high because the second-viewing gaze points are not as good of a predictor of the ground-truth first-viewing gaze points.  

A similar set of results quantifying the models' performance according to NSS$'$ is shown in Fig.~\ref{fig:NSS}.\footnote{Results for other metrics are provided in the supplementary material~\cite{CSEsupplementary}.} As seen in Figs.~\ref{fig:AUC} and~\ref{fig:NSS}, the models that are able to handle I-frames (top parts of the figures) achieve similar average scores on the I-frames as they do on the P-frames (bottom parts of the figures). For this reason, and to save space, in the remainder of the paper the results for I- and P-frames will sometimes be reported jointly. That is, in such cases, all scores will be the averages across all frames that the model is able to handle. Since the number of I-frames is much smaller than the number of P-frames, for the models that are able to handle I-frames, the effect of I-frame scores on the combined score is relatively small.

\begin{figure}
\centering
\includegraphics[width=.8\linewidth]{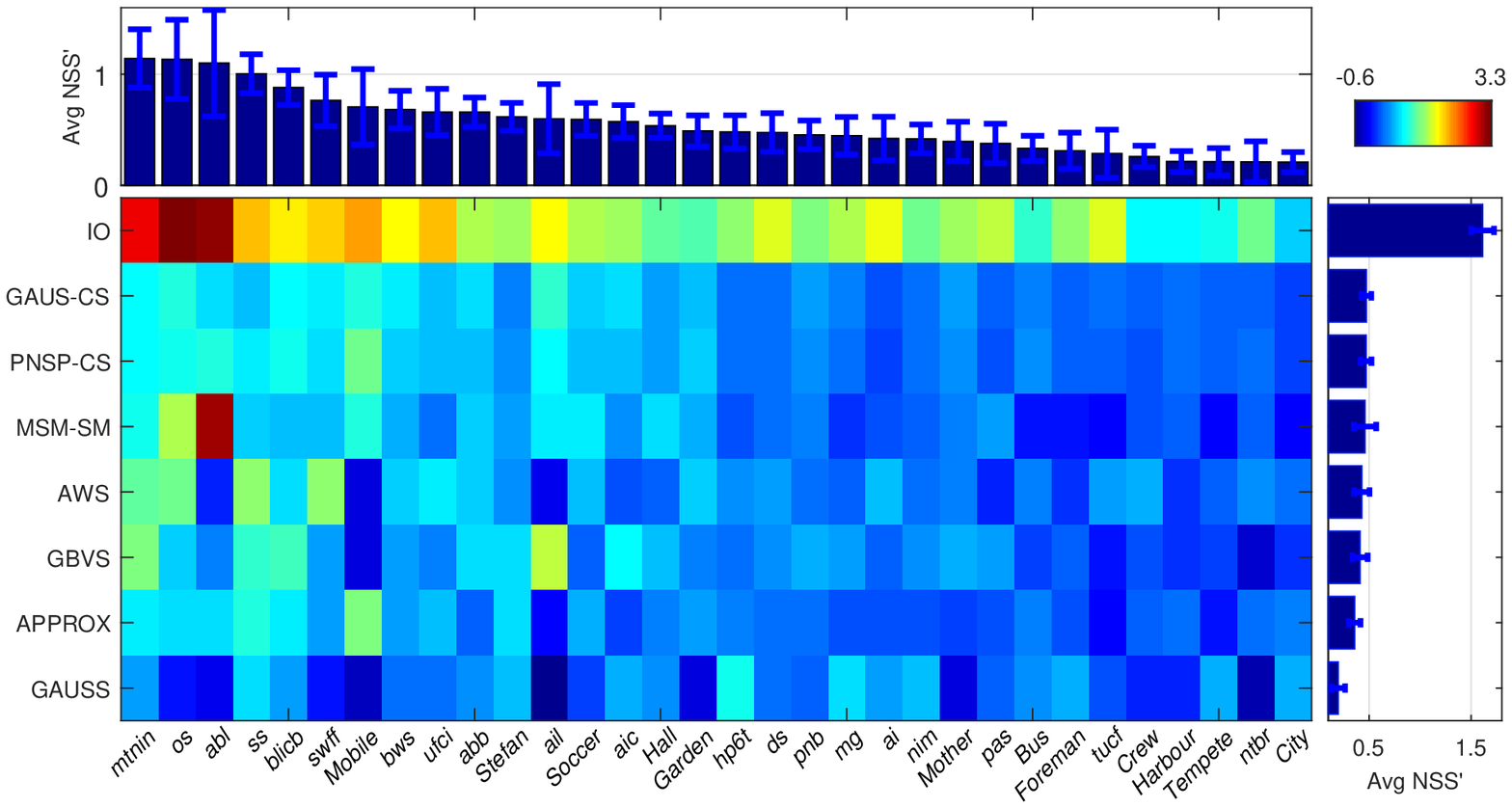} \\
\includegraphics[width=.8\linewidth]{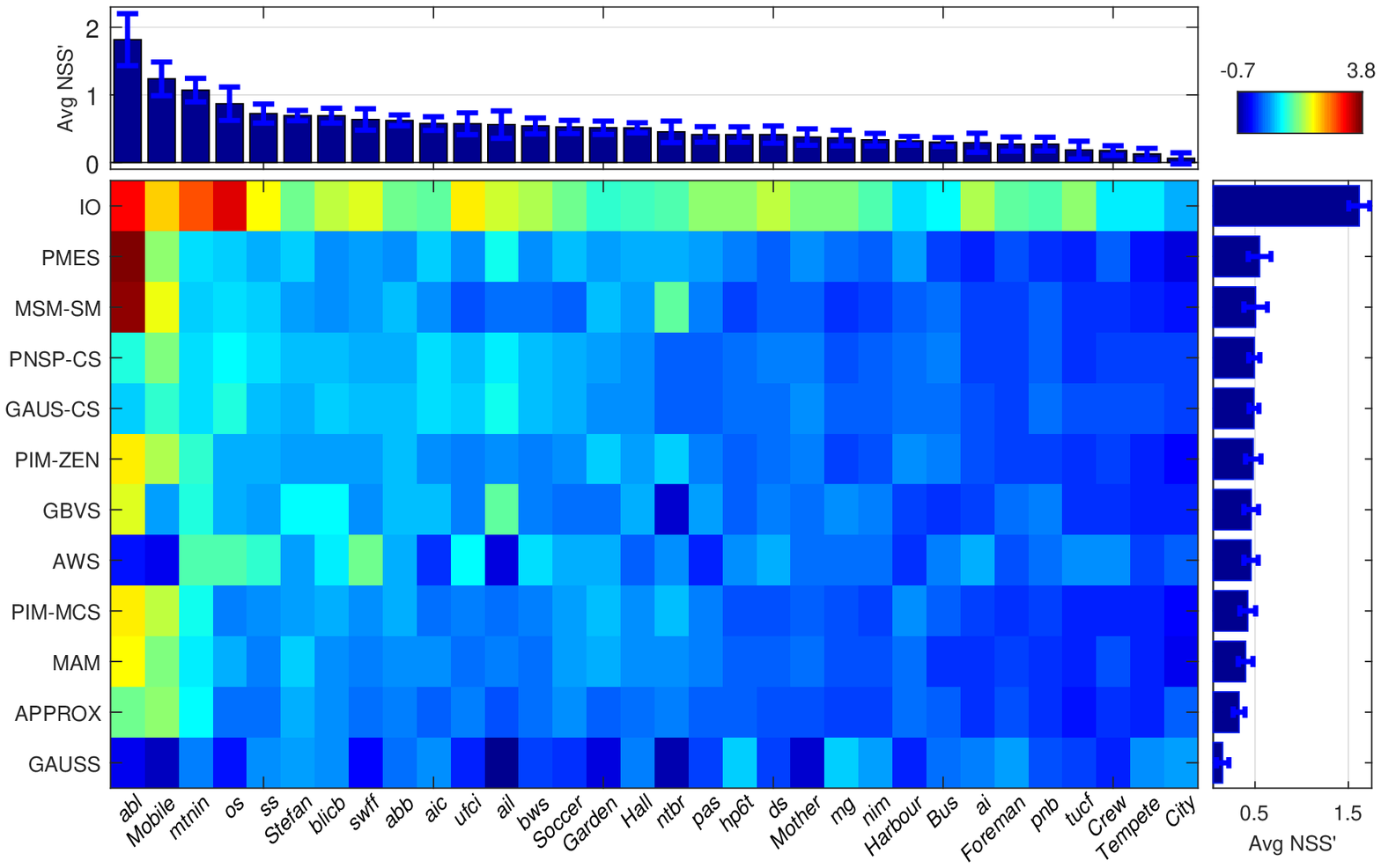} \\
\caption{Accuracy of various saliency models over MPEG-4 ASP encoded sequences according to NSS$'$.}
\label{fig:NSS}
\end{figure}

Table~\ref{tab:CMRanked} shows the ranking of test sequences according to the average scores across all models except IO and GAUSS.\footnote{The full ranking across the computational models and the IO model are separately provided in the supplementary material~\cite{CSEsupplementary}.} The sequences are ranked in decreasing order of average scores -- the highest-ranked sequences are those for which the average scores are highest, and therefore seem to be the easiest for saliency prediction. Meanwhile, the lowest-ranked sequences are those for which saliency prediction seems the most difficult. Although the ranking differs somewhat for different metrics, overall, \textit{one-show}, \textit{advert-bbc4-library}, \textit{Stefan} and \textit{Mobile Calendar} seem to be among the easiest sequences for saliency prediction, while \textit{City} and \textit{Tempete} are among the hardest. \textit{one-show}, \textit{advert-bbc4-library} and \textit{Stefan} have only one salient object, and all models are generally able to correctly identify them. \textit{Mobile Calendar} contains several moving objects, including a ball and a train. The motion of each of these is sufficiently strong and different from the surroundings that almost all models are able to correctly predict viewers' gaze locations. It should be noted that the background of this sequence involves many static colorful regions that, in the absence of motion, would have the potential to attract attention. It is encouraging that the compressed-based models are generally able to identify the salient moving objects against such colorful and potentially attention-grabbing background. Meanwhile, the two pixel-domain models show a relatively poor performance on this sequence. 

On the other hand, \textit{City} and \textit{Tempete} do not contain salient moving objects. In fact, \textit{City} does not contain any moving objects; all the motion in this sequence is due to camera movement. \textit{Tempete} also contains significant camera motion (zoom out) and in addition shows falling yellow leaves that act like motion noise, as they do not attract viewers' attention. While all models get confused by the falling leaves in \textit{Tempete}, APPROX achieves a decent performance on \textit{City} due to its use of global motion compensation (GMC). APPROX is the only model in the study that employs GMC and its success on \textit{City} is an indication that other models could be improved by incorporating GMC. Note that AWS also scores well on \textit{City} because, as a spatial saliency model, it ignores motion and therefore does not get confused by it in this sequence.

\begin{table}[t]
	\centering
	%\scriptsize 
	\caption{\small Ranking of MPEG-4 ASP encoded test sequences according to average scores across all models excluding IO and GAUSS.}
	\begin{tabular}{ccccc}
	\hline \hline
	\textbf{Rank} & \textbf{AUC$'$} & \textbf{JSD$'$} & \textbf{NSS$'$} & \textbf{PCC} \\
	\hline \hline
1	&	\textit{os}	&	\textit{os}	&	\textit{abl}	&	\textit{Stefan}	\\
2	&	\textit{abl}	&	\textit{Mobile}	&	\textit{Mobile}	&	\textit{mtnin}	\\
3	&	\textit{Mobile}	&	\textit{Stefan}	&	\textit{mtnin}	&	\textit{abl}	\\
4	&	\textit{Stefan}	&	\textit{Hall}	&	\textit{os}	&	\textit{Mobile}	\\
\vdots & \vdots & \vdots & \vdots & \vdots \\
30	&	\textit{tucf}	&	\textit{ufci}	&	\textit{tucf}	&	\textit{City}	\\
31	&	\textit{Tempete}	&	\textit{nim}	&	\textit{Tempete}	&	\textit{pnb}	\\
32	&	\textit{City}	&	\textit{pnb}	&	\textit{City}	&	\textit{Tempete}	\\
		\hline \hline
	\end{tabular}
	\label{tab:CMRanked}
\end{table}

\begin{figure}
\centering
\includegraphics[width=.8\linewidth]{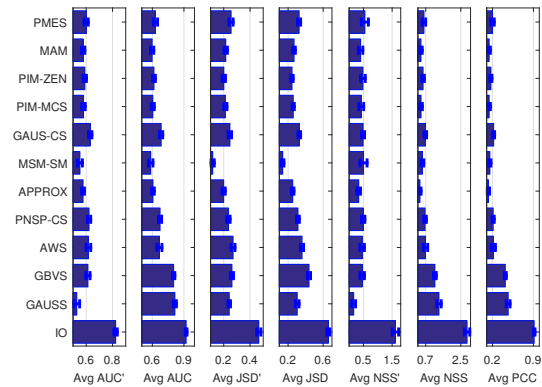} 
\caption{Evaluation of models depending on MPEG-4 ASP video bitstream using various metrics.}
\label{fig:AvgModels}
\end{figure}

The average scores of MPEG-4 ASP based saliency models across all sequences in both datasets are shown in Fig.~\ref{fig:AvgModels} for various accuracy metrics. Please note that the horizontal axis has been focused on the relevant range of scores. Not surprisingly, IO achieves the highest scores regardless of the metric. At the same time, the effect of center bias is easily revealed by comparing AUC and NSS scores to their center bias-corrected versions AUC$'$ and NSS$'$. For example, the AUC measures the accuracy of saliency prediction of a particular model against a control distribution drawn uniformly across the frame. Since the uniform distribution is a relatively poor control distribution for saliency and easy to outperform, all models achieve a higher AUC score compared to their AUC$'$ score, which uses a control distribution fitted to the empirical gaze points shown in Fig.~\ref{fig:shuffle}. This effect is most visible in the GAUSS benchmark model, which has the AUC score of around $0.8$ (higher than all the models except IO), but the AUC$'$ score of only slightly above $0.5$ (lower than all other models). This over-exaggeration of the accuracy of a simple scheme such as GAUSS when plain AUC used was the reason why~\cite{parkhurst03scene,kanan09sun} suggest center bias correction via non uniform control sampling. The center bias-corrected AUC$'$ score is a better reflection of the models' performance. Center bias also has a significant effect on NSS, but a less pronounced effect on JSD. It can also be observed that GAUSS (and then GBVS) achieves a higher PCC score than any other method except IO, due to the accumulation of fixations near the center of the frame. 

Thus far, we showed the results for saliency models that accept MPEG-4 ASP encoded bitstream. The accuracy assessment according to AUC$'$ and NSS$'$ over the saliency models that accept H.264/AVC-encoded data is shown in Fig.~\ref{fig:H264-Accuracy}.\footnote{The same results according to JSD$'$ and PCC are shown in supplementary material~\cite{CSEsupplementary}.} 
Two recent compressed-domain methods, MVE+SRN and OBDL-MRF, top all other methods, including pixel-domain ones, on both metrics. Based on these results, MVE+SRN seems like the best saliency predictor, while OBDL-MRF comes a closes second. Both of these models have been built upon compressed-domain features that are highly correlated with human gaze, and are therefore able to compete even with high-performing pixel-domain models such as GBVS and AWS.

\begin{figure}
\centering
\includegraphics[width=.8\linewidth]{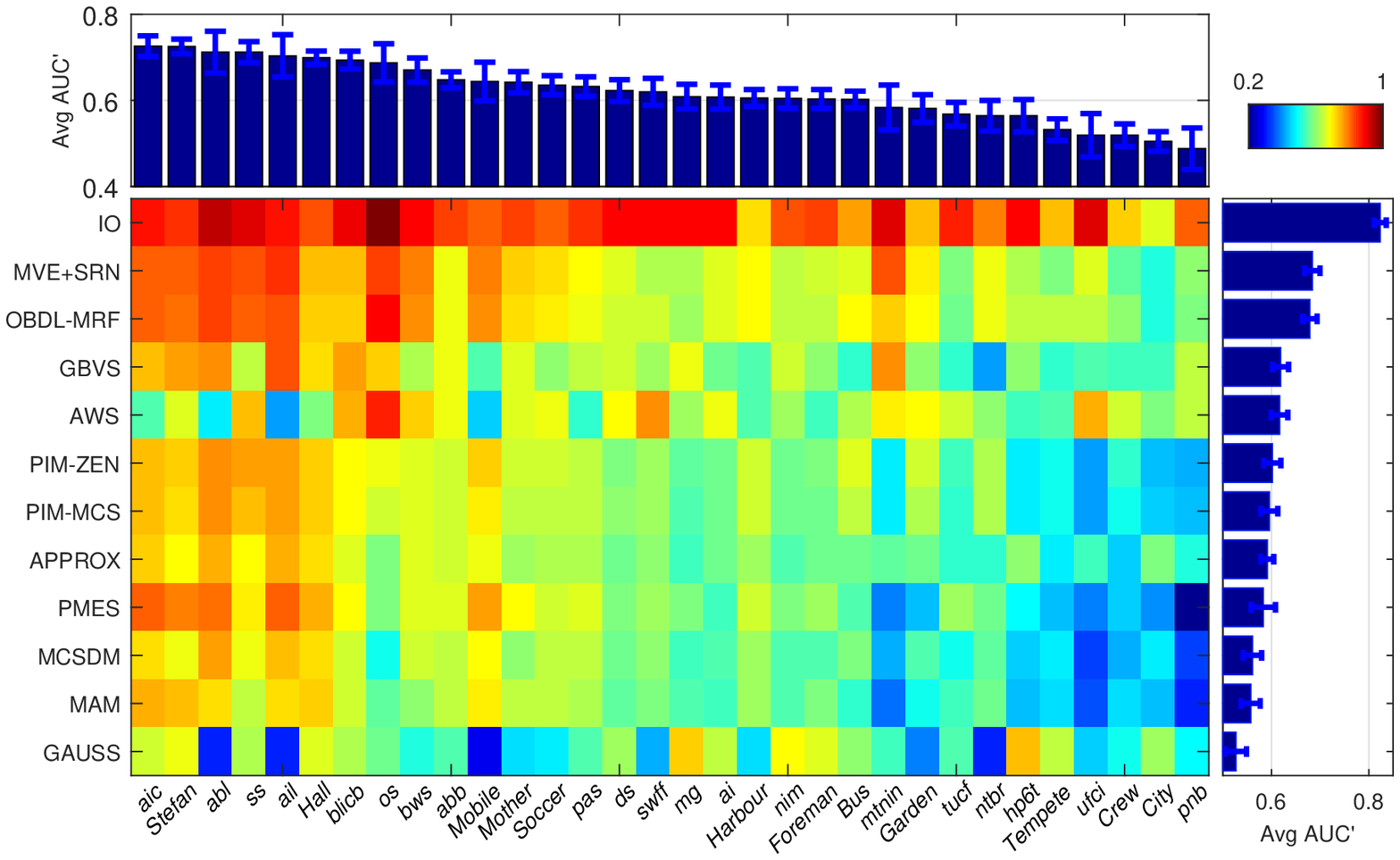} \\
\includegraphics[width=.8\linewidth]{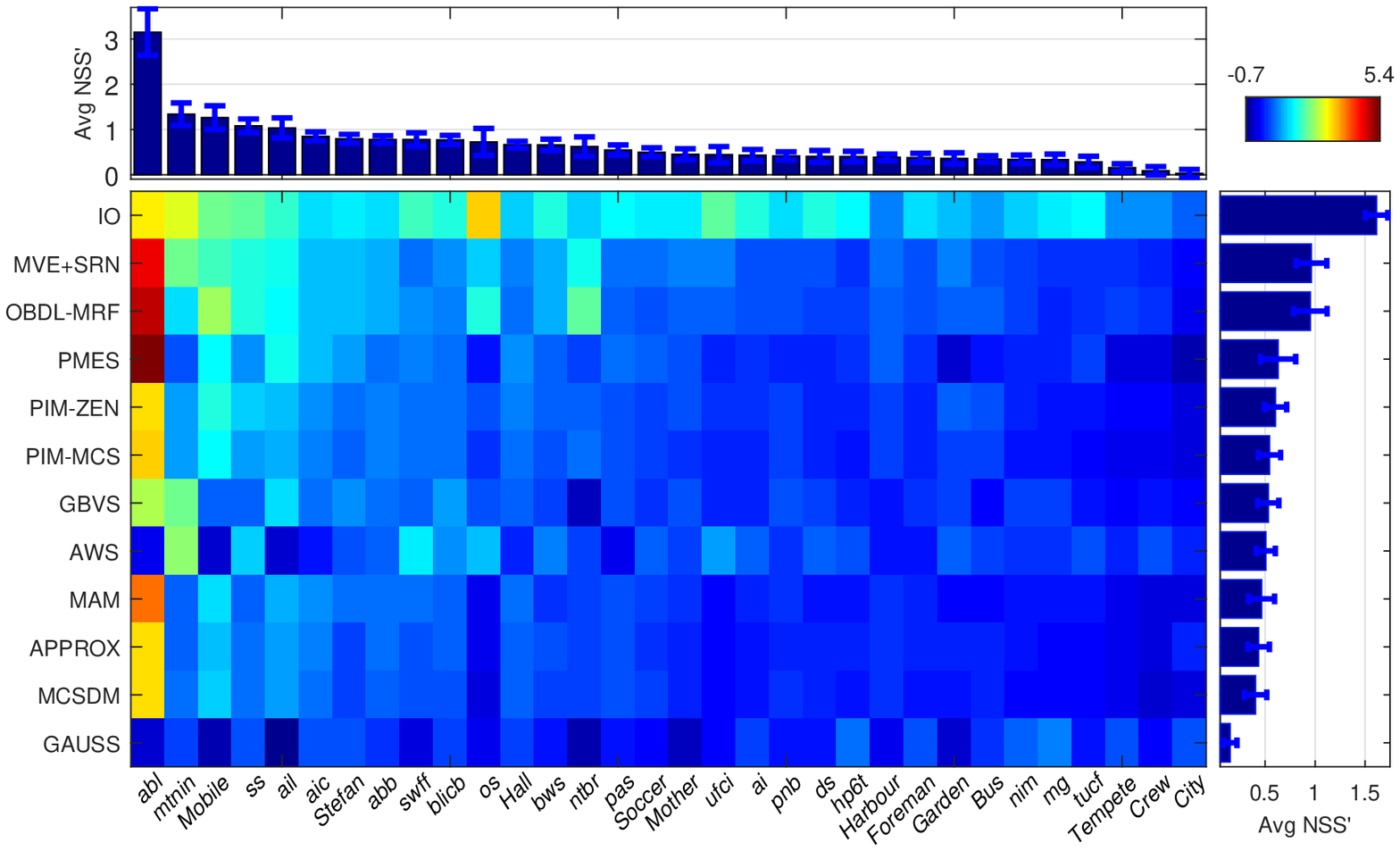} \\
\caption{Accuracy of various saliency models over H.264/AVC encoded bitstream of SFU and DIEM dataset according to AUC$'$ and NSS$'$.}
\label{fig:H264-Accuracy} 
\end{figure}

\begin{figure}
\centering
\includegraphics[width=.8\linewidth]{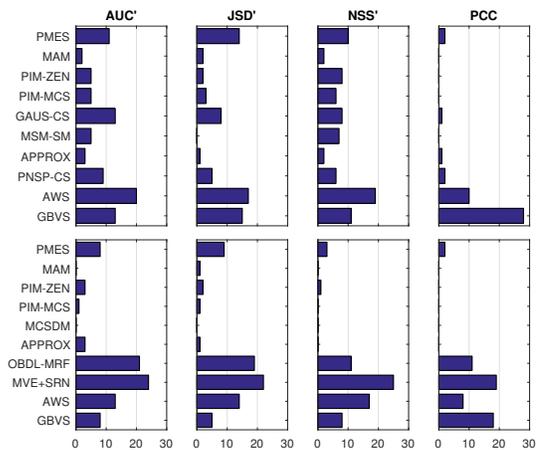} 
\caption{The number of appearances among top performers, using various evaluation metrics. Results based on MPEG-4 ASP are shown at the top, those based on H.264/AVC at the bottom.}
\label{fig:bestScores}
\end{figure}

In addition to the average scores, another type of assessment of a model's performance is counting its number of appearances among top performing models for each sequence~\cite{mateescu12evaluation}. To this end, a multiple comparison test is performed using Tukey's honestly significant difference as the criterion~\cite{hochberg87multiple}. Specifically, for each sequence, we compute the average score of a model across all frames, as well as the 95$\%$ confidence interval for the average score. Then we find the model with the highest average score (excluding IO), and find all the models whose 95$\%$ confidence interval overlaps that of the highest-scoring model. All such models are considered top performers for the given sequence. The number of appearances among top performers for each model is shown in Fig.~\ref{fig:bestScores}. These results show similar trends as average scores, with MVE+SRN, OBDL-MRF, AWS, GBVS, PMES, GAUS-CS and PNSP-CS often being among top performers, while MCSDM, MAM and APPROX rarely offering top scores. 

\subsection{Sensitivity to Compression}
\label{sec:sensitivity_result}
\begin{figure}
\centering
\includegraphics[width=.8\linewidth]{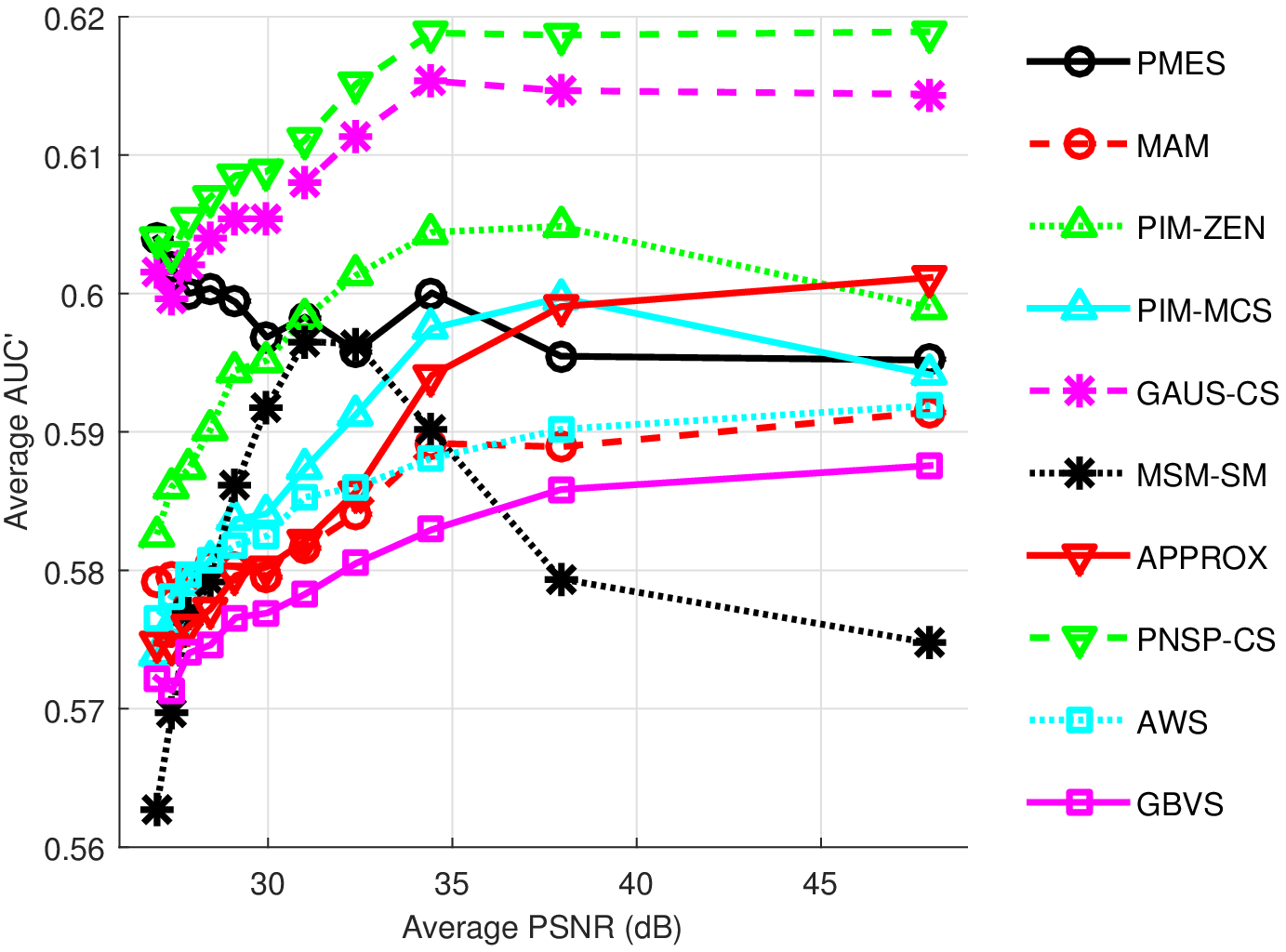}
\includegraphics[width=.8\linewidth]{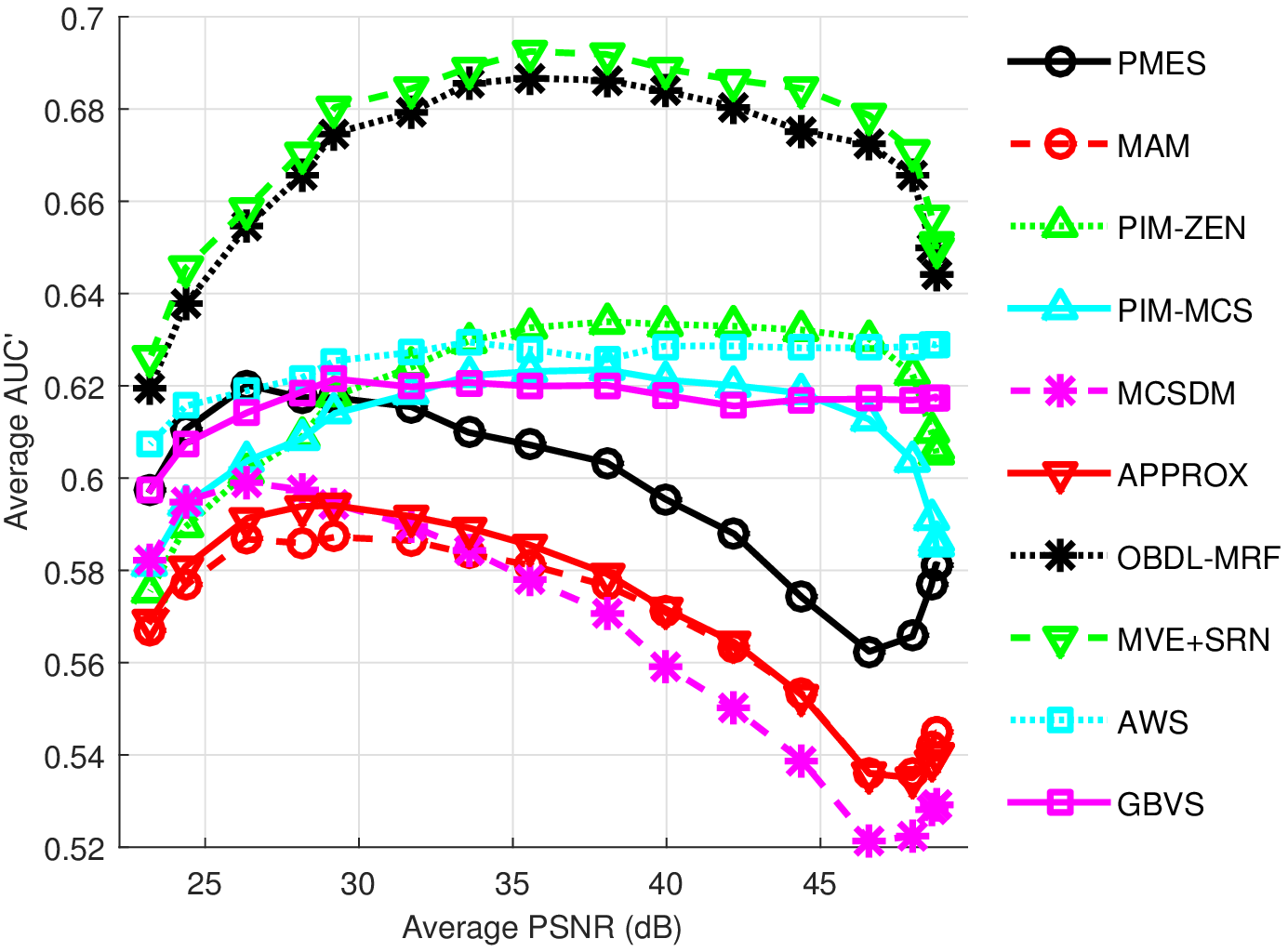}
\caption{The relationship between the average PSNR and the models' accuracy over SFU and DIEM dataset. Top: the sensitivity over MPEG-4 ASP encoded bitstream, Bottom: the sensitivity over H.264/AVC encoded bitstream.}
\label{fig:Per-PSNR}
\end{figure}

In the assessments presented thus far, the QP value was set to a constant value (in MPEG-4 ASP, QP~=~16 and in H.264/AVC, QP~=~36). The quality of encoded video drops as the QP increases due to the larger amount of compression. Fig.~\ref{fig:Per-PSNR} shows how the average AUC$'$ score changes as a function of the average PSNR by varying QP$\in$\{1, 4, 7, ..., 31\} for MPEG-4 ASP (top figure) and varying QP$\in$\{3, 6, 9, ..., 51\} for H.264/AVC (bottom figure).\footnote{The relationship between the average NSS$'$ score and the average PSNR is provided in the supplementary material~\cite{CSEsupplementary}.} The results in Fig.~\ref{fig:Per-PSNR} indicate the sensitivity of the models' saliency prediction relative to encoding parameters. In this experiment, AWS and GBVS were applied to the decoded video, hence they effectively used the same data as compressed-domain models, but in the pixel domain after full video reconstruction. 

The figure shows that pixel-domain models GBVS and AWS score lower at low video qualities, while their accuracy improves as the video quality increases. Their accuracy is fairly consistent beyond a certain level of video quality, around 35~dB, suggesting that so long as video quality is sufficiently high, compression does not affect the models' ability to estimate saliency. This observation is consistent with studies undertaken by Le Meur~\cite{le11robustness}, and Milanfar and Kim~\cite{kim13visual}.

Compressed-domain models exhibit a somewhat different behavior. Their accuracy is also generally low at low video qualities, because MVs are less accurate and there is a large amount of quantization noise present in prediction residuals. But unlike pixel-domain models, compressed-domain models also seem to suffer at high video qualities. As the quality increases, compressed domain features become less informative. Small quantization step size makes most transform coefficients non-zero, which makes some of the models predict high spatial saliency throughout the frame. At the same time, MVs may become too noisy, since rate-distortion optimization does not impose sufficient constraints on motion estimation. The results suggest that the PSNR range in which most compressed-domain saliency models tend to be most accurate is 30-40 dB, which also happens to be a range in which a good trade-off is thought to be achieved between video quality and the required bitrate.

\subsection{Complexity}
\label{sec:complexity_result}
The average processing time per frame on the SFU dataset (CIF resolution videos at 30 fps) using two different input formats is listed in Table~\ref{tab:complexity}. The time taken for extracting MVs and DCT values from the bitstream is excluded. Please note that these results correspond to MATLAB implementations of the models and the processing time can be significantly decreased by implementation in a low-level programming language such as C/C++. Despite this, some of the models are fast enough for real time performance (under 33~ms per frame) even when implemented in MATLAB. Discussion of accuracy and complexity of the models is presented in the next section.

\begin{table*}
%\scriptsize
\renewcommand{\arraystretch}{1.3}
	\centering
	\caption{\small Average processing time in milliseconds per frame.}
	\begin{tabular}{c|ccccccccccccc}
	\hline \hline
	\rotatebox{90}{\textbf{Model}}  & \rotatebox{90}{AWS}  & \rotatebox{90}{GBVS}  & \rotatebox{90}{MAM}  & \rotatebox{90}{PMES}  & \rotatebox{90}{GAUS-CS~}  & \rotatebox{90}{PNSP-CS~}  & \rotatebox{90}{PIM-ZEN~}  & \rotatebox{90}{OBDL-MRF~}  & \rotatebox{90}{MVE+SRN}  & \rotatebox{90}{APPROX}  & \rotatebox{90}{MCSDM}  & \rotatebox{90}{PIM-MCS~}  & \rotatebox{90}{MSM-SM} \\
	\hline

	\textbf{MPEG-4}  & 1650  & 864  & 155  & 94  & 64  & 64  & 20  &  &  & 12  &  & 9  & 7 \\	

	\textbf{H.264}  & 1468  & 806  & 600  & 477  &  &  & 42  & 35  & 18  & 16  & 8  & 5  & \\	
	\hline \hline
	\end{tabular}
	\label{tab:complexity}
\end{table*}

%%%%%%%%%%%%%%%%%%%%
% Section V
%%%%%%%%%%%%%%%%%%%%
\section{Discussion}
\label{sec:discussion}

Considering the results in Fig.~\ref{fig:AvgModels} and Fig.~\ref{fig:bestScores}, MVE+SRN, OBDL-MRF, AWS, GBVS, PMES and GAUS-CS consistently achieve high scores across different metrics. It is encouraging that the performance of some compressed-domain models is superior to that of high-performing pixel-domain models. Note that, in general, achieving a high score with one metric does not guarantee a high score with other metrics. As an example, MSM-SM achieves a relatively high average scores across several metrics, but the lowest JSD and JSD$'$ score. Hence, the fact that MVE+SRN, OBDL-MRF, AWS, GBVS, PMES and GAUS-CS perform consistently well across all metrics considered in this study lends additional confidence in their accuracy.

PMES was the first compressed-domain saliency model, proposed in 2001, and it only uses MVs to estimate saliency. It is well known that motion is a strong indicator of saliency in dynamic visual scenes~\cite{milanese95attentive,mahadevan10spatiotemporal,itti09bayesian}, so it is not surprising that MVs would be a powerful cue for saliency estimation. PMES estimates saliency by considering two properties: large motion magnitude in a spatio-temporal region, and the lack of coherence among MV angles in that region. These two properties seem to describe salient objects reasonably well in most cases, as demonstrated by the results. Taken together, they resemble a center-surround mechanism where a region is considered salient if it sufficiently ``stands out'' from its surroundings.  

GAUS-CS and PNSP-CS show high performance in both I- and P-frames. Both models are based on the center-surround difference mechanism, and both employ MVs for saliency estimation in P-frames and DCT of pixel values in I-frames. The capability of center-surround difference mechanism to predict where people look has been discussed extensively~\cite{itti98model}, so their success is also not surprising. 

Although PIM-MCS and MSM-SM also attempt to employ the center-surround difference mechanism, their scores are not as consistently high as those of GAUS-CS and PNSP-CS. The reason may be that in GAUS-CS and PNSP-CS models, the contrast is inversely proportional to the distance between the current DCT block and all other DCT blocks in the frame, which means that they consider not only the contrast between blocks, but also the distance between them. This seems to be a good strategy for compressed-domain saliency estimation. 

OBDL-MRF and MVE-SRN are two of the most recent compressed-domain saliency models. Taking advantage of the availability of gaze point data for video, which was not the case when earliest models such as PMES were developed, both OBDL-MRF and MVE-SRN were built upon compressed-domain features that have been shown to be highly correlated with gaze points in video. Their advantage over other compressed-domain models is therefore not surprising. What is perhaps surprising is their ability to go toe-to-toe with the best pixel-domain models, and be more accurate in many cases. Their success lends further support to the hypotheses that relate saliency to compressibility~\cite{bruce06saliency,itti09bayesian}, although their operational realization is quite different from these earlier works.

According to the results in the previous section, the lowest-scoring models on most metrics were APPROX and MCSDM. Incidentally, APPROX was originally developed for a different type of input data and had to be modified for this comparison, which may have had a negative impact on its performance.
%It was developed for MVs corresponding to $4 \times 4$ blocks, whereas the evaluation in this work employed MVs corresponding to $8 \times 8$ blocks in case the input data is MPEG-4 ASP bitstream. Additionally, APPROX originally assumed DCT coefficients of $16 \times 16$ blocks from a raw (uncompressed) frame, whereas in this evaluation, $16 \times 16$ DCT was computed from four $8 \times 8$ DCTs of compressed frames, which involved quantization noise.}
%Looking at the results in Figs.~\ref{fig:AUC} and~\ref{fig:NSS}, the gap between GAUS-CS (PNSP-CS) and APPROX is smaller in the top parts of the figures (I-frames) than in the bottom parts (P-frames), which indicates that the effect of quantization noise was not as detrimental to the performance of APPROX as the switch from $4 \times 4$ MVs to $8 \times 8$ MVs. 

The influence of global (camera) motion on visual saliency is still a fairly open research problem, with limited work in the literature addressing this issue. Reference~\cite{Abdollahian08camera} studied separately the effect of pan/tilt and zoom-in/-out. It was found that in the case of pan/tilt, the gaze points tend to shift towards the direction of pan/tilt, in the case of zoom-in, they tend to concentrate near the frame center, and in the case of zoom-out, they tend to scatter further out. On the other hand, according to~\cite{Baudrier09camera}, the presence of camera motion tends to concentrate gaze points around the center of the frame ``according to the direction orthogonal to the tracking speed vector.'' 

Among the models tested in the present study, only APPROX took global motion into account by removing it prior to the analysis of MVs. This paid off in the case of \textit{City}, which was overall the most difficult sequence for other spatio-temporal saliency models in Figs.~\ref{fig:AUC} and~\ref{fig:NSS}. However, global motion compensation (GMC) did not help much in the case of \textit{Tempete} or \textit{Flower Garden}. In fact, \textit{Tempete} contains strong zoom-out, which, according to~\cite{Abdollahian08camera}, would tend to scatter the gaze points around the frame. However, Figs.~\ref{fig:AUC} and~\ref{fig:NSS} show that GAUSS, with its simple center-biased saliency map, scores well here (even with center-bias-corrected metrics), suggesting that the gaze points are still located near the center of the frame. This is due to the presence of a yellow bunch of flowers in the center of the frame, which turns out to be highly attention-grabbing. Apparently, the key to accurate saliency estimation in \textit{Tempete} is not in the motion, but rather in the color present in the scene. \textit{Flower Garden} is another example where GMC did not pay off. The viewers' gaze in this sequence is attracted to the objects in the background, specifically the windmill and the pedestrians, whose motion tends to be zeroed out after GMC on $8 \times 8$ MVs. Overall, the results suggest that global motion is not sufficiently well handled by current compressed-domain methods, and that further research is needed to make progress on this front.  

Considering models' complexity and processing time in Table~\ref{tab:complexity}, MSM-SM, PIM-MCS and MCSDM are the fastest while AWS is the most demanding. Note that the smallest block size is 4$\times$4 in H.264/AVC encoded bitstream and 8$\times$8 in MPEG-4 ASP encoded bitstream, and therefore more data typically needs to be processed in the H.264/AVC case, which is why compressed-domain models that are able to accept both input formats tend to take more time when applied on H.264/AVC bitstreams. MSM-SM, PIM-MCS and MCSDM are the least complex models, but unfortunately not the most accurate.

While MVE+SRN and OBDL-MRF scored the highest in terms of accuracy, this did not come at a cost of high complexity. In fact, according to complexity, they are in the middle of the pack, with processing times below those of other high-performing saliency models.
 %-- only around half that of GAUS-CS and one twentieth that of PMES. 
MVE+SRN appears twice as fast as OBDL-MRF because entropy decoding time of MVs and DCT residuals was not taken into account (as with other compressed-domain models). But OBDL-MRF does not require any such decoding and would therefore likely end up being faster in a real-world scenario.

%%%%%%%%%%%%%%%%%%%%
% Section VI
%%%%%%%%%%%%%%%%%%%%
\section{Conclusions}
\label{sec:conclusions}

In this study we attempted to provide a comprehensive comparison of eleven compressed-domain visual saliency models for video. All methods were reimplemented in MATLAB and tested on two eye-tracking datasets using several accuracy metrics. Care was taken to correct for center bias and border effects in the employed metrics, which were issues found in earlier studies on visual saliency model evaluation. 
The results indicate that reasonably accurate visual saliency estimation is possible using only a limited set of data from the compressed bitstream, such as motion vectors, prediction residuals, or even just the number of bits per block, without further decoding. 
Several compressed-domain saliency models showed competitive accuracy with some of the best currently known pixel-domain models. On top of that, some of the compressed-domain methods are fast enough for real-time saliency estimation on CIF video even with a relatively inefficient MATLAB implementation, which suggests that their optimized implementation could be used for online saliency estimation in a variety of applications, even for higher-resolution video.

Many sequences that have turned out to be difficult for models to handle contain global (camera) motion. The influence of global motion on visual saliency is not very well understood, and most models in the study did not account for it. A number of compressed-domain global motion estimation methods, based on motion vectors alone, have been developed recently, so it is reasonable to expect that compressed-domain saliency models should be able to benefit from these developments.   

\bibliographystyle{IEEEtran}
\bibliography{CSE_arXiv}

\end{document}